\documentclass[iop]{emulateapj}
\usepackage{graphicx}
\usepackage{color}

\shorttitle{Slipping Reconnection, Chromospheric Evaporation, Implosion, and Precursors in X-class Flare}
\shortauthors{Dud\'ik et al.}

\begin{document}
\title{Slipping Magnetic Reconnection, Chromospheric Evaporation, Implosion, and Precursors in the 2014 September 10 X1.6-Class Solar Flare}
\author{Jaroslav Dud\'ik\altaffilmark{1,2}}\email{dudik@asu.cas.cz}
\author{Vanessa Polito\altaffilmark{3}}
\author{Miho Janvier\altaffilmark{4}}
\author{Sargam M. Mulay\altaffilmark{3}}
\author{Marian Karlick\'y\altaffilmark{1}}
\author{Guillaume Aulanier\altaffilmark{5}}
\author{Giulio Del Zanna\altaffilmark{3}}
\author{Elena Dzif\v{c}\'akov\'a\altaffilmark{1}}
\author{Helen E. Mason\altaffilmark{3}}
\author{Brigitte Schmieder\altaffilmark{5}}

\altaffiltext{1}{Astronomical Institute of the Academy of Sciences of the Czech Republic, Fri\v{c}ova 298, 251 65 Ond\v{r}ejov, Czech Republic}
\altaffiltext{2}{RS Newton Alumnus}
\altaffiltext{3}{Department of Applied Mathematics and Theoretical Physics, CMS, University of Cambridge, Wilberforce Road, Cambridge CB3 0WA, United Kingdom}
\altaffiltext{4}{Institut d'Astrophysique Spatiale, Centre Universitaire d’Orsay, Bât 120 – 121, 91405 Orsay Cedex, France}
\altaffiltext{5}{LESIA, Observatoire de Paris, PSL Research University, CNRS, Sorbonne Universités, UPMC Univ. Paris 06, Univ. Paris-Diderot, Sorbonne Paris Cité, 5 place Jules Janssen, F-92195 Meudon, France}

\begin{abstract}
 We investigate the occurrence of slipping magnetic reconnection, chromospheric evaporation, and coronal loop dynamics in the 2014 September 10 X-class flare. The slipping reconnection is found to be present throughout the flare from its early phase. Flare loops are seen to slip in opposite directions towards both ends of the ribbons. Velocities of 20--40 km\,s$^{-1}$ are found within time windows where the slipping is well resolved. The warm coronal loops exhibit expanding and contracting motions that are interpreted as displacements due to the growing flux rope that subsequently erupts. This flux rope existed and erupted before the onset of apparent coronal implosion. This indicates that the energy release proceeds by slipping reconnection and not via coronal implosion. The slipping reconnection leads to changes in the geometry of the observed structures at the \textit{IRIS} slit position, from flare loop top to the footpoints in the ribbons. This results in variations of the observed velocities of chromospheric evaporation in the early flare phase. Finally, it is found that the precursor signatures including localized EUV brightenings as well as non-thermal X-ray emission are signatures of the flare itself, progressing from the early phase towards the impulsive phase, with the tether-cutting being provided by the slipping reconnection. The dynamics of both the flare and outlying coronal loops is found to be consistent with the predictions of the standard solar flare model in 3D.

\end{abstract}
\keywords{Magnetic reconnection -- Sun: UV radiation -- Sun: flares -- Sun: transition region}

%
\section{Introduction}
\label{Sect:1}

Solar flares are local energetic and even explosive phenomena within the solar atmosphere, exibiting a fast increase of radiation throughout the electromagnetic spectrum \citep[e.g.,][]{Kane74,Fletcher11,White11} as well as a wealth of dynamic phenomena including ejections of material into the interplanetary space \citep[e.g.,][]{vanB89,Dere99,Amari00,Moore01,Green09,Green11,Zhang12,Patsourakos13,vanDriel14}. A flare typically involves nearly all local regions of the solar atmosphere, from the chromosphere to the transition region and the corona. The flare emission originates dominantly in hot flare loops with temperatures above 10\,MK, anchored in bright chromospheric ribbons \citep[e.g.,][]{Schmieder96,Warren01,Fletcher11,Graham11,Reid12,Young13,Inglis13,Dudik14a,Doschek15}.

The energy powering solar flares is believed to be released via the mechanism of magnetic reconnection \citep[e.g.,][]{Dungey53,Parker57,Sweet58,Priest00,Zweibel09}, which is a process involving mutual annihilation of oppositely-oriented components of magnetic field lines. In the standard solar flare model in 2D, sometimes called the CSHKP model \citep{Carmichael64,Sturrock66,Hirayama74,Kopp76} the reconnection happens at the magnetic null-point, where the magnetic field is locally zero. Oppositely oriented field lines are brought together by flows into the vicinity of the null-point, where they reconnect. The post-reconnected magnetic field forms a growing system of flare loops and the erupting flux rope. This model succeeds in explaining many flare phenomena \citep[e.g.,][]{Shibata95,Tsuneta97,Shibata01}, such as plasma heating and particle acceleration. Nevertheless, this 2D model fails to explain the inherently 3D properties of flares, such as the shear of flare loops and its strong-to-weak evolution during the course of the flare \citep{Aulanier12}, movements of EUV or X-ray sources along chromospheric flare ribbons \citep[e.g.,][]{Tripathi06,Chifor07,Inglis13}, or the morphology of flare ribbons that are often seen as a J-shaped structure \citep[e.g.,][]{Chandra09}.

To remedy this, the standard solar flare model has been extended into 3D in the recent years \citep{Aulanier12,Aulanier13,Janvier13}. This standard solar flare model in 3D is a presureless MHD model that includes a torus-unstable magnetic flux rope \citep{Torok04,Aulanier10} located in a generic sigmoidal solar active region \citep[see also][]{Green09,Tripathi09,Savcheva12a,Savcheva12b,Savcheva14}. Since the flux rope is unstable, it undergoes an eruption as the result of the torus instability. The restructuring of the magnetic field during the rise and eruption of the flux rope involves the development of a current sheet beneath the flux rope, where the magnetic reconnection can proceed. However, in this 3D model the current layer does not originate in the vicinity of a magnetic null-point (or any other topological discontinuity) as in the 2D CSHKP model. Rather, the current layer is associated with the quasi-separatrix layers \citep[QSLs;][]{Priest95,Demoulin96a,Titov02}, where the magnetic connectivity has strong gradients, but is still continuous. It has been found that the photospheric traces of such QSLs correspond well with the observed flare ribbons \citep[e.g.,][]{Demoulin97,Savcheva15,Zhao16}. The distortion of the magnetic field within the QSLs gives rise to electric currents parallel to the magnetic field \citep{Masson09,Wilmot09} thereby fulfilling the necessary condition for magnetic reconnection in 3D \citep{Hesse88}. Reconnection within the QSLs is of a slipping nature: local rotation of the magnetic field and the continuous exchange of connectivities of neighbouring field lines within the coronal diffusive region induce apparent velocity of the entire reconnecting field line \citep{Priest95,Priest03,Aulanier06}. This process is exhibited as an apparent slipping motion of the field line footpoints in the photospheric QSL traces. Since the QSLs generalize the concept of separatrices, which are true topological discontinuities within the magnetic field, the slipping reconnection in QSLs is a generalized mechanism of magnetic reconnection in 3D.

The standard solar flare model in 3D, featuring the slipping reconnection, has so far withstood several important observational tests. \citet{Aulanier12} found that the observed strong-to-weak shear transition of the flare loop arcade can be explained by the original shear of the reconnecting coronal loops, as well as the stretching effect of the erupting flux rope that decreases the shear of field lines reconnecting higher up in the solar atmosphere. Using photospheric vector magnetograms, \citet{Janvier14} identified the photospheric footprints  of the 3D coronal electric current layer and found that these match well the observed flare ribbons, in agreement with the model. The slipping motion of individual flare loops, reported first by \citet{Dudik14a}, was found to match the model-predicted morphology and dynamics of slipping magnetic field lines. The slipping motion was however observed predominantly in one direction, toward the hook (elbow) of the flare ribbon away from the inversion line. This slipping motion was also found to contribute to the buildup of the erupting flux rope, as well as possibly to smaller magnetic structures formed in the current layer such as plasmoids that are manifested in the radio emission \citep[e.g.,][]{Kliem00,Kolomanski07,Karlicky02,Karlicky10,Karlicky14,Nishizuka15}. Since the first report, occurrence of the slipping reconnection was reported in several flares \citep{Li14,Li15}. However, the standard solar flare model in 3D predicts slipping motion of both the flare loops and the flux rope field lines. Therefore, the slipping motion motion should be observed in both directions along both flare ribbons: flux rope field lines slipping towards the hooks, while the flare loops slip in the opposite direction toward the straight part of the ribbons. The latter motion in the opposite direction has however been largely absent in the observational reports so far \citep{Dudik14a,Li14,Li15}.

In addition to slipping motion, solar flares exhibit many other dynamic phenomena such as the occurrence of precursors before the impulsive phase \citep[e.g.,][]{Bumba59,Harrison85,Harrison86,Farnik98,Farnik03,Sterling05,Chifor06,Chifor07}, the evaporation of chromospheric plasma filling post-reconnection field lines \citep[e.g.,][]{Neupert68,Raftery09,Milligan09,Brosius10,DelZanna11b,Ning11,Graham11,Doschek13,Young13,Young15,Polito15a,Tian15,Graham15}, and also the dynamics of neighbouring coronal loops, including loop expansion, contraction, and oscillations \citep[e.g.,][]{Liu09a,Liu09b,Liu10,Liu12a,Sun12,Gosain12,Simoes13a,Shen14,Imada14,Kushwaha15,Zhang15}. These phenomena have not been studied in relation to the standard solar flare model in 3D. The dynamics of coronal loops is of special interest, since contracting motions are interpreted in terms of the \textit{coronal implosion} conjecture \citep{Hudson00}. Coronal implosion is a proposed mechanism for the release of magnetic energy, proportional to $\int B^2 dV$, by the decrease of the associated coronal volume $V$ and the associated drop of magnetic pressure. Since the rate of coronal loop contraction was found to be closely associated to the hard X-ray and microwave emission of the flare \citep{Simoes13a}, the coronal implosion can be interpreted as an energy release mechanism \citep{Hudson00,Simoes13a} alternative to magnetic reconnection. We note however that the implosion is sometimes interpreted only as a \textit{response} to the energy release \citep[][]{Russell15}.

In this paper, we investigate the slipping reconnection and chromospheric evaporation in the flare from its onset well before the impulsive phase, together with their connection to the precursors and the expanding/contracting motions of the overlying corona. To do this, we examine observations of the 2014 September 10 X-class solar flare (event SOL2014-09-10T17:45). This event has been studied already by several authors \citep[][]{Li15,Cheng15,Tian15,Graham15,Zhao16}. They focused on its impulsive phase including the slipping reconnection \citep{Li15}, chromospheric evaporation \citep{Tian15,Graham15}, as well as the presence of magnetic flux ropes being built by tether-cutting reconnection before the flare \citep{Cheng15}, and the calculation of the QSLs including their photospheric traces and comparison to observed flare ribbons \citep{Zhao16}. However, each of these authors studied only separate aspects of the flare.

This paper is organized as follows. In Sect. \ref{Sect:2}, we report on the observations of the flare performed mainly by the \textit{Solar Dynamics Observatory}, including an analysis of the coronal loop dynamics and the apparent slipping motion of flare loops. Section \ref{Sect:3} deals with the characteristics and evolution of chromospheric evaporation during the flare as observed by \textit{IRIS} in connection to the slipping reconnection. The observational results are discussed with respect to the standard solar flare model in 3D in Sect. \ref{Sect:4}. There, the relation of precursor signatures to the slipping reconnection is discussed. Conclusions are summarized in Sect. \ref{Sect:5}. 

%
\begin{figure*}[ht]
	\centering
	\includegraphics[width=8.60cm,clip,bb= 0  0 495 265]{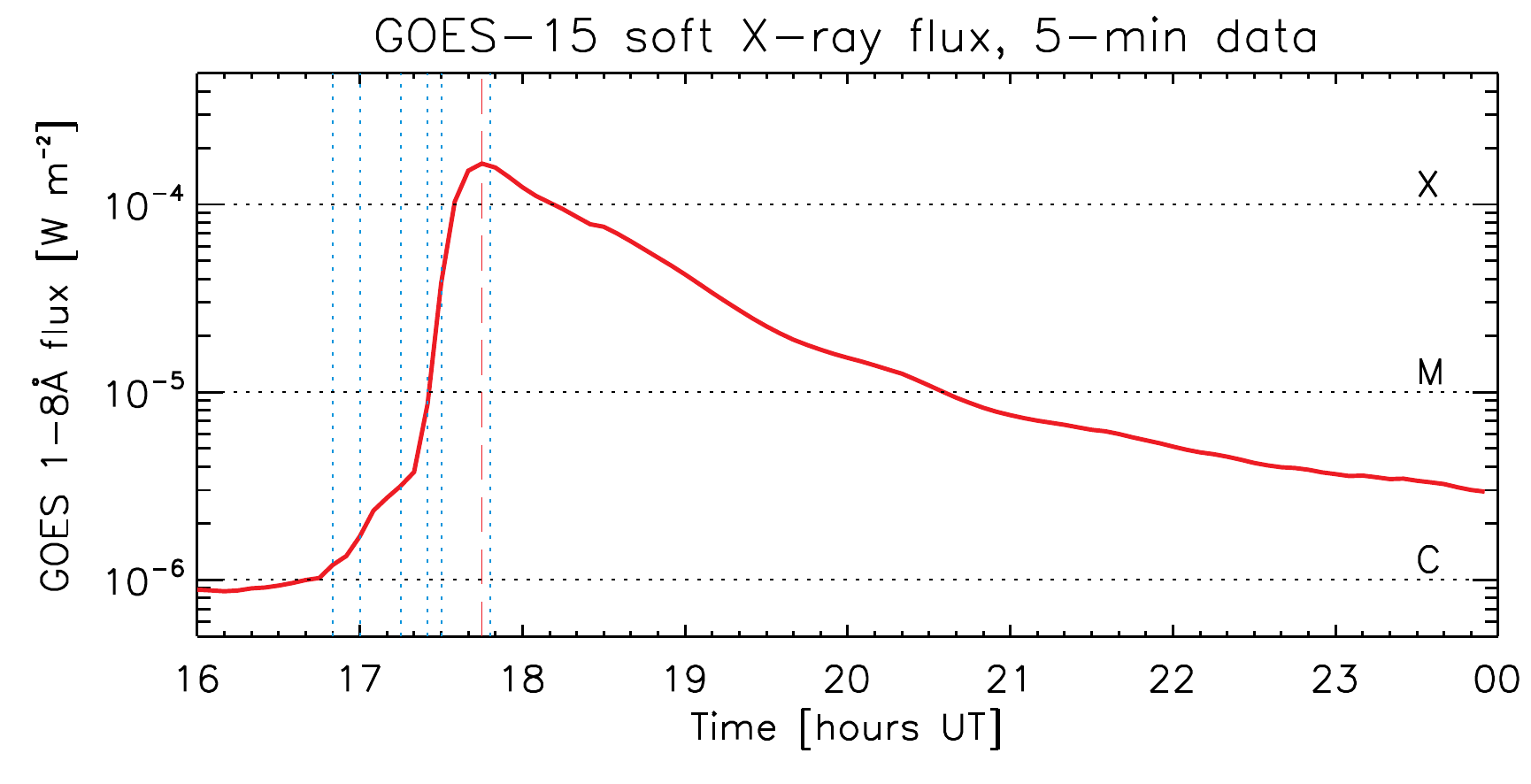}
	\includegraphics[width=8.60cm,clip,bb= 0 42 495 310]{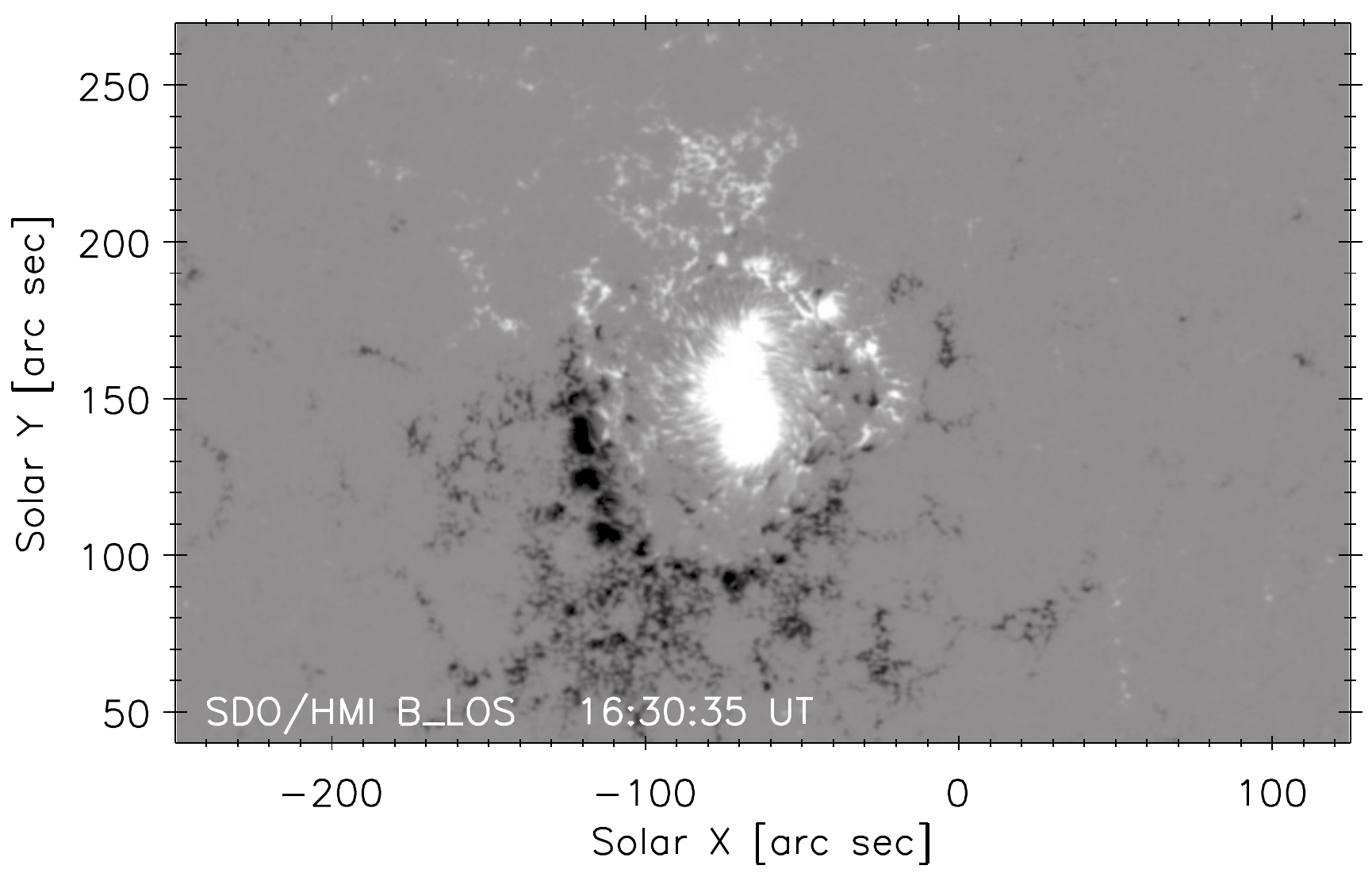}
	\includegraphics[width=8.60cm,clip,bb= 0  0 495 310]{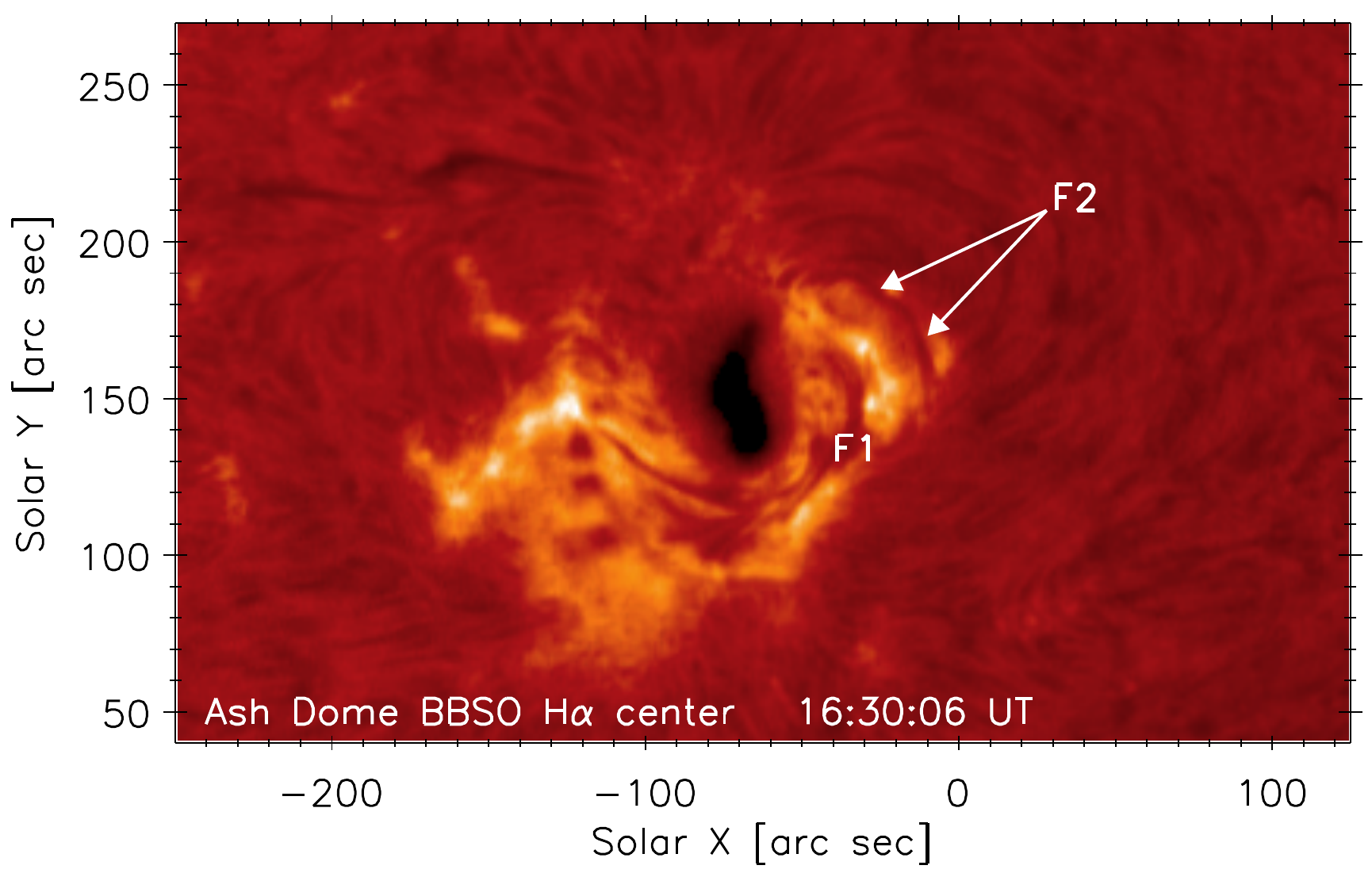}
	\includegraphics[width=8.60cm,clip,bb= 0  0 495 310]{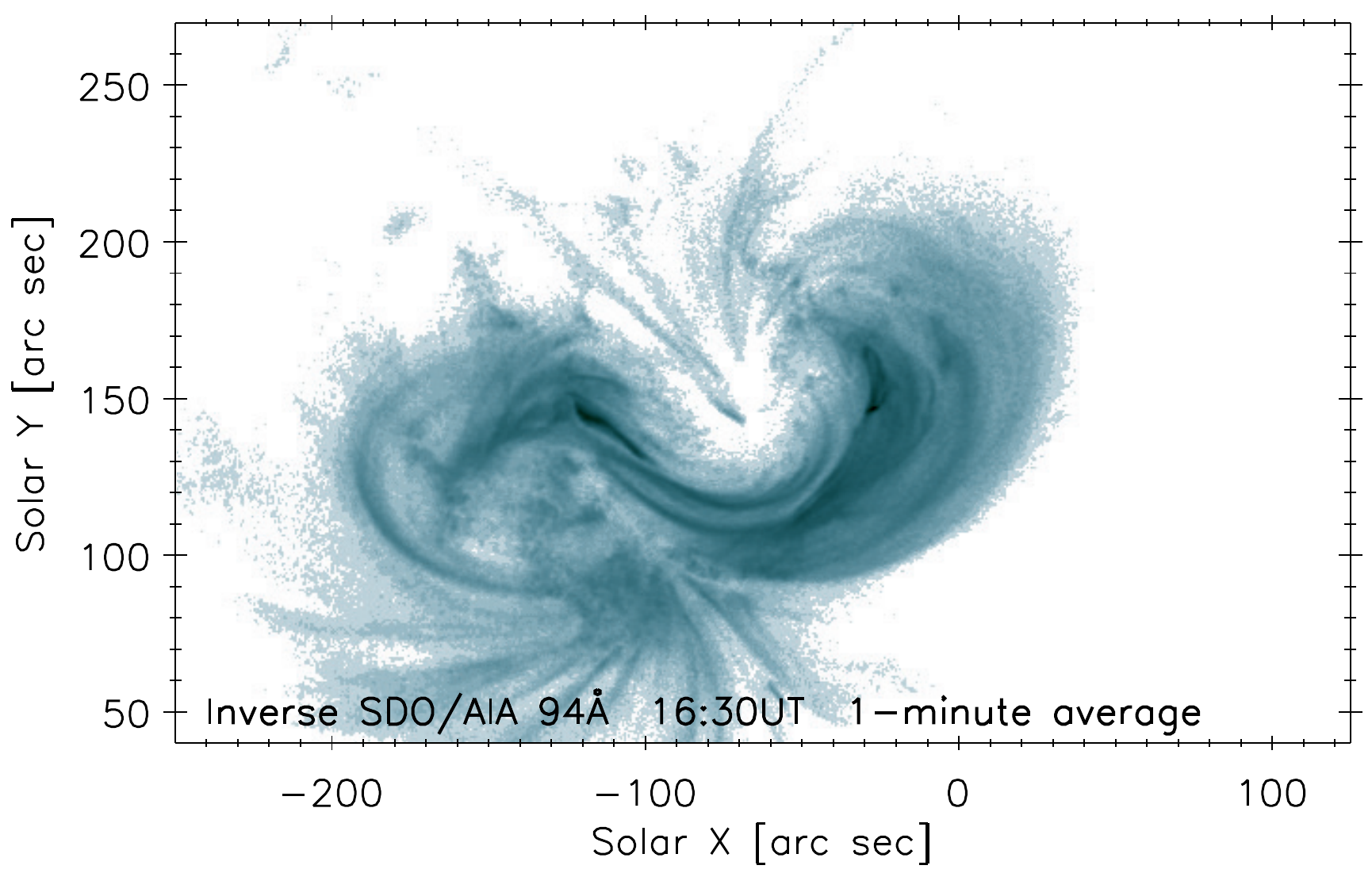}
\caption{GOES 1--8\AA~flux (\textit{top left}) and the pre-flare state of AR 12158 at 16:30 UT: the longitudinal component of the magnetic field as measured by SDO/HMI (\textit{top right}), including BBSO H$\alpha$ showing two filaments, F1 and F2 (\textit{bottom left}), and the inverted SDO/AIA 94~\AA~(\textit{bottom right}). Observed wavelengths or physical quantities are indicated on each frame of the image. The dotted light blue lines in the GOES plot (\textit{top left}) denote the times corresponding to the AIA snapshots shown in Fig.\,\ref{Fig:Overview_AIA}. \\
A color version of this image is available in the online journal.
\label{Fig:Preflare}}
\end{figure*}
%

\begin{figure*}[!Ht]
	\centering
	\includegraphics[width=6.26cm,clip,bb= 0  0 495 85] {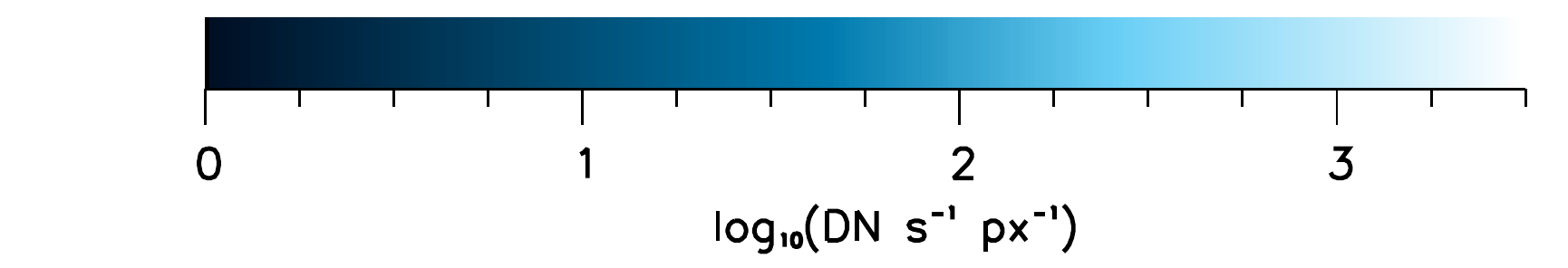}
	\includegraphics[width=5.57cm,clip,bb=55  0 495 85] {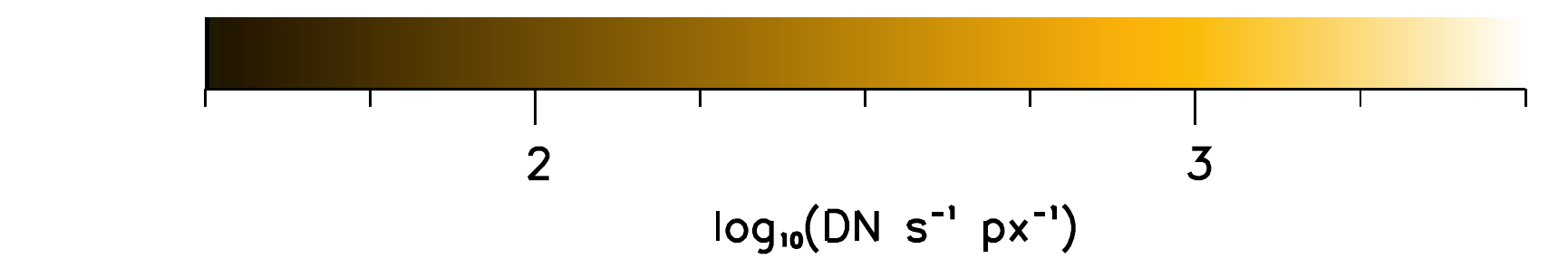}
	\includegraphics[width=5.57cm,clip,bb=55  0 495 85] {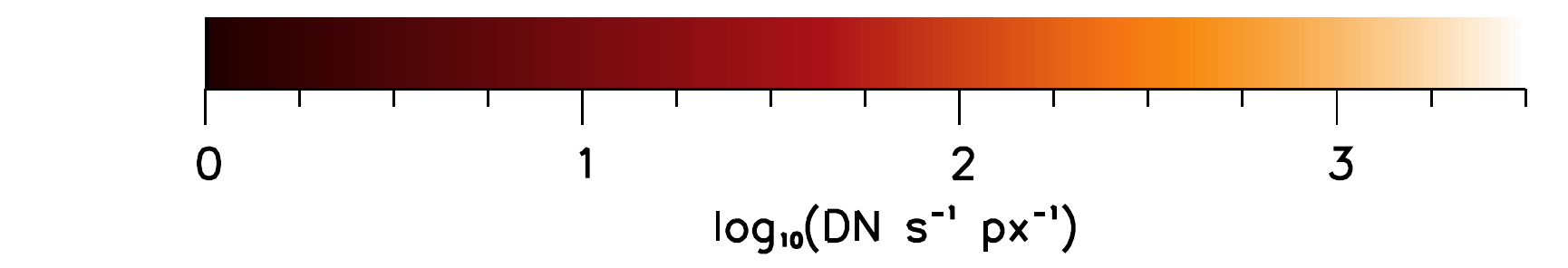}
	\includegraphics[width=6.26cm,clip,bb= 0 42 495 312]{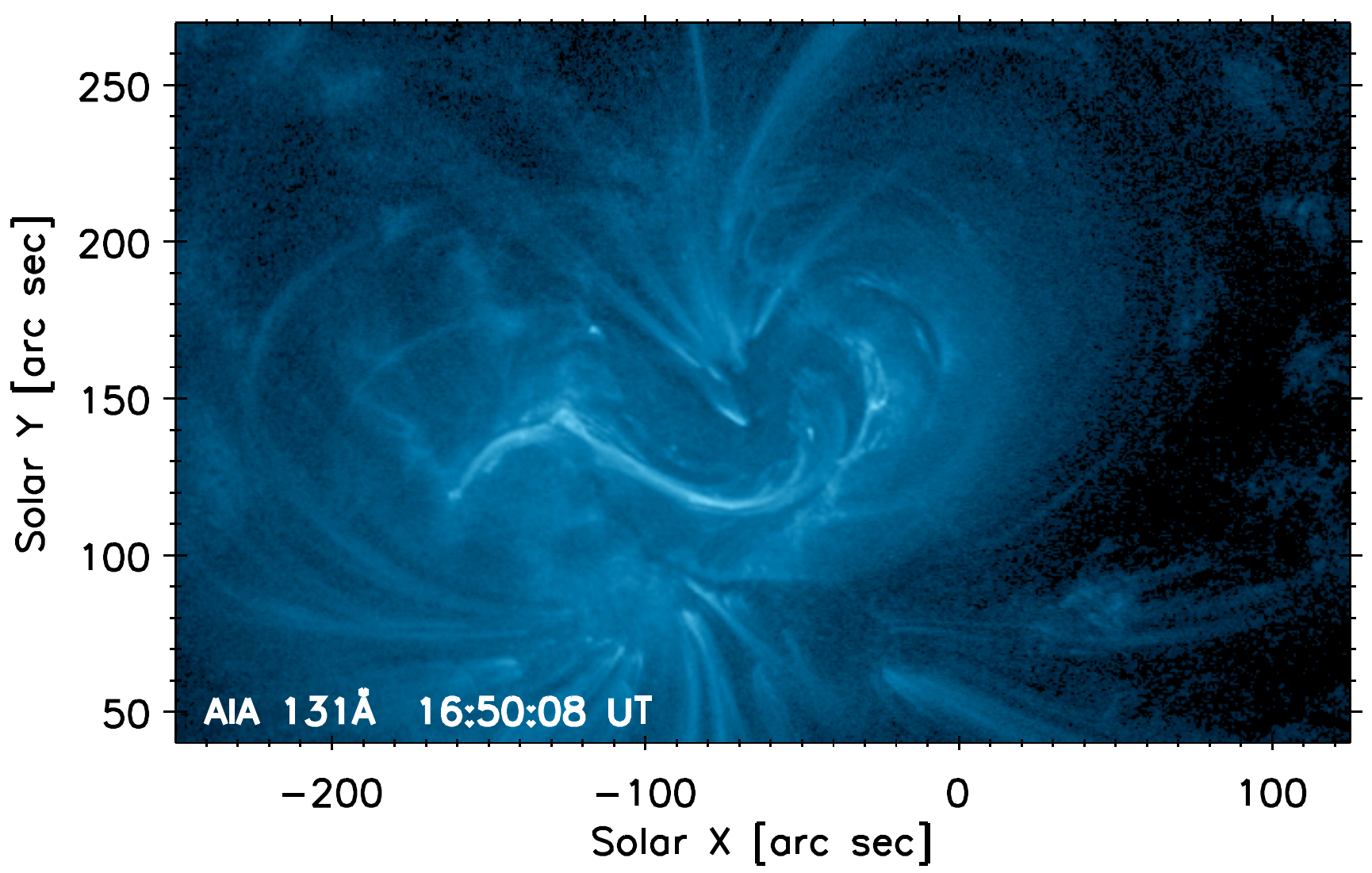}
	\includegraphics[width=5.57cm,clip,bb=55 42 495 312]{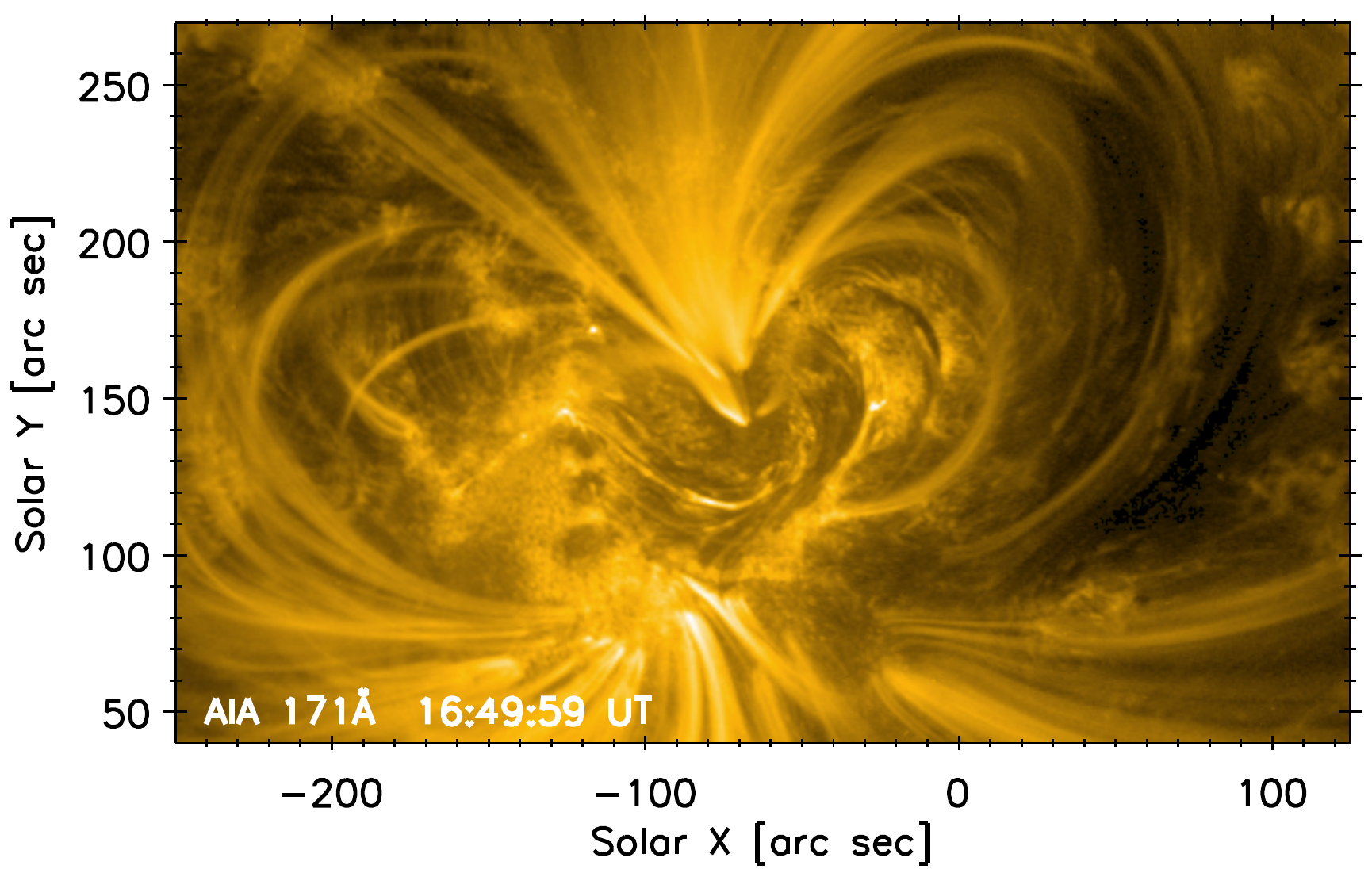}
	\includegraphics[width=5.57cm,clip,bb=55 42 495 312]{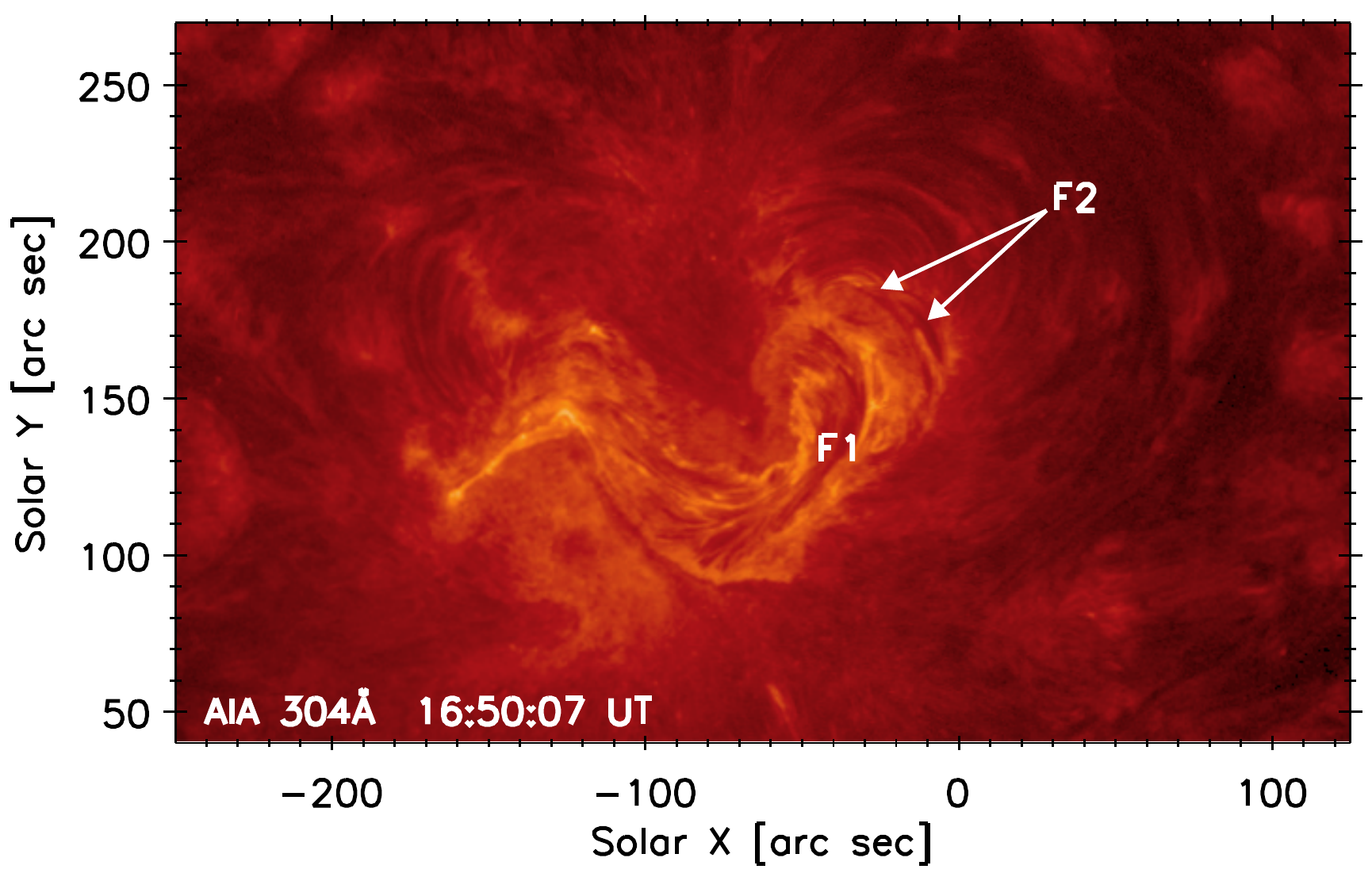}
	\includegraphics[width=6.26cm,clip,bb= 0 42 495 312]{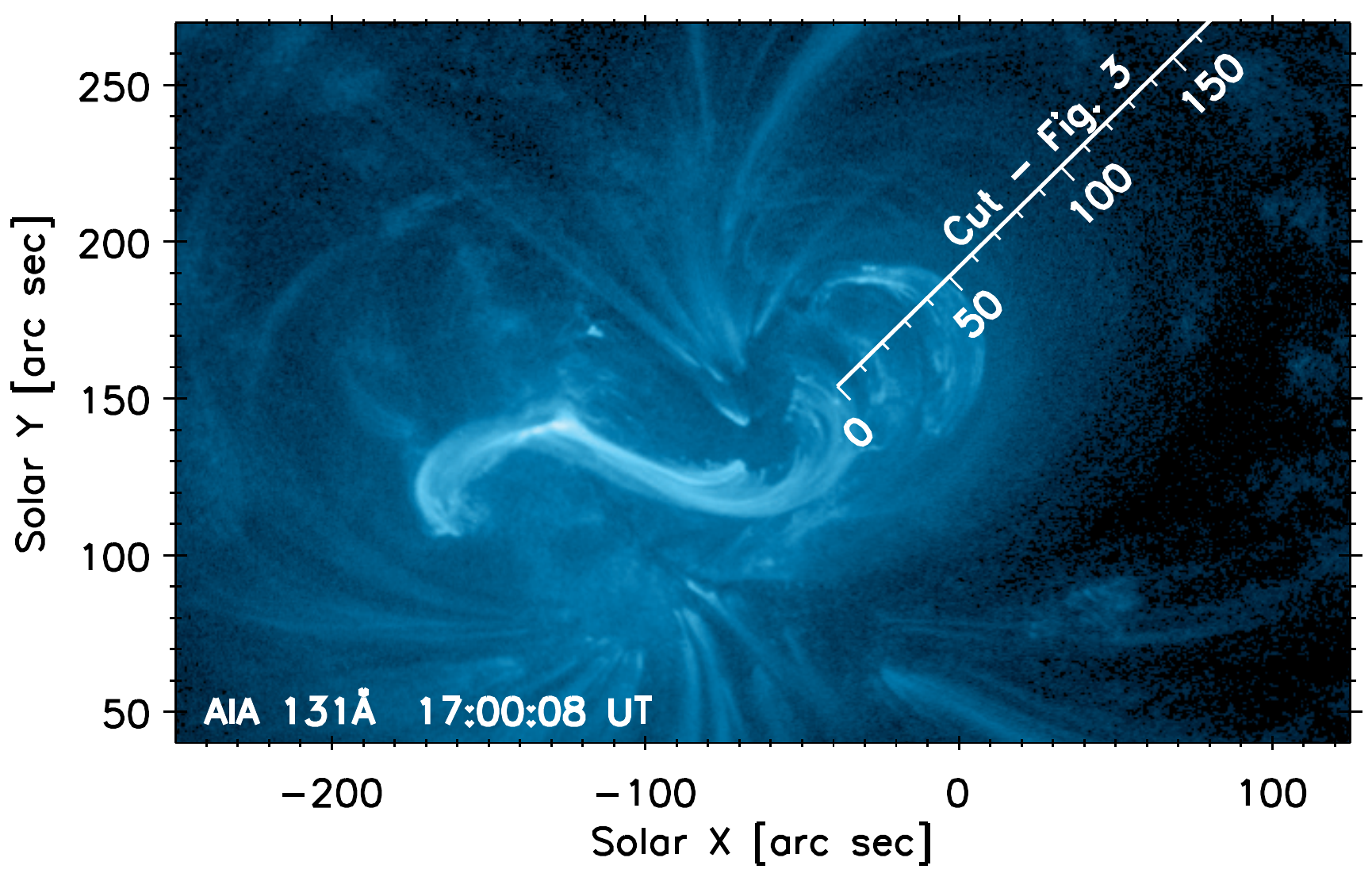}
	\includegraphics[width=5.57cm,clip,bb=55 42 495 312]{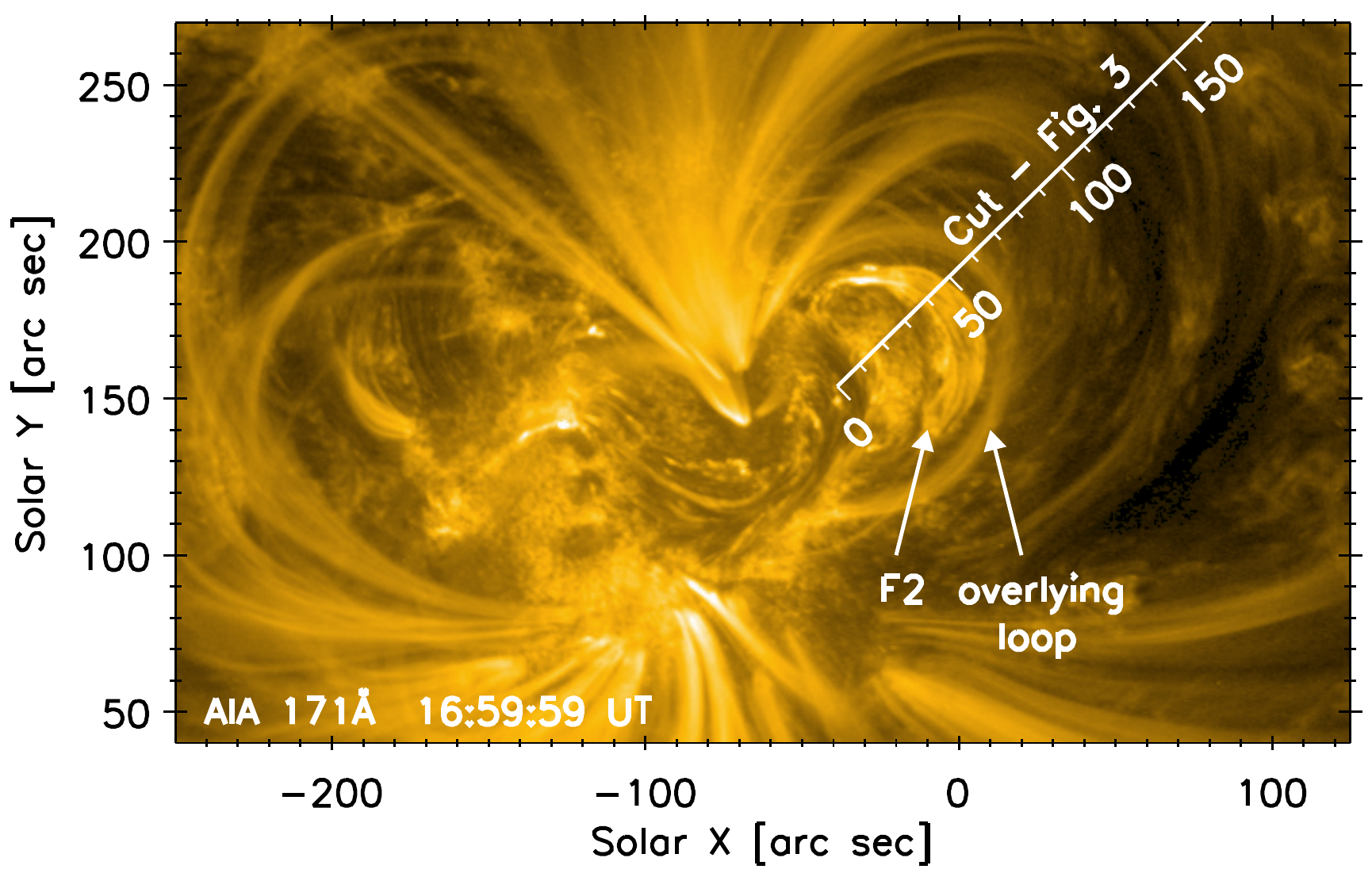}
	\includegraphics[width=5.57cm,clip,bb=55 42 495 312]{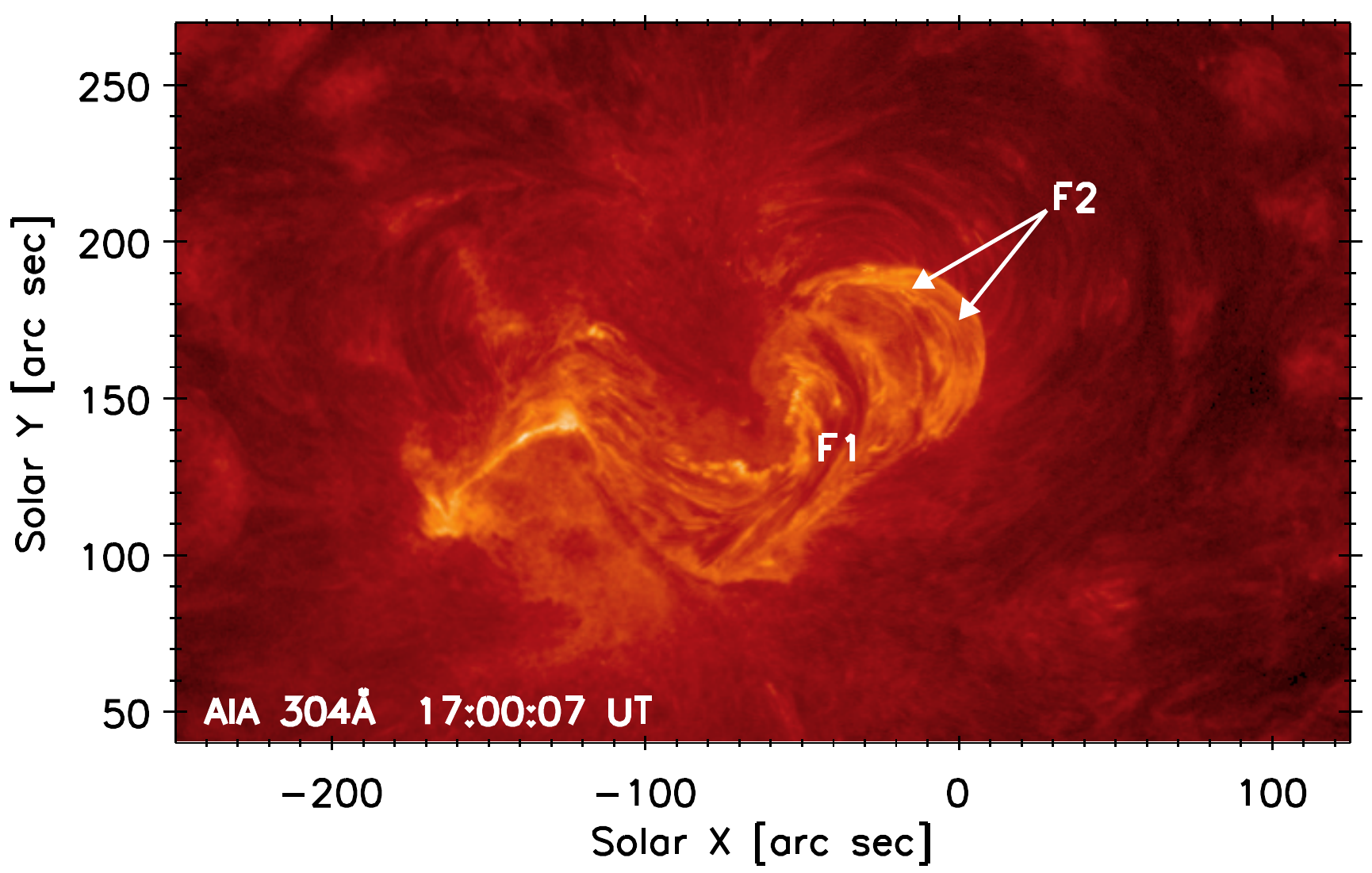}
	\includegraphics[width=6.26cm,clip,bb= 0 42 495 312]{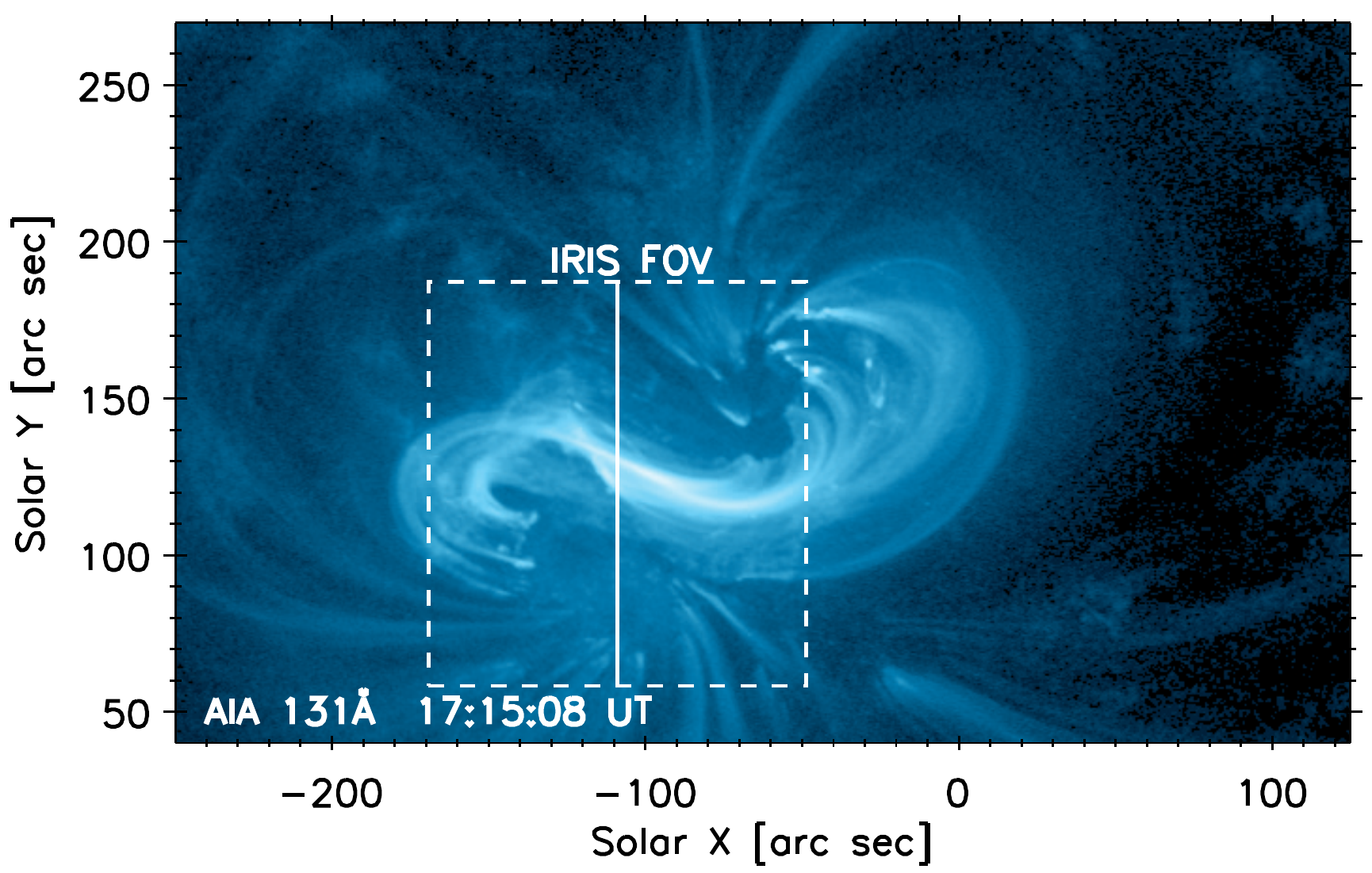}
	\includegraphics[width=5.57cm,clip,bb=55 42 495 312]{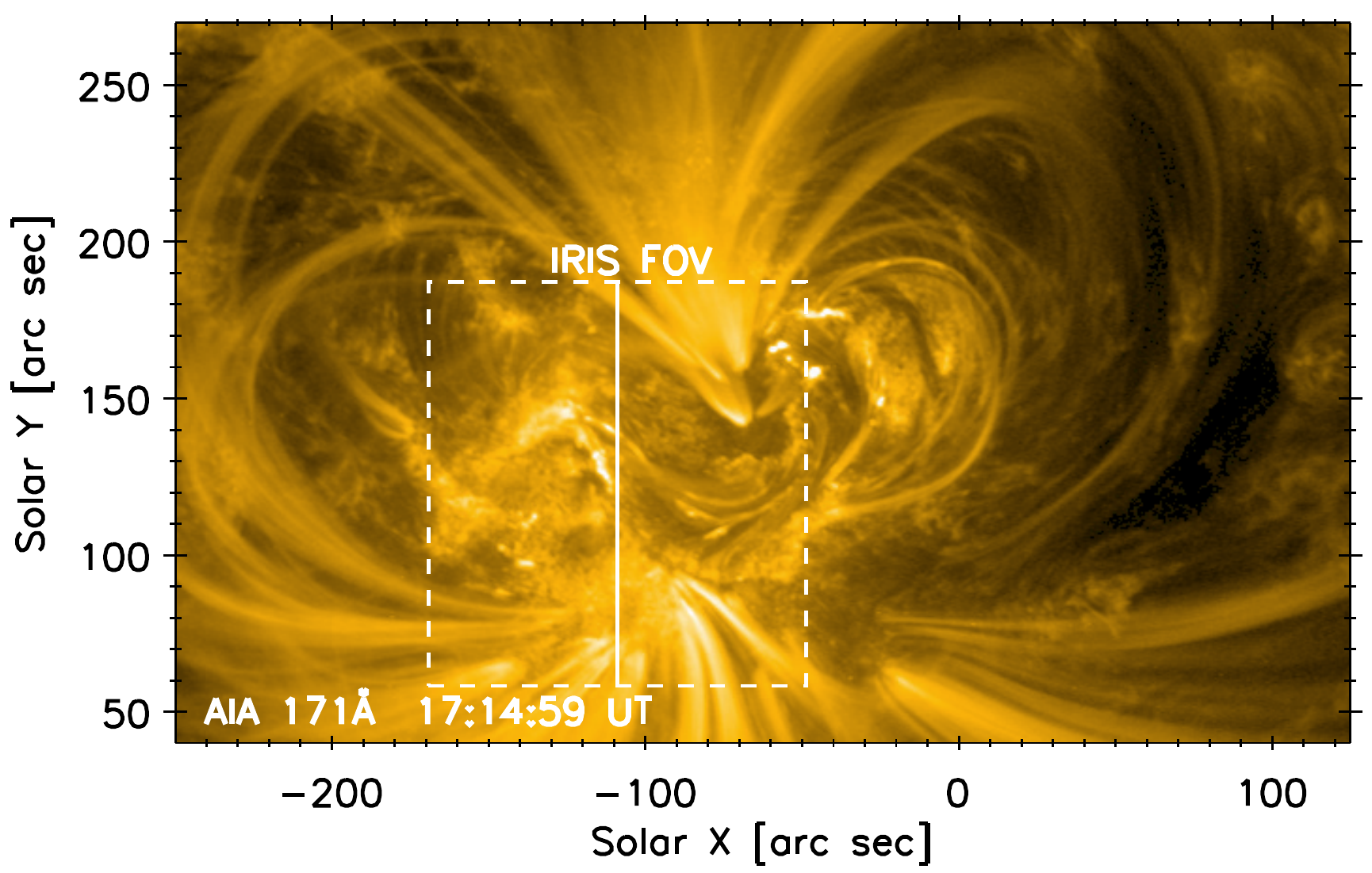}
	\includegraphics[width=5.57cm,clip,bb=55 42 495 312]{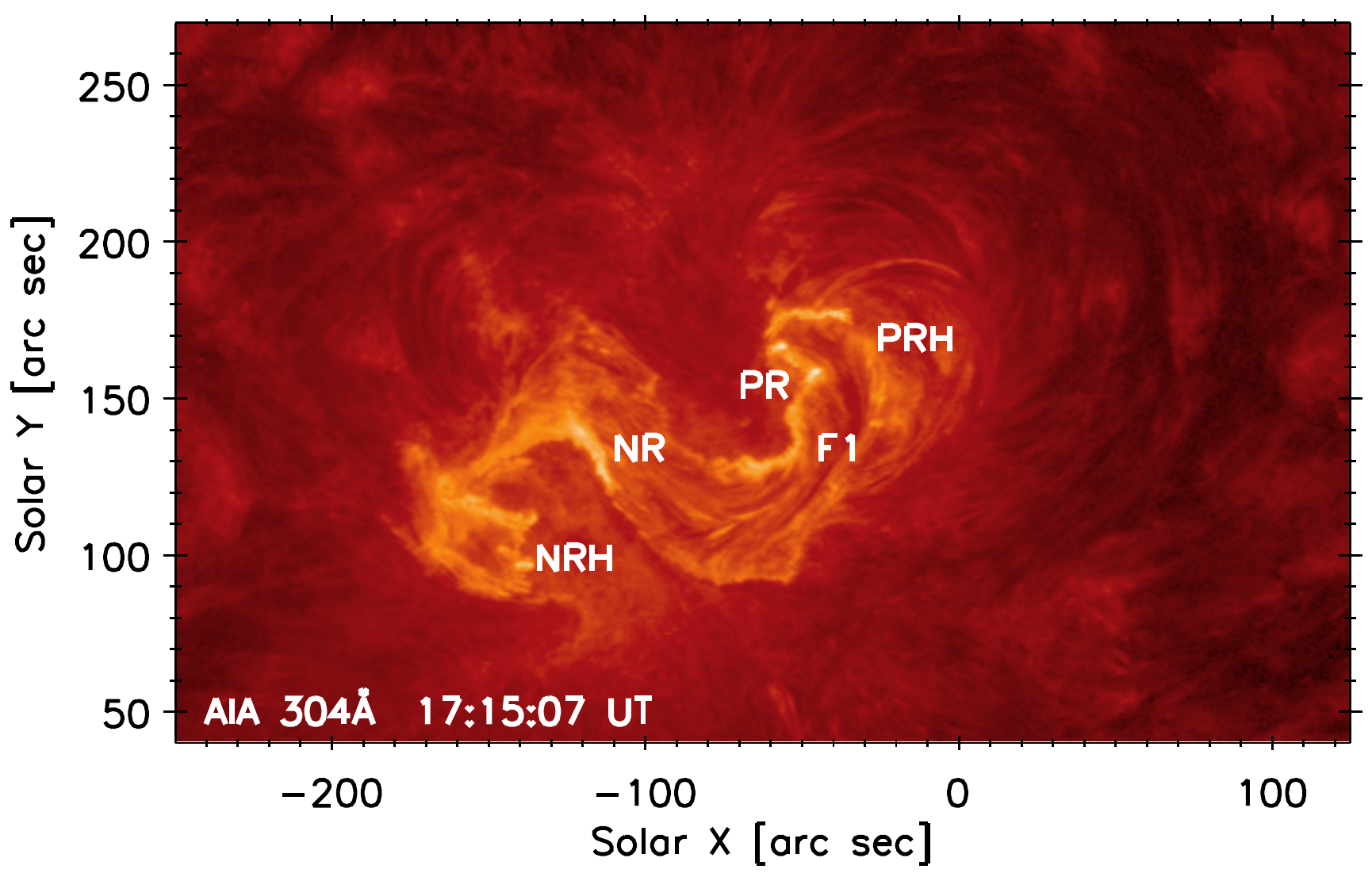}
	\includegraphics[width=6.26cm,clip,bb= 0 42 495 312]{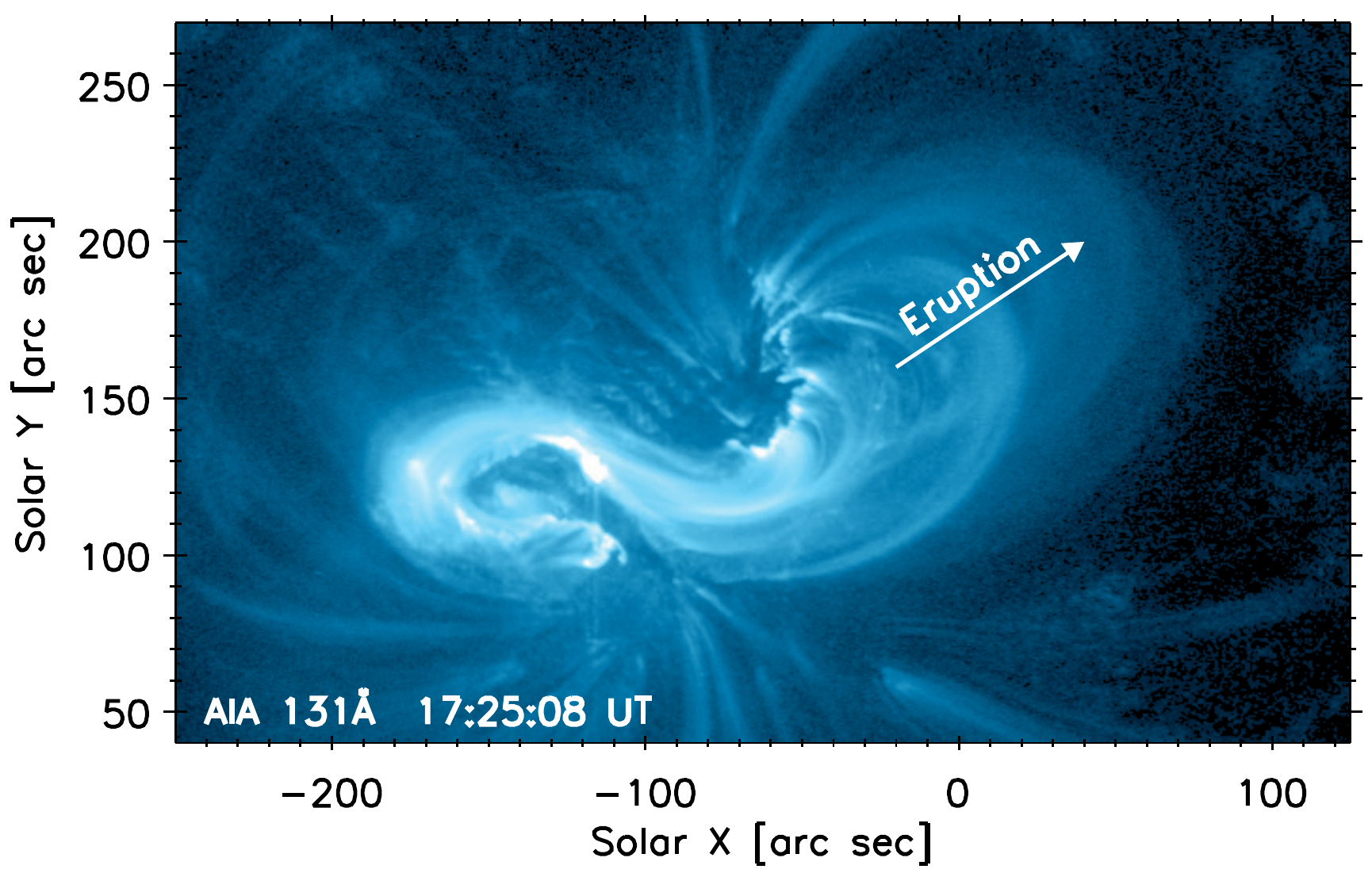}
	\includegraphics[width=5.57cm,clip,bb=55 42 495 312]{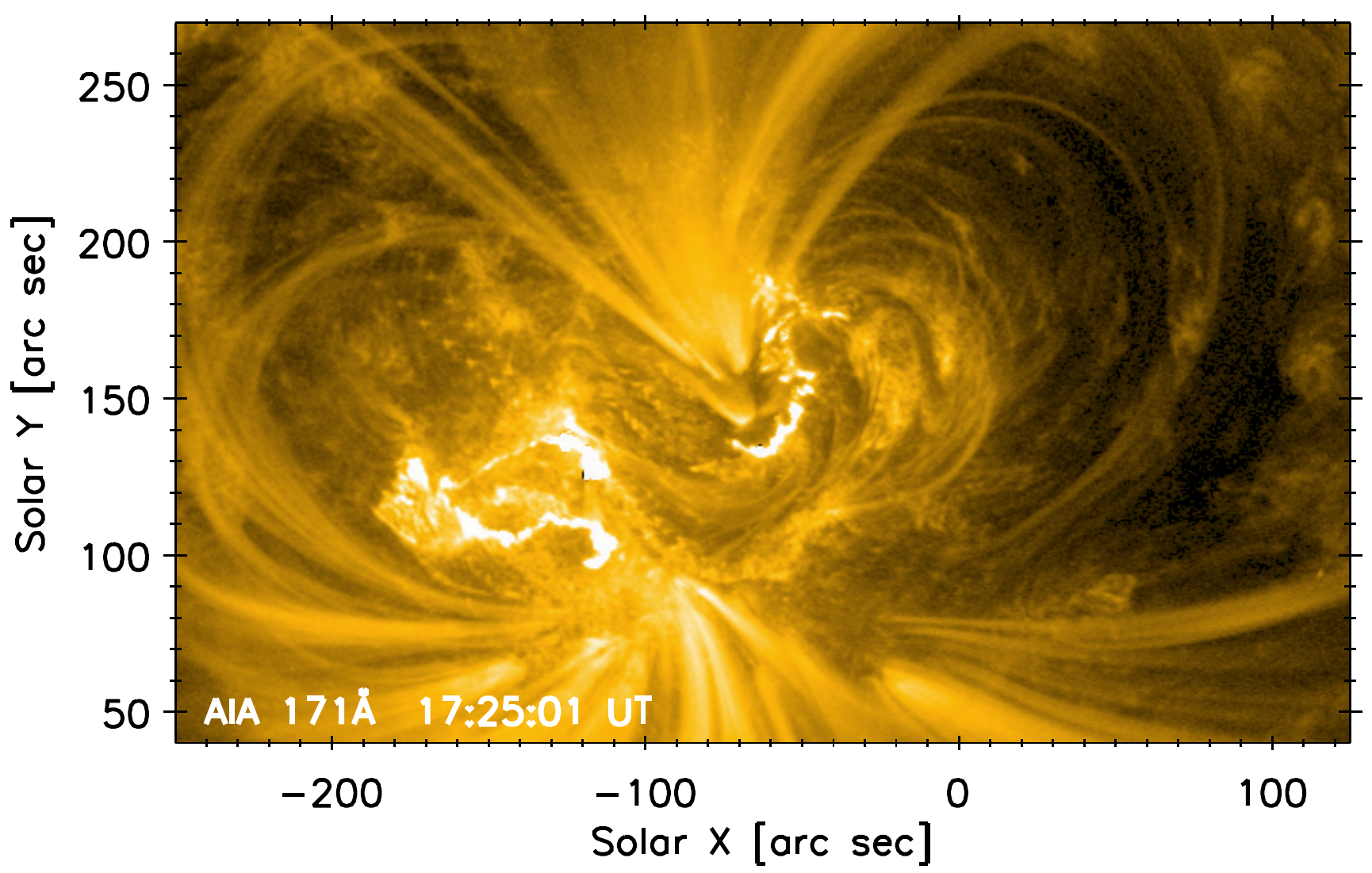}
	\includegraphics[width=5.57cm,clip,bb=55 42 495 312]{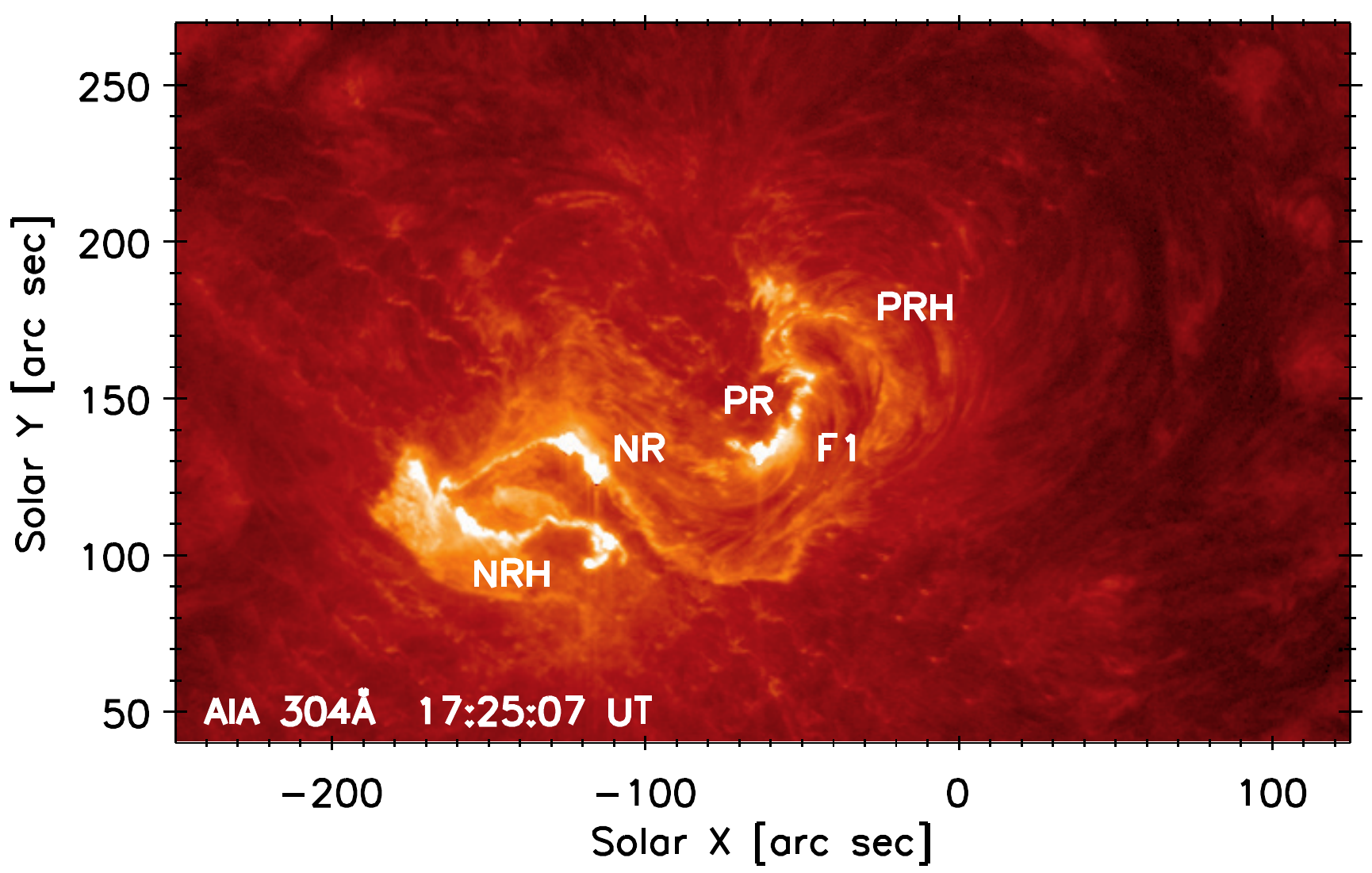}
	\includegraphics[width=6.26cm,clip,bb= 0 42 495 312]{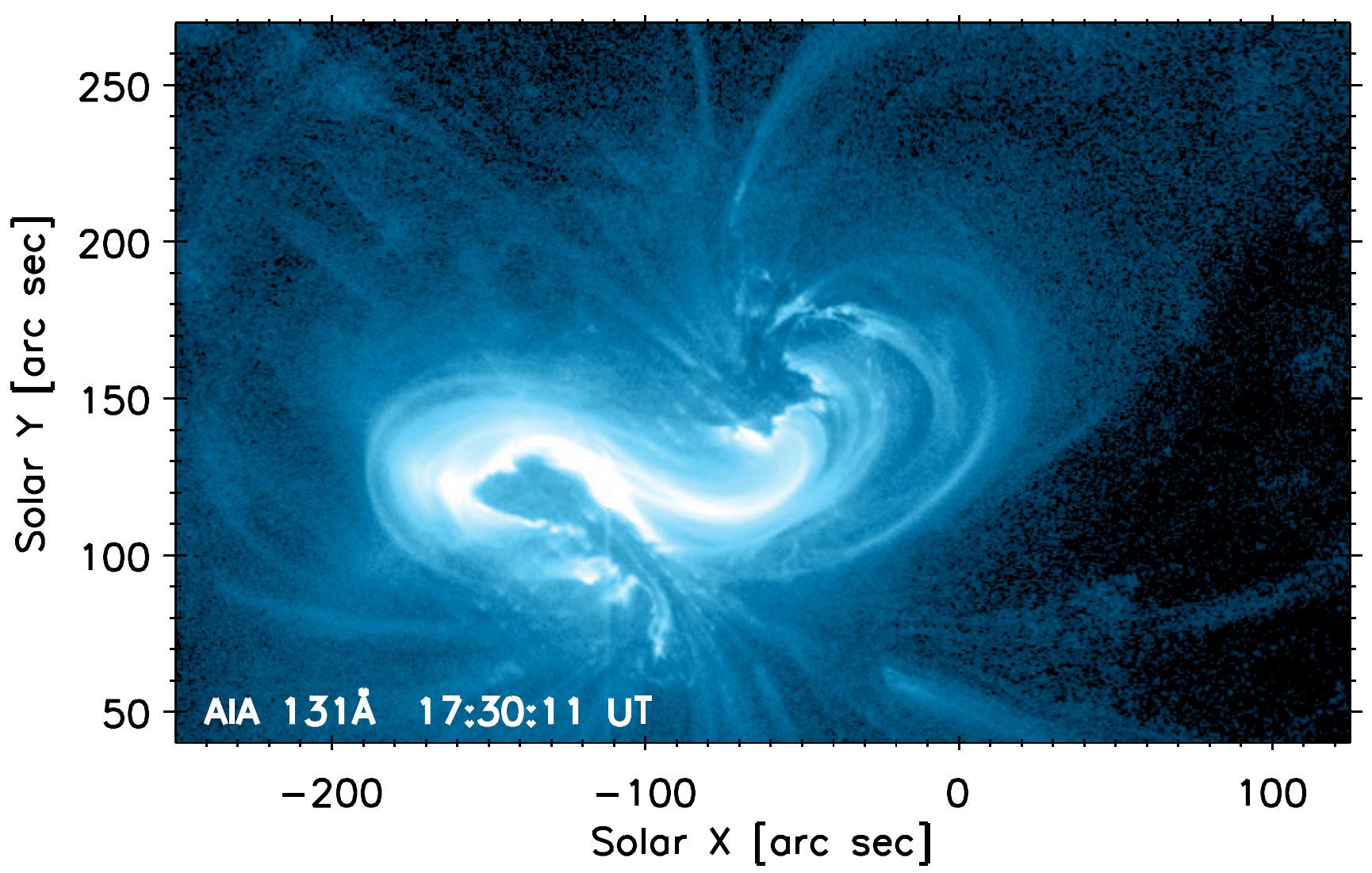}
	\includegraphics[width=5.57cm,clip,bb=55 42 495 312]{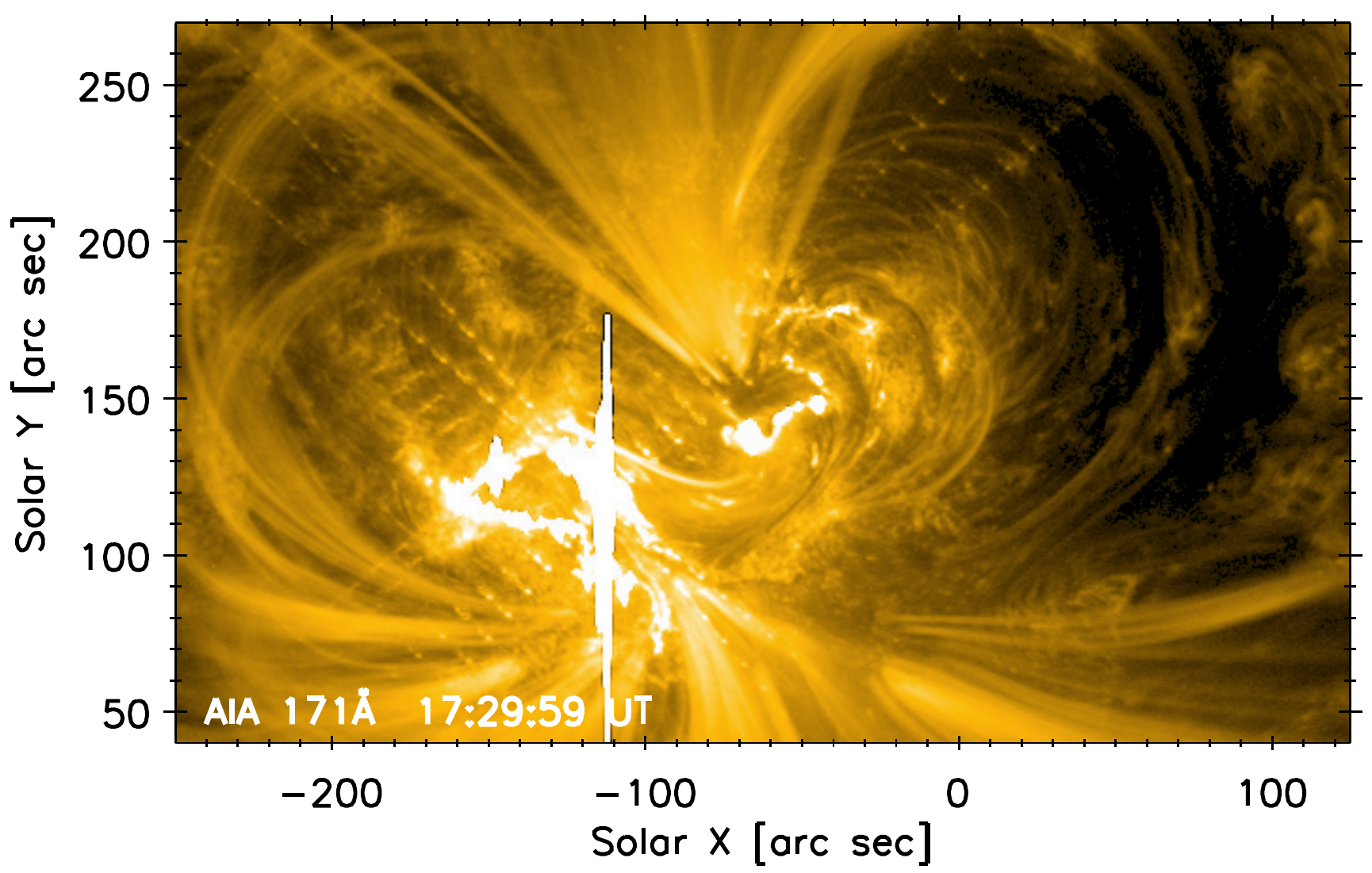}
	\includegraphics[width=5.57cm,clip,bb=55 42 495 312]{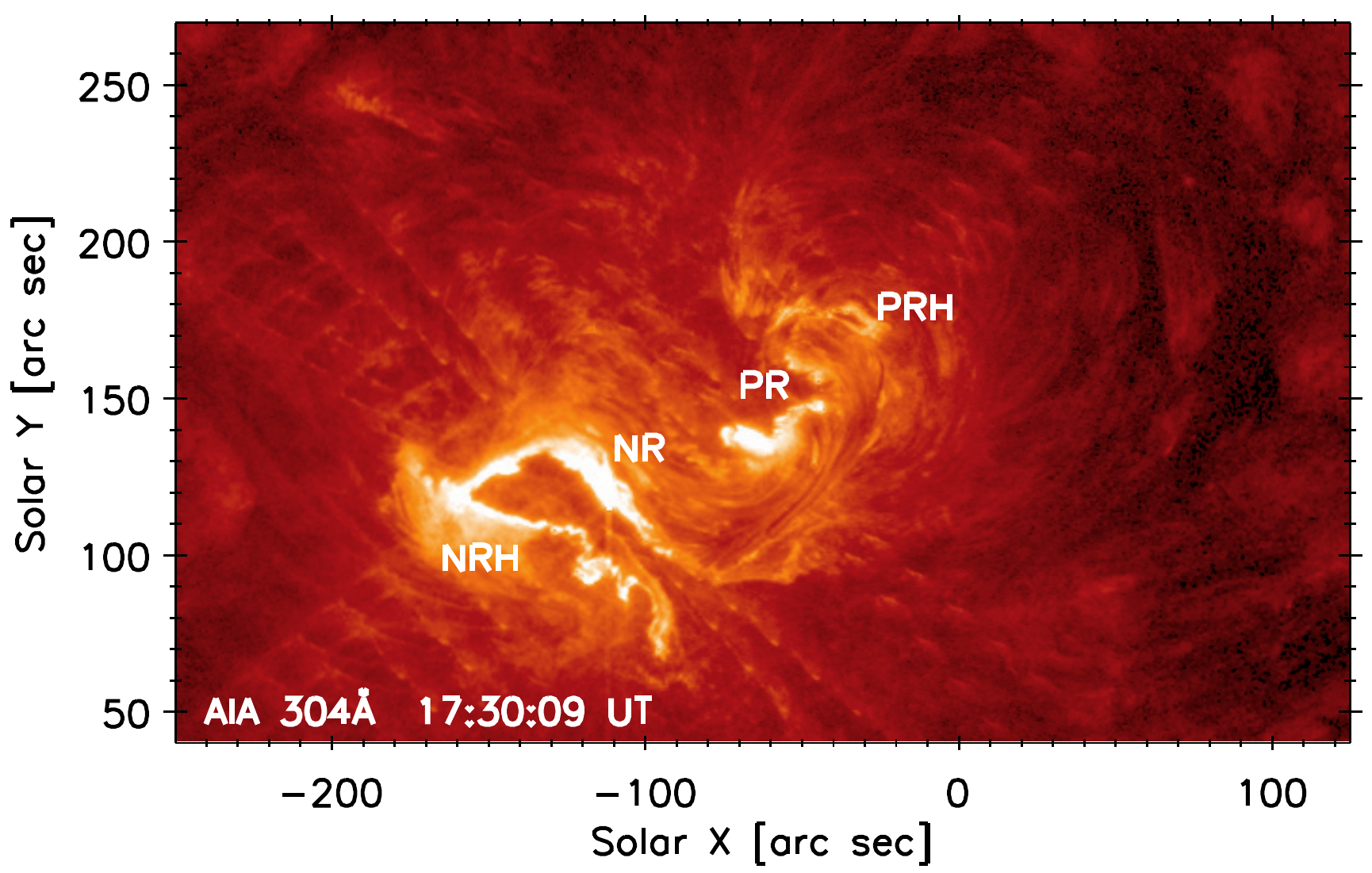}
	\includegraphics[width=6.26cm,clip,bb= 0  0 495 312]{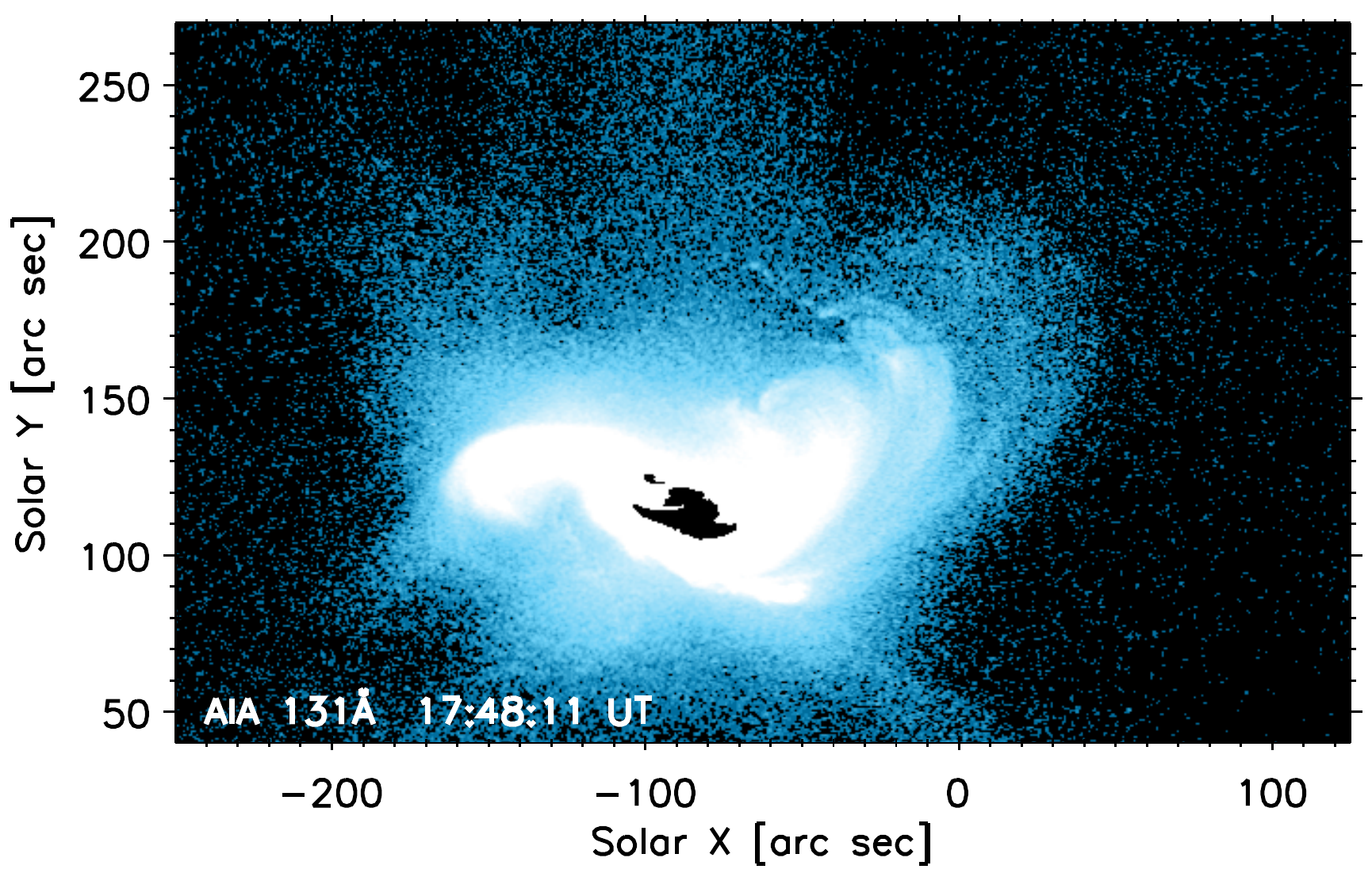}
	\includegraphics[width=5.57cm,clip,bb=55  0 495 312]{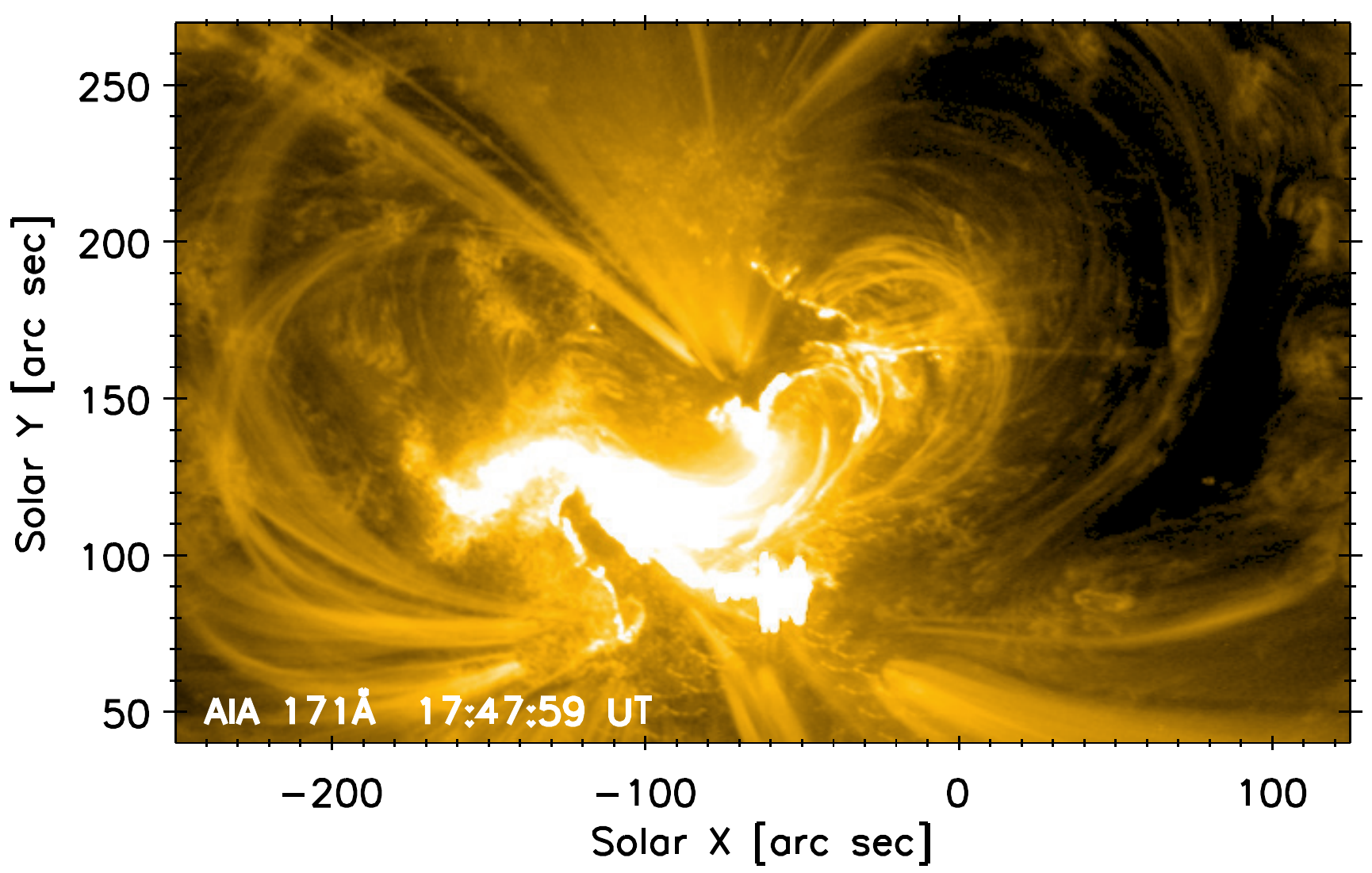}
	\includegraphics[width=5.57cm,clip,bb=55  0 495 312]{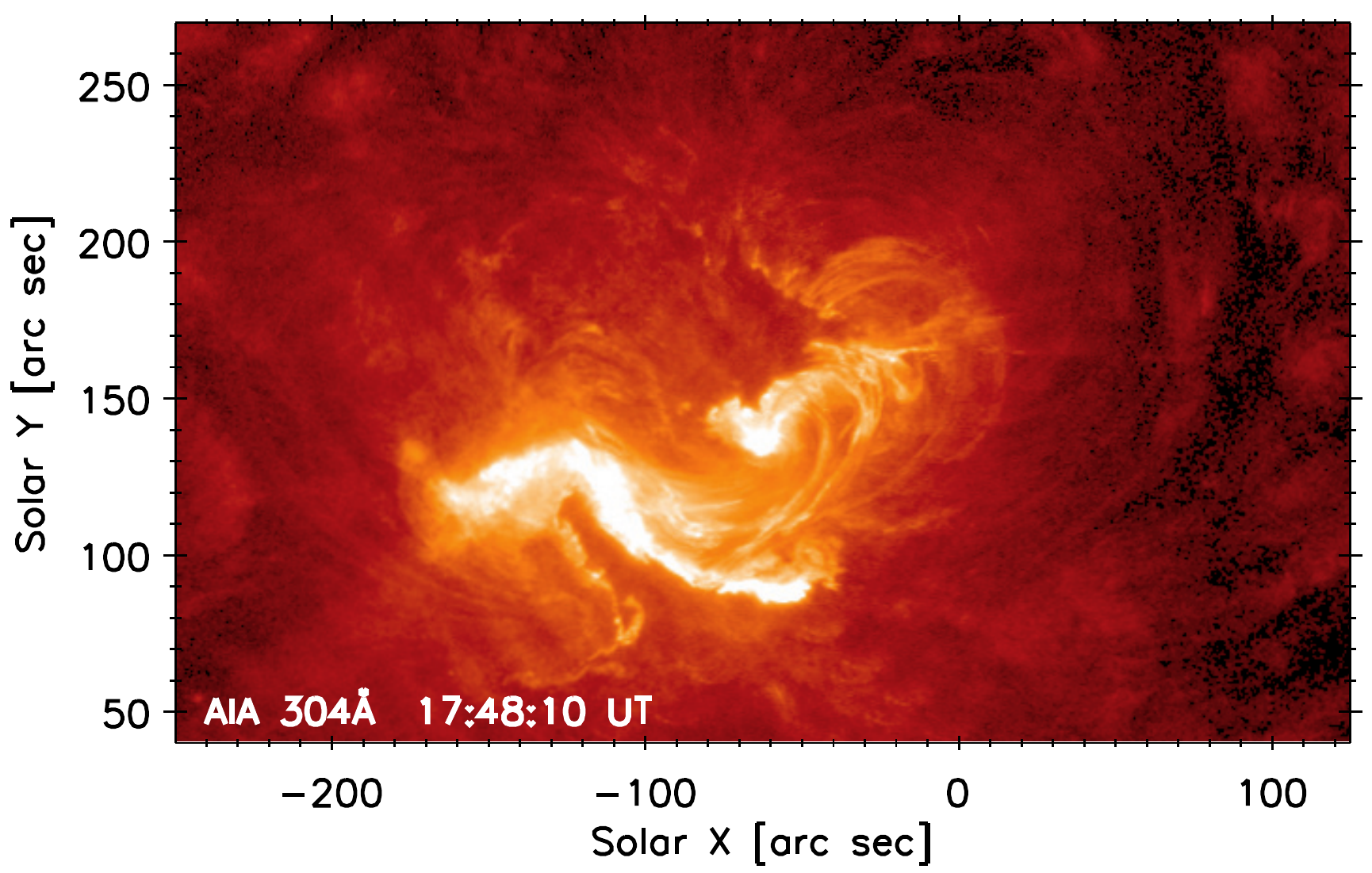}
\caption{Overview of the X1.6 flare evolution, as observed by the \textit{SDO}/AIA instrument in the AIA bandpasses at 131~\AA~(\textit{left}), 171~\AA~(\textit{middle}), and 304~\AA~(\textit{right}). \\ \\
A color version of this image and animations are available in the online journal.
\label{Fig:Overview_AIA}}
\end{figure*}
%
%
\begin{table*}
\begin{center}
\scriptsize
\caption{Summary of individual events during the 2014 September 10 flare. Times and locations given are approximate, as the events are dynamic, or have a spatial extension or temporal duration. 
\label{Table:1}}
\tabletypesize{\normalsize}
\begin{tabular}{llll}
\tableline\tableline
Approx. Time [UT]	& Event 							& Location [Solar $X$, $Y$]			& Notes											\\
\tableline
16:47 UT		& Onset of the early flare phase in the sigmoid			& $[-100\arcsec, 130\arcsec]$			& Figs. \ref{Fig:Preflare}, \ref{Fig:Overview_AIA}					\\
16:50 UT onwards	& Growing system of flare loops, precursors			& $[-100\arcsec, 130\arcsec]$			& Figs. \ref{Fig:Overview_AIA} \textit{top}, Sect. \ref{Sect:4.3}			\\
16:50--17:24 UT		& Expanding warm coronal loops on the AR periphery		& $[60\arcsec, 200\arcsec]$			& Sects. \ref{Sect:2.2.1} and \ref{Sect:4.1}, Figs. \ref{Fig:Overview_AIA}--\ref{Fig:Eruption_stackplots}, \ref{Fig:Topotr3D} \\
16:51--17:05 UT		& Failed F2 eruption						& $[0\arcsec, 170\arcsec]$, F2			& Fig.\,\ref{Fig:Overview_AIA} \textit{second row}; Fig.\,\ref{Fig:Eruption_stackplots}	\\
16:58--17:26 UT		& NR hook extension, squirming and slipping motions		& $[-150\arcsec, 100\arcsec]$, NRH		& Sect. \ref{Sect:2.3.1}, Figs. \ref{Fig:Overview_AIA}, \ref{Fig:NR_AIA_slit2a}--\ref{Fig:NR_AIA_slit2a_TR}	 	\\
17:00--17:30 UT		& slipping motion well visible in PR				& $[-50\arcsec, 140\arcsec]$, PR		& Sect. \ref{Sect:2.3.2}, Figs. \ref{Fig:Overview_AIA}, \ref{Fig:PR_AIA}, and \ref{Fig:PR_stackplots}		 	\\
17:03--17:14 UT		& weak blue-shifts detected in \ion{Fe}{21}~			& NR						& Sect. \ref{Sect:3.2}, Figs. \ref{Fig:Overview_IRIS}--\ref{Fig:Fits_FeXXI}		\\
17:10--17:27 UT		& Growing system of S-shaped loops, hot eruption		& $[0\arcsec, 170\arcsec]$			& Sect. \ref{Sect:2.2.2}, Fig.\,\ref{Fig:Overview_AIA}					\\
17:24 UT		& Impulsive phase onset, strong blue-shifts in \ion{Fe}{21}~	& NR						& Figs. \ref{Fig:Overview_AIA}, \ref{Fig:Overview_IRIS}, and \ref{Fig:Fits_FeXXI}	\\
17:24 UT	& Onset of fast eruption (velocities $> 270$ km\,s$^{-1}$) &					& Fig.\,\ref{Fig:Eruption_stackplots}						\\
17:25--17:40 UT		& Coronal loop oscillations following the hot eruption		& $[60\arcsec, 200\arcsec]$			& Sect. \ref{Sect:2.2.3}, Figs. \ref{Fig:Overview_AIA}--\ref{Fig:Eruption_stackplots}	\\
17:28, 17:32 UT		& Peaks of the \textit{Callisto} radio flux at 350\,MHz		&						& Fig.\,\ref{Fig:Eruption_stackplots} \textit{bottom}, Sect. \ref{Sect:2.2.3}		\\
17:45 UT		& Peak of the GOES 1--8\AA\,X-ray flux, onset of gradual phase	& 						& Fig.\,\ref{Fig:Preflare} \textit{top left}						\\
17:58 UT		& End of IRIS sit-and-stare observations, \ion{Fe}{21}~nearly at rest	& NR					& Fig.\,\ref{Fig:Overview_IRIS}								\\
\tableline\tableline
\end{tabular}
\end{center}
\end{table*}
%
%

%
%
\section{The X1.6-Class Flare as Observed by \textit{SDO}/AIA}
\label{Sect:2}

\subsection{AR 12158 and the pre-flare state}
\label{Sect:2.1}

The Active Region NOAA 12158 (hereafter, AR 12158) was visible on the solar disk during 2014 September 03--16. During this time, the AR 12158 produced several flares, including 12 C-class ones, an M4.6-class flare on 2014 September 7, and an X1.6-class long-duration flare on 2014 September 10. It is this X-class flare that is studied in this paper.

The X-class flare itself started at about 16:47 UT, reached maximum of its X-ray flux at 17:45, as measured in the 1--8\AA~passband by the \textit{GOES-15} satellite at Earth, and exhibited a long gradual phase (Fig.\,\ref{Fig:Preflare}, \textit{top}). Several specific aspects of the flare have already been reported on by different authors \citep{Li15,Tian15,Cheng15}. The flare evolution is detailed in Sect. \ref{Sect:2.2}.

The pre-flare state of the AR 12158 is shown in Fig.\,\ref{Fig:Preflare}. The leading positive-polarity sunspot is encircled by a wreath of several smaller ones of both polarities, with the negative-polarity pores located to the S, SW, and W of the spot. The magnetic configuration of this active region is peculiar, since the leading sunspot is of positive and not negative polarity, as would be expected of an active region in the northern solar hemisphere during the cycle 24.

The AR 12158 contains two filaments, F1 and F2, as shown by the chromospheric H$\alpha$ observations obtained at the Big Bear Solar Observatory (\textit{BBSO}), see Fig.\,\ref{Fig:Preflare}. The filaments are of the same chirality \citep[see, e.g.,][]{Martin98,Devore05}. Both these filaments are overlaid by a sigmoid visible only in the 94~\AA~channel \citep{Cheng15} of the Atmospheric Imaging Assembly \citep[AIA,][]{Lemen12,Boerner12} onboard the Solar Dynamics Observatory \citep[\textit{SDO},][]{Pesnell12}. This suggests that the temperature of the sigmoid corresponds to the formation of \ion{Fe}{18} \citep[i.e., about 7\,MK;][]{ODwyer10,DelZanna13}. The X1.6-class flare occurs within this sigmoid (Fig.\,\ref{Fig:Overview_AIA}).

%
\subsection{Flare evolution}
\label{Sect:2.2}

In this section, we describe the evolution of the flare, as observed by the \textit{SDO}/AIA imager. AIA acquires full-Sun images in 10 EUV and UV filters with high spatial resolution (1\farcs5, 0\farcs6 pixel size) as well as high temporal resolution (12\,s). The bandpasses of the AIA filters are centered on several strong lines in the solar EUV/UV spectrum. These emission lines originate at different plasma temperatures, with some filter bandpasses containing more than one strong emission line \citep[e.g.,][]{ODwyer10,DelZanna11c}. This means that the AIA temperature responses are multithermal in general \citep[see also, e.g.,][]{DelZanna13,Schmelz13}. That is, the signal observed within a particular AIA filter can originate at several different temperatures. Using combinations of AIA filters however, it is possible to identify the approximate temperature of the emitting plasma, as well as to perform the differential emission measure reconstruction \citep[][]{Schmelz11a,Schmelz11b,Warren12,Hannah12,Hannah13,DelZanna13,Schmelz13,Dudik14a,Dudik15}. This makes the AIA instrument an excellent tool to study the morphology of plasma emission at different temperatures.

As an example of the multithermality of the AIA filters relevant to the present study, we point out that the 131~\AA~bandpass has two dominant contributions, from \ion{Fe}{8} and \ion{Fe}{21}, arising at about 0.5 and 10 MK respectively in equilibrium conditions \citep[see, e.g.,][]{Petkaki12,DelZanna13,Dudik14a}. The \ion{Fe}{8} emission can be discerned visually by comparison with the 171~\AA~bandpass, dominated by \ion{Fe}{9} formed at around 0.8\,MK. This is because of the significant overlap of the peaks of the relative ion abundances of \ion{Fe}{8} and \ion{Fe}{9} under equilibrium conditions \citep{Dere09,Bryans09}. Under non-equilibrium conditions characterized by the presence of $\kappa$-distributions with high-energy power-law tails, as is the case in flaring plasma \citep{Kasparova09,Oka13,Oka15}, the AIA temperature responses become more multithermal, and the peaks of the temperature responses are shifted towards higher temperatures \citep{Dzifcakova15}.

A timeline of individual events during the flare is given in Table \ref{Table:1}. An overview of AIA observations of the flare is given in Fig.\,\ref{Fig:Overview_AIA}, and in the online Movies 1, 2, and 3, corresponding to the filters 131~\AA, 171~\AA, and 304~\AA, respectively. The 131~\AA~and 171~\AA~bandpasses are chosen since they allow us to distinguish the 10\,MK flare emission from the warm coronal loops. The 304~\AA~bandpass is shown to depict the evolution of the filaments F1 and F2, as well as that of the flare ribbons. The ribbons are denoted PR and NR for the positive-polarity and negative-polarity ribbons, while PRH and NRH stand for the respective hooks of these ribbons. Note that the presence of hooks is a signature of the presence of a flux rope as also evidenced by the sigmoid \citep{Aulanier12,Janvier13,Janvier14,Dudik14a}, and also F1 and F2 \citep[see also][]{Cheng15}. 

\subsubsection{Early flare phase}
\label{Sect:2.2.1}

\begin{figure}
	\centering
	\includegraphics[width=8.20cm,clip,bb= 0  0 495  85]{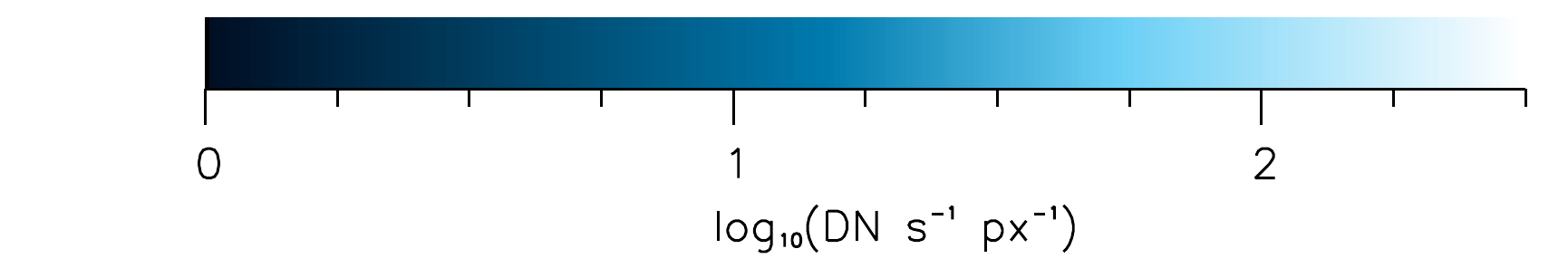}
	\includegraphics[width=8.20cm,clip,bb= 0  0 495  85]{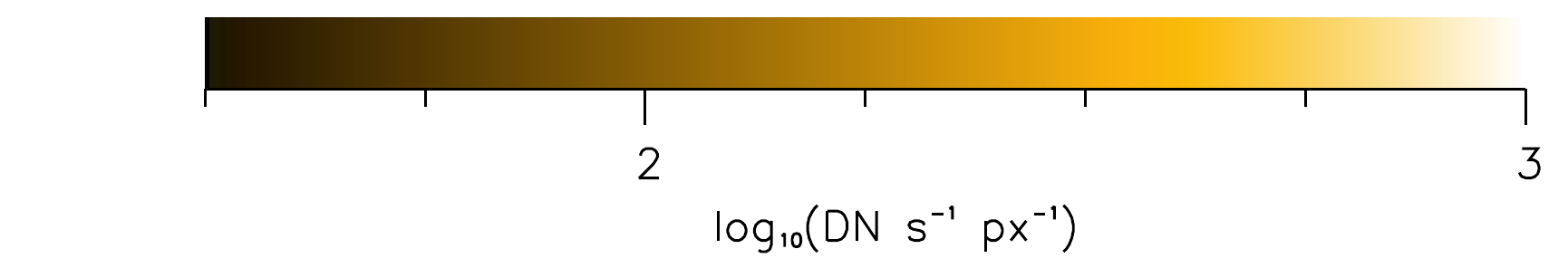}
	\includegraphics[width=8.20cm,clip,bb= 0 40 495 385]{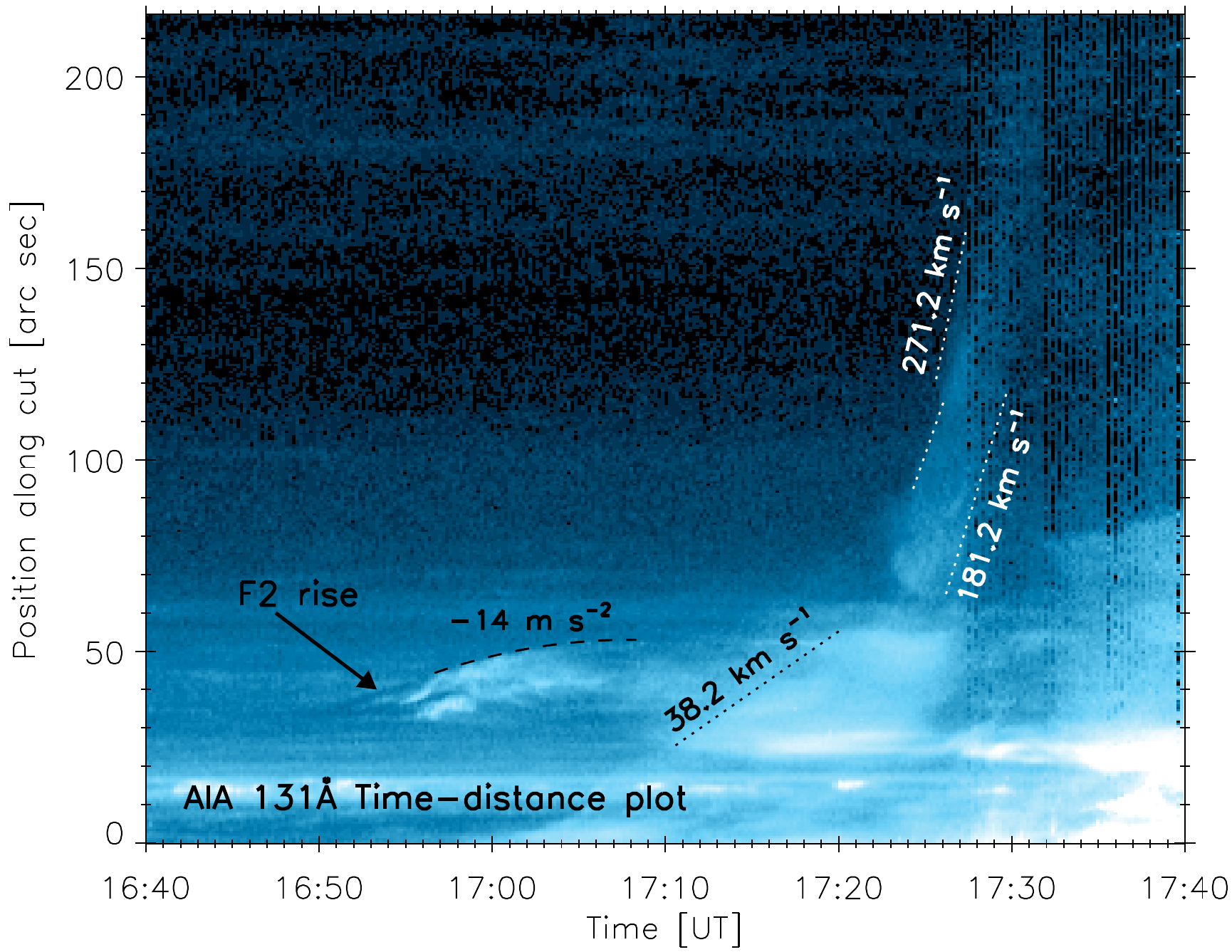}
	\includegraphics[width=8.20cm,clip,bb= 0 40 495 385]{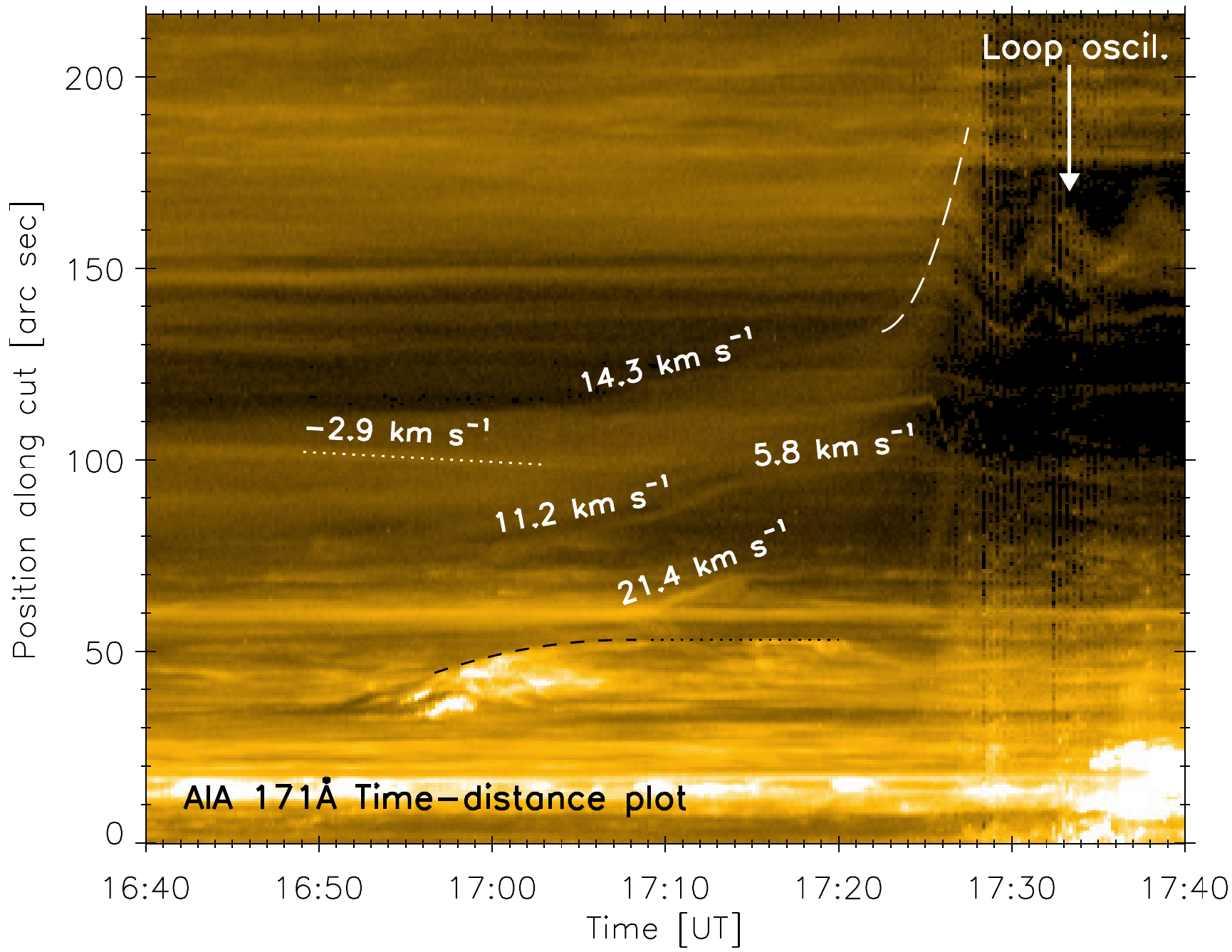}
	\includegraphics[width=8.20cm,clip,bb= 0  0 495 288]{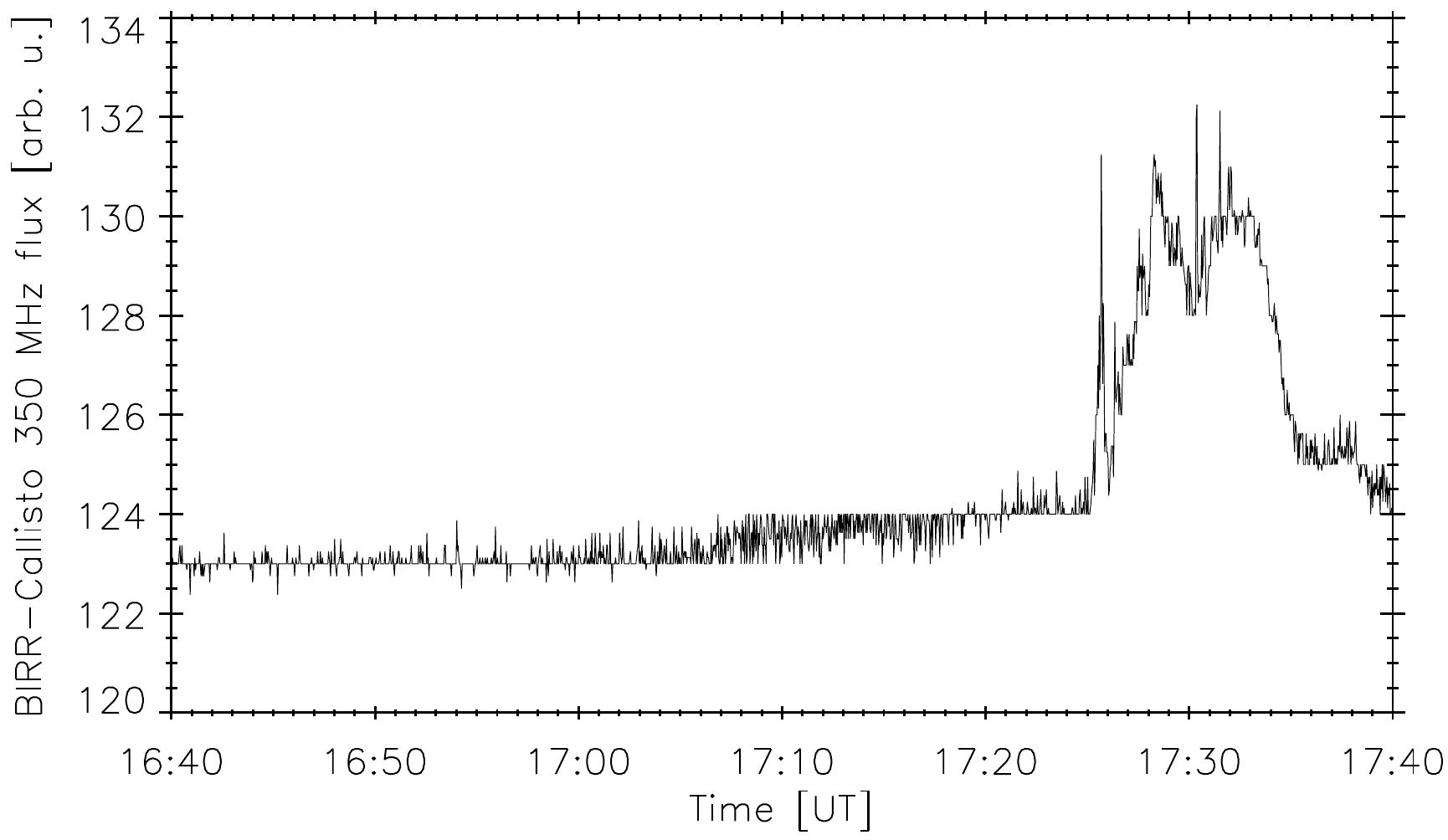}
\caption{AIA time-distance plots along the cut shown in Fig.~\ref{Fig:Overview_AIA}. Velocities corresponding to individual features are indicated.  See text for details. \textit{Bottom}: \textit{BIR--Callisto} radio flux at 350\,MHz.\\
A color version of this image is available in the online journal.
\label{Fig:Eruption_stackplots}}
\end{figure}

The X1.6-class flare starts with a loop-like brightening within the sigmoid above F1. The brightening is observable in 131~\AA~and 94~\AA~bandpasses and develops into a series of flare loops. This behaviour is very similar to the one described for another X-class flare by \citet{Dudik14a}. At 16:50\,UT, these loops are clearly visible (Fig.~\ref{Fig:Overview_AIA}, \textit{top left}). At this time, F2 starts to rise and brighten in both 131~\AA~and 171~\AA, indicating that it is heated to at least several times 10$^5$\,K. The F2 subsequently decelerates as it is stopped by the overlying coronal loops seen in AIA 171~\AA~and 193~\AA~(see Fig.~\ref{Fig:Overview_AIA} at 17:00\,UT and online Movie 2). To study the evolution of F2 and the neighbouring coronal loops, we place an artificial cut across a direction of F2 rise (see Fig.~\ref{Fig:Overview_AIA}). The time-distance plots obtained along the direction of this cut are shown in Fig.~\ref{Fig:Eruption_stackplots}. Using these time-distance plots, we measure the deceleration of F2 to be about $-14$\,m\,s$^{-2}$ by approximating a parabola to the profile on the time-distance plot (dashed black line in Fig.~\ref{Fig:Eruption_stackplots}). The stopping of F2 is denoted by the black horizontal dotted line in the 171\,\AA~time-distance plot.

The time-distance plots constructed along the artificial cut also reveal rich dynamics of the overlying warm coronal loops observed in the 171\,\AA~bandpass as well as in 193\,\AA~(not shown). These loops are highly likely to have been inclined with respect to the local vertical; note also the similar pattern of fibrils observed in H$\alpha$ and AIA 304~\AA (Figs. \ref{Fig:Preflare}, \textit{bottom left} and \ref{Fig:Overview_AIA}, \textit{right}). One of the coronal loops observed in 171\,\AA~is contracting with a speed of about $-2.9$\,$\pm0.9$\,km\,s$^{-1}$. It is located at the approximate position of 100$\arcsec$ along the artificial cut, and denoted by the white dotted line in the respective 171~\AA~time-distance plot (Fig.\,\ref{Fig:Eruption_stackplots}). The contraction is discernible even before the onset of F2 rise, an important fact discussed in terms of the coronal implosion mechanism in Sect. \ref{Sect:2.2.3}.

The rise of F2 is also accompanied by widespread dynamics of the coronal loops. A series of these loops start to rise at various times with different velocities, ranging from 5.8\,$\pm1.7$ to 21.4\,$\pm2.1$\,km\,s$^{-1}$ (see Fig.\,\ref{Fig:Eruption_stackplots}). The velocities are determined by calculating the slope of the respective line on a time-distance plot. The location of the line is determined by trial-and-error method repeated many times. The uncertainty is then calculated by error propagation from the uncertainties of the line endpoints. We consider that the uncertainty in position is equal to half of the AIA resolution (0\farcs75), while the uncertainty in time is half of the AIA cadence (0.6\,s).

The rising of the warm coronal loops in our flare continues until the eruption and the associated disturbance during the impulsive phase. We however note that the occurrence of a contracting loop before the presence of rising structures in our flare is contrary to the behaviour reported for five different flares by \citet{Liu12a}.

During the rise of F2, the system of flare loops continue to widen and brighten (Fig.\,\ref{Fig:Overview_AIA}), while individual flare loops exhibit apparent slipping motion in both their footpoints. The slipping motion is clearly visible in the online Movie 1 and is analyzed in Sect. \ref{Sect:2.3}.

%
%
\subsubsection{Eruption of the flux rope}
\label{Sect:2.2.2}

After 17:10\,UT, a system of growing S-shaped flare loops occurs in the same location as the previous failed eruption of F2 ($X$\,$\approx$\,$0\arcsec$, $Y$\,$\approx$\,180$\arcsec$, Figs. \ref{Fig:Overview_AIA} and \ref{Fig:Eruption_stackplots}.) These loops are observed in AIA 131\,\AA~and 94\,\AA, but not in 171\,\AA, indicating that their temperature reaches 10\,MK, similarly as reported in \citet{Dudik14a}. This system of hot S-shaped loops subsequently accelerates to about 270 km\,s$^{-1}$ in the projected velocity as measured by the time-distance technique along the cut (Fig.\,\ref{Fig:Eruption_stackplots}, \textit{top}), and erupts in the NW direction after 17:24\,UT (Table \ref{Table:1}, Fig.\,\ref{Fig:Overview_AIA}). A growing halo of an EIT wave is observed in the other AIA channels (see Movie 2). The eruption also manifests itself as the impulsive phase, a common behaviour for the eruptive flares \citep[e.g.,][]{Moore01,Zhang01,Zhang12,Cheng13,Cheng14a,Cheng14b,Dudik14a}. Subsequently, the flare reaches its maximum X-ray flux at 17:45 and enters into the gradual phase (Fig.~\ref{Fig:Preflare}), during which the flare loops exhibit the strong-to-weak shear transition \citep{Aulanier12}.

During the eruption, the neighbouring warm coronal loops are pushed and accelerated by the eruption of the hot S-shaped loops, (white long-dashed lines in Fig.\,\ref{Fig:Eruption_stackplots}), a behaviour reported also by \citet{Zhang12} and \citet{Cheng13}. The interesting feature here is that the acceleration is non-linear \citep[see also][]{Cheng14a}, as we were unable to approximate the observed profile with a parabola. The white long-dashed line shown in the 171\,\AA~time-distance plot in Fig.\,\ref{Fig:Eruption_stackplots} correspond to a third-order polynomial approximation of the accelerated front.

We interpret this eruption of hot loops as the eruption of a flux rope \citep[see][]{Zhang12,Cheng13,Cheng14a,Cheng14b,Dudik14a}. In terms of the standard solar flare model in 3D, the hot eruption represents the post-reconnection envelope of the torus-unstable flux rope whose core is invisible in AIA bands \citep[see Sect. 4.4 of][for another such case]{Dudik14a}. In this flare, the situation is further complicated by the presence of two distinct, underlying filaments, F1 and F2, one of which (F2) undergoes a preceding failed eruption, while the filament F1 stays unperturbed, i.e., does not erupt during the entire flare similar to cases discussed in \citet{Dudik14a} and \citet{Dalmasse15}. During the impulsive phase, F1 is bordered by both flare ribbons, NR and PR (see Fig.\,\ref{Fig:Overview_AIA}, \textit{rows 4--5}), while during the gradual phase later on, F1 is overlaid by the cooling flare loops, already brightening in the AIA 304~\AA~channel. 

Based on the observed morphology with two filaments and an S-shaped hot envelope of the erupting magnetic flux rope, the magnetic configuration during the eruption is likely that of a double-decker flux rope \citep{Liu12b,Kliem14,Cheng14b}. The upper deck consists of the flux rope undergoing the torus instability together with its post-reconnection envelope constituted by the S-shaped hot loops, similar to the MHD model with a single flux rope, see \citet[][Fig.\,10 therein]{Aulanier10}, \citet[][]{Aulanier12}, \citet{Savcheva12a}, and \citet{Dudik14a}. The lower deck consists of F2 and possibly also F1. The F2 erupts before the upper deck, similar to the model of \citet{Kliem14}, while F1 stays unperturbed and is later overlaid by the flare loops. The failed eruption of F2 suggests that F2 is a flux rope rather than a sheared arcade. The magnetic structure of F1 is less certain. It could be either another flux rope or simply located in dips induced by the upper deck. Nevertheless, the present situation is more complex than the previous observational reports of the double-decker flux ropes in \citet{Liu12b} and \citet{Cheng14b} or the modeling performed by \citet{Kliem14}, where the two flux ropes corresponding to both decks are located in the same plane. In our event, this might arise as a consequence of the complex photospheric flux distribution (see Sect. \ref{Sect:2.1} and Fig.\,\ref{Fig:Preflare}, \textit{top right}). The situation is further complicated by the fact that only F2, but not F1, undergoes a failed eruption.

The temporal evolution of the emission of F2 and the following hot eruption suggest that F2 is first heated to coronal temperatures. This is indicated by F2 being a bright feature in both AIA 131~\AA~and 171~\AA~(Fig.\,\ref{Fig:Eruption_stackplots}) from about 16:56\,UT, when the F2 decelerates during the failed eruption. The brightening then subsequently fades away. Following that, at about 17:10\,UT, the growing system of S-shaped hot loops is observed with a projected velocity of $\approx$38.2\,km\,s$^{-1}$. At about 17:20\,UT, the system of the growing S-shaped loops reaches the position of the halted F2 indicated by the black horizontal dotted line in the 171\,\AA~time-distance plot. Since at this time F2 is faint, we are not able to discern whether the filament material returned to lower heights by flowing along its legs, or whether F2 merged with the erupting hot S-shaped loops of the same chirality \citep[see also,][]{Devore05}, i.e., the upper deck. The overlap of the dotted line at the position of about 50$\arcsec$ in the AIA 171~\AA~time-distance plot with the growing S-shaped loops in 131~\AA~at about 17:15 -- 17:20\,UT (Fig.\,\ref{Fig:Eruption_stackplots}) indicates that at least a partial merging of F2 with the erupting upper-deck flux rope is a possibility \citep[c.f.,][]{Kliem14}. We note however that in another flare, merging of two flux ropes and the associated presence of the hot plasma has already been reported by \citet{Joshi14}, while \citet{Karlicky11} have shown that the merging of two plasmoids (representing flux ropes in 2D) can lead to particle acceleration, heating, and X-ray emission.

The evolution of the flare emission and its morphology suggests that the breakout-type reconnection \citep[see, e.g.,][]{Antiochos99,Sterling04a,Sterling04b,Lynch04,Lynch08} is not the primary cause of the rise of F2 and the subsequent hot eruption. This is since (1) the flare and the associated rearranging of the neighbouring warm coronal loops starts even before the initiation of F2 rise, and that (2) the three lobes required in breakout in a quadrupolar geometry are not identifiable in our flare. Therefore, we propose that the flare is driven by a tether-cutting reconnection building the upper, erupting flux rope, \citep[as argued already by][]{Cheng15}, with a transition to an ideal MHD (torus) instability of the erupting flux rope \citep[c.f.,][]{Aulanier10,Aulanier12,Inoue14a,Inoue15}, and a feedback between the instability and reconnection \citep{Savcheva12b}. It is possible that this transition occurs after the merging of F2 and the hot flux rope. Such merging, entailing reconnection or flux transfer between the lower and upper flux ropes, could lead to an increase of flux and twist in the upper flux rope, which may thereby render it unstable. This scenario is supported by the fact that the time 17:20\,UT, as well as the corresponding position of 50$\arcsec$ on the time-distance plots in Fig.\,\ref{Fig:Eruption_stackplots} are both located closely to the onset of fast eruption, visible after 17:24\,UT.



%
\subsubsection{Apparent implosion and oscillations of the warm coronal loops}
\label{Sect:2.2.3}

After the onset of the fast eruption, the warm coronal loops exibit oscillations with a typical period of several minutes (Fig.\,\ref{Fig:Eruption_stackplots}). These oscillating loops are observed in 171~\AA, 193~\AA, and 211~\AA. They occur after about 17:30\,UT, i.e., after the onset of the impulsive phase, and extend into the flare maximum at 17:45\,UT. At this time, after several periods, the oscillations have mostly damped. In our case, the onset of oscillations following the hot eruption is observed without a preceding strong contraction phase, reported for other event by \citet{Simoes13a}, and interpreted as a coronal implosion \citep[see][]{Hudson00}.

Here, we point out that the oscillations and the associated change in loop position occur only \textit{after} the onset of the fast eruption (see Fig.\,\ref{Fig:Overview_AIA}). Together with the rising of some of the coronal loops observed prior to the eruption, this indicates that the oscillations are a result of the displacement of the position of the coronal loops due to the large-scale dynamics of the magnetic field during the ongoing slipping magnetic reconnection (Sect. \ref{Sect:2.3}) and the flux rope eruption. Since the hot flux rope (10\,MK as evidenced by AIA 131~\AA) exists and is built before and during the eruption, the loop oscillations are \textit{not} the result of a coronal implosion, a mechanism for energy release proposed by \citet{Hudson00}. Rather, the apparent ``implosion'' accompanying the loop oscillations \citep[see also][]{Simoes13a} is a behaviour driven by the large-scale dynamics of the magnetic field during the flux rope eruption.

Instead, the observation of a contracting loop with a speed of $-2.9$\,$\pm0.9$\,km\,s$^{-1}$ may be a better candidate for the coronal implosion, since this contraction starts around the beginning of the flare (but before F2 rise). However, the 12\,s cadence of the AIA observations together with its spatial resolution do not allow us to unambiguously determine the exact time of the onset of this contraction. Similarly, determination of the onset of flare-related reconnection from AIA 131\AA~data is precluded by the limited cadence and the presence of background emission at coronal temperatures. We note that the timescales needed for filling of the flare loops with dense enough hot plasma in AIA 131~\AA~due to limited velocities chromospheric evaporation, of the order of several hundreds of km\,s$^{-1}$ \citep[see, e.g.,][]{Graham11,Young13,Young15,Doschek13,Brosius13,Polito15a,Tian14,Tian15,Graham15}, represent an additional complication.

Because of these limitations, it is not possible to determine whether this contraction of a single loop starts before the onset of the flare reconnection, or vice versa. This is in contrast to the behaviour reported by \citet{Shen14}, where the peripheral coronal loops started contracting only after the onset of their flare. Nevertheless, if the contraction is a signature of a coronal implosion, in our flare it is a very weak signature of only a single loop structure. It seems unlikely that this single and quite localised structure can account for the overall flare dynamics; furthermore, the energy release during the flare increases strongly with time, but there are no corresponding signatures of accelerating implosion. On the contrary, the contraction of this loop is no longer detectable at 17:10\,UT, well before the impulsive phase.

The onset of loop oscillations after the eruption is connected with a modulated radio signal detected by the \textit{Callisto} radio spectrometer \citep{Benz09,Monstein13} network station at the Birr castle in Ireland. This network station measures linearly polarized solar radio flux in the horizontal direction. In Fig.\,\ref{Fig:Eruption_stackplots} \textit{bottom}, we show the lightcurve at the frequency of 350\,MHz, smoothed with a 2\,s boxcar to reduce the noise. The signal shows an increase from about 17:07\,UT, a spike at 17:25:40\,UT, and then two strong maxima at about 17:28:20\,UT and 17:32:30\,UT. A weaker third maximum occurs at about 17:38\,UT. The spike occurs approximately at the onset of the fast eruption. The following maxima appear to be in phase with the oscillating loops detected in the AIA 171~\AA~time-distance plot at the cut position of $\approx$160$\arcsec$, see Fig.\,\ref{Fig:Eruption_stackplots}. To our knowledge, this is a first possible detection of loop oscillations modulating the solar radio flux. It is however not clear at present why the amplitudes of these maxima are different.


\begin{figure*}
	\centering
	\includegraphics[width=9.39cm,clip,bb= 0  0 495 85]{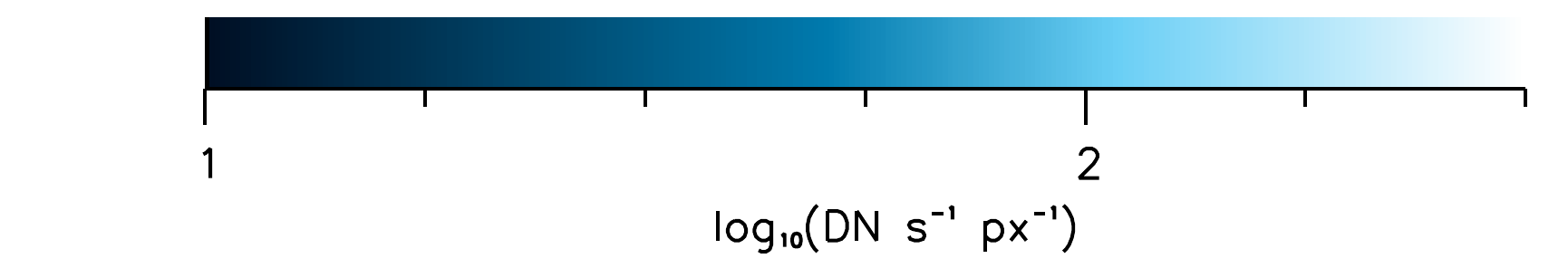}
	\includegraphics[width=8.21cm,clip,bb=62  0 495 85]{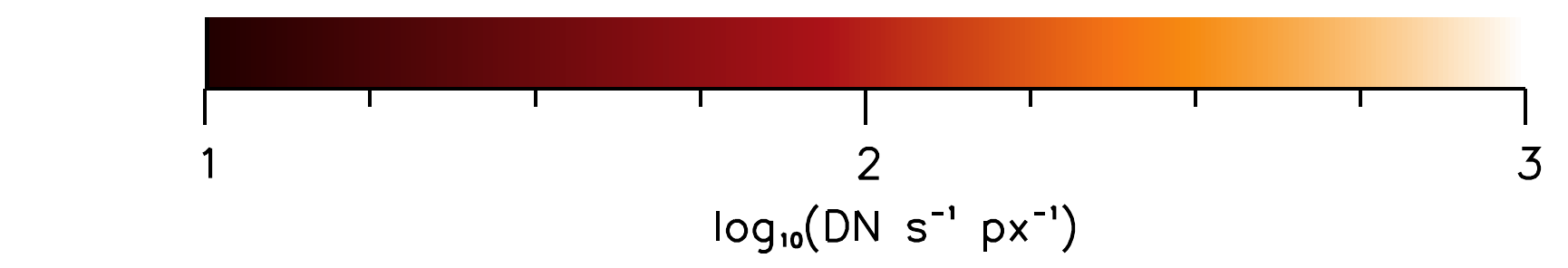}
	\includegraphics[width=4.42cm,clip,bb= 0 40 243 225]{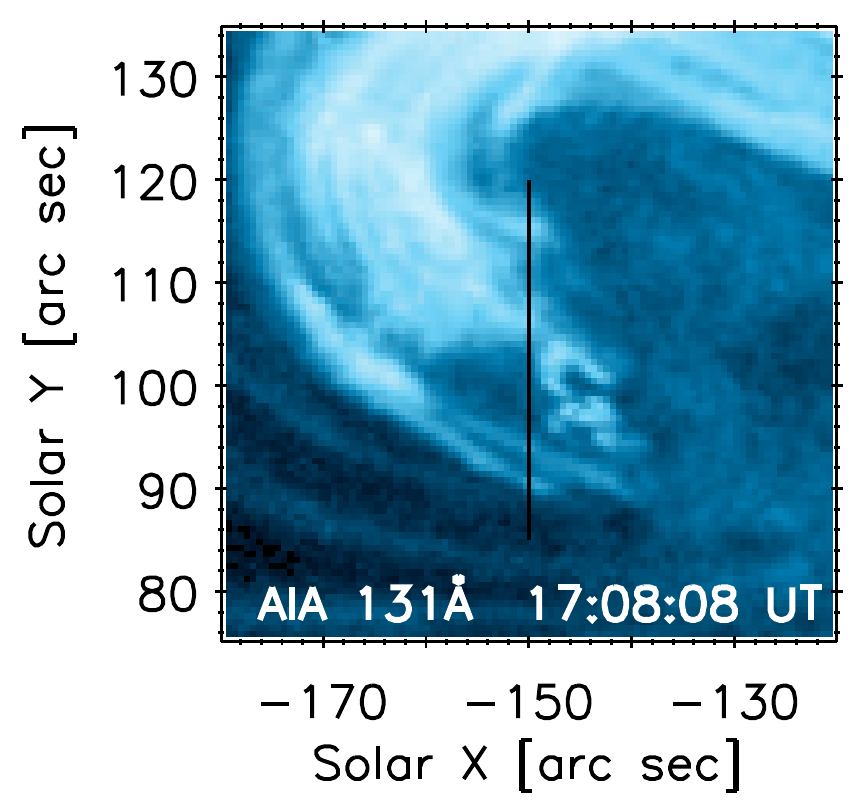}
	\includegraphics[width=3.29cm,clip,bb=62 40 243 225]{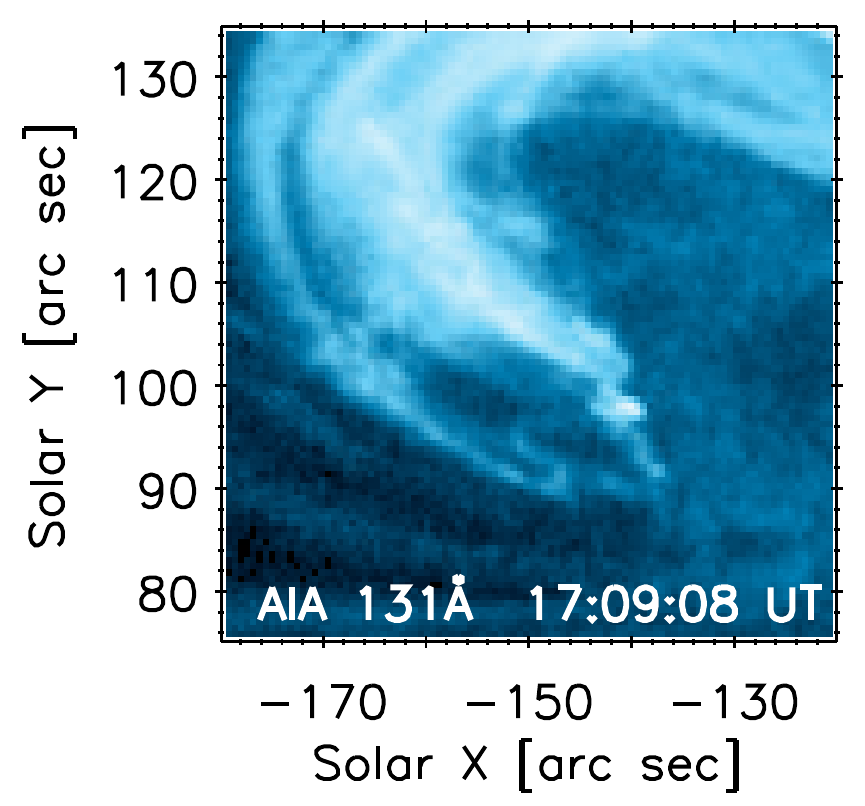}
	\includegraphics[width=3.29cm,clip,bb=62 40 243 225]{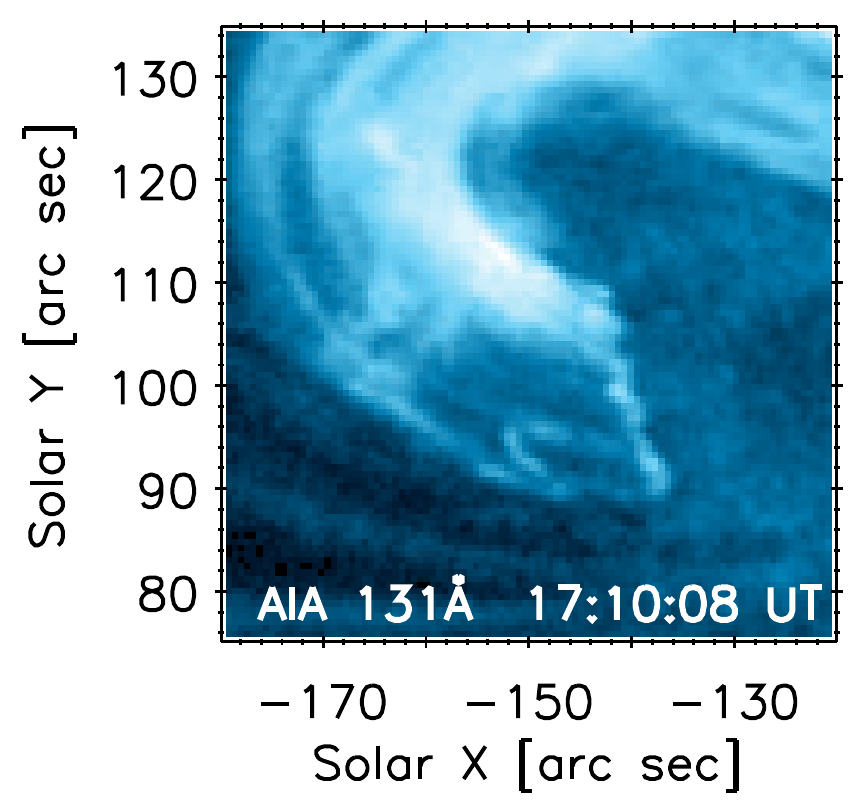}
	\includegraphics[width=3.29cm,clip,bb=62 40 243 225]{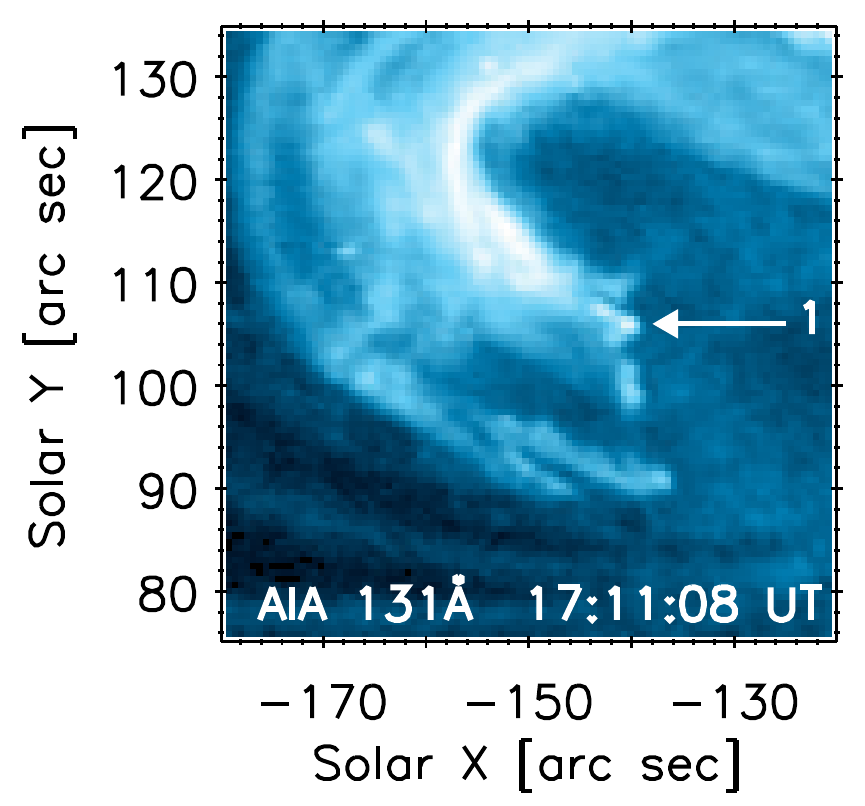}
	\includegraphics[width=3.29cm,clip,bb=62 40 243 225]{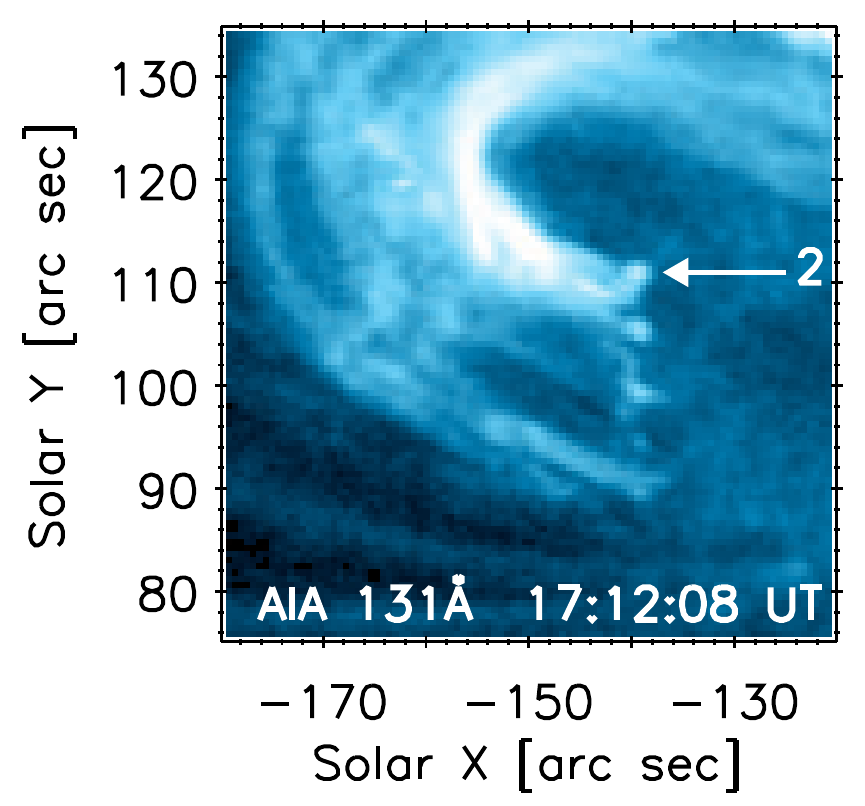}
	\includegraphics[width=4.42cm,clip,bb= 0  0 243 225]{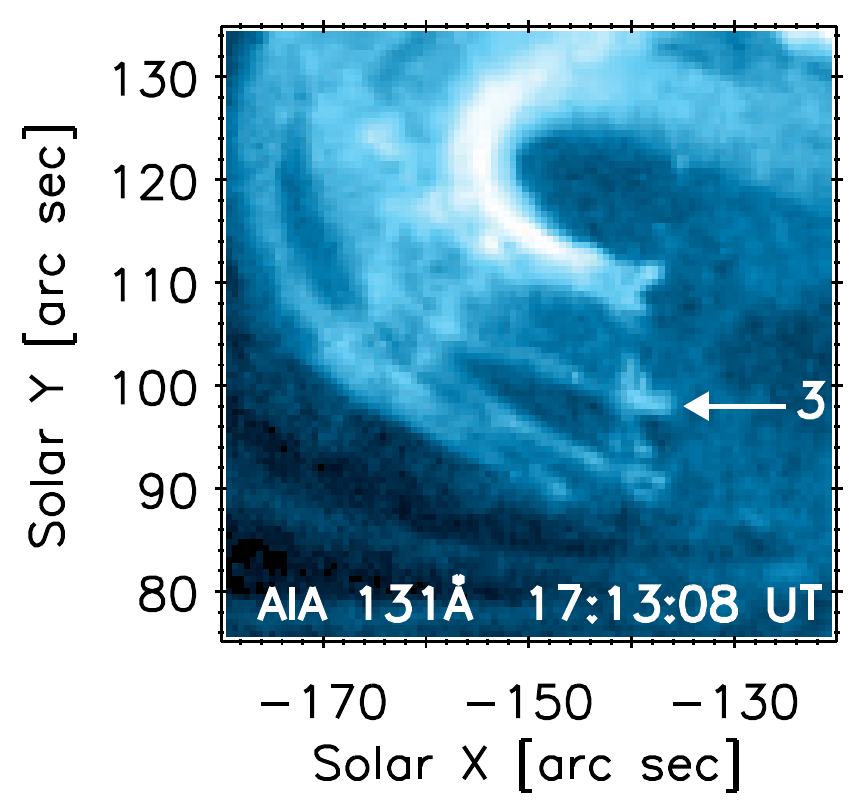}
	\includegraphics[width=3.29cm,clip,bb=62  0 243 225]{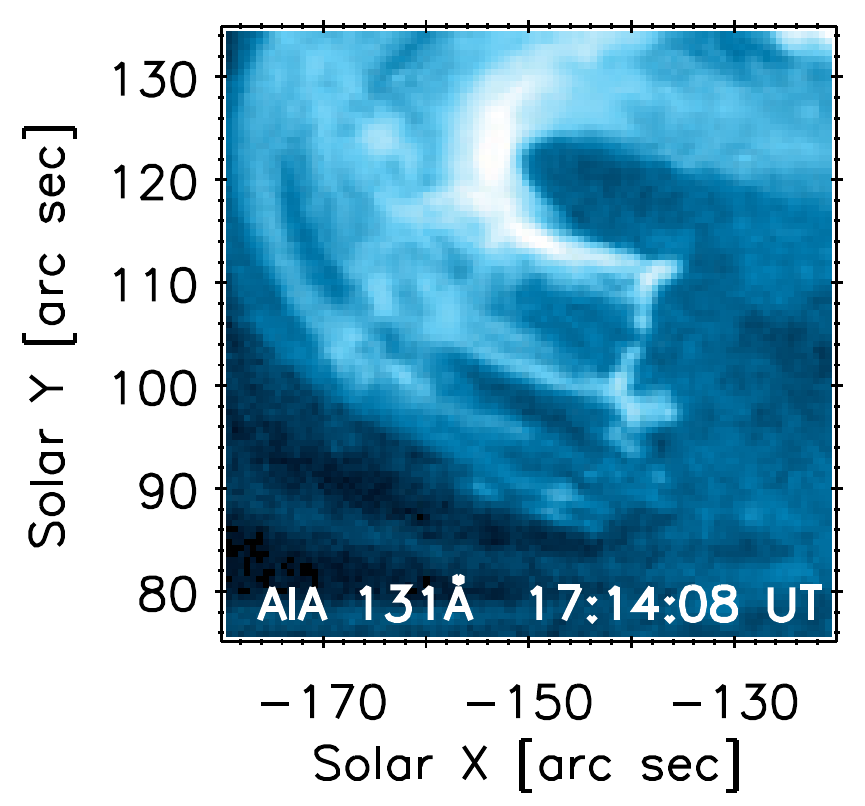}
	\includegraphics[width=3.29cm,clip,bb=62  0 243 225]{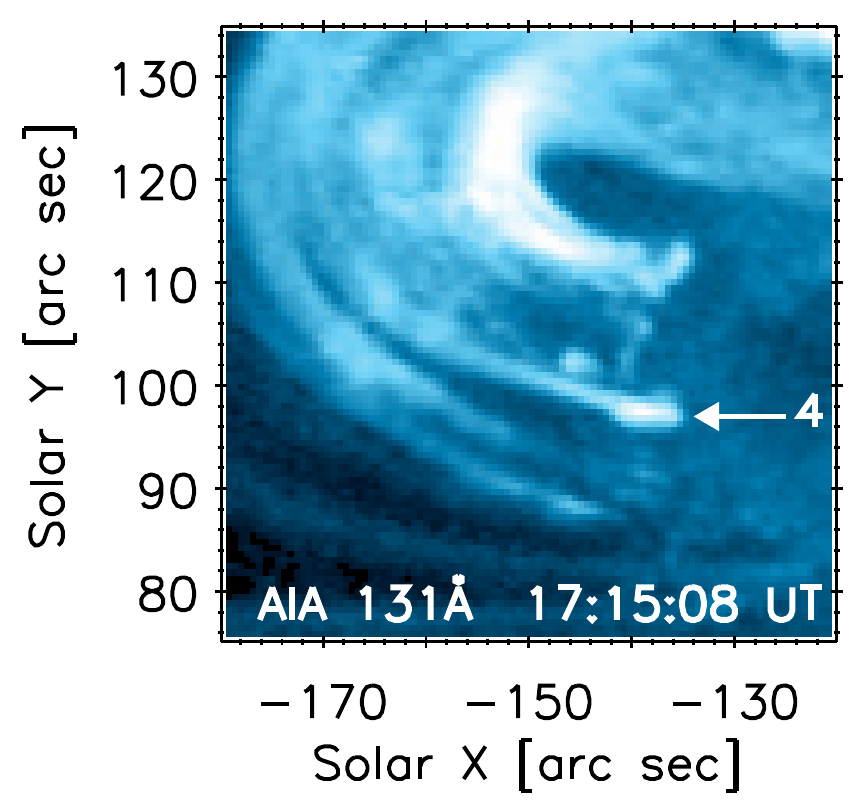}
	\includegraphics[width=3.29cm,clip,bb=62  0 243 225]{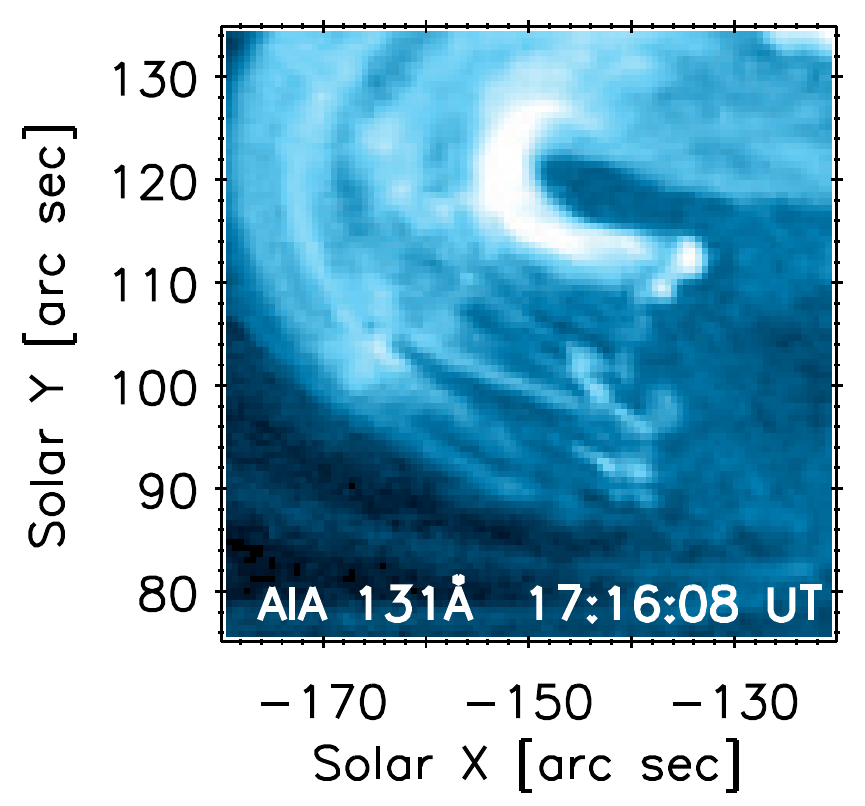}
	\includegraphics[width=3.29cm,clip,bb=62  0 243 225]{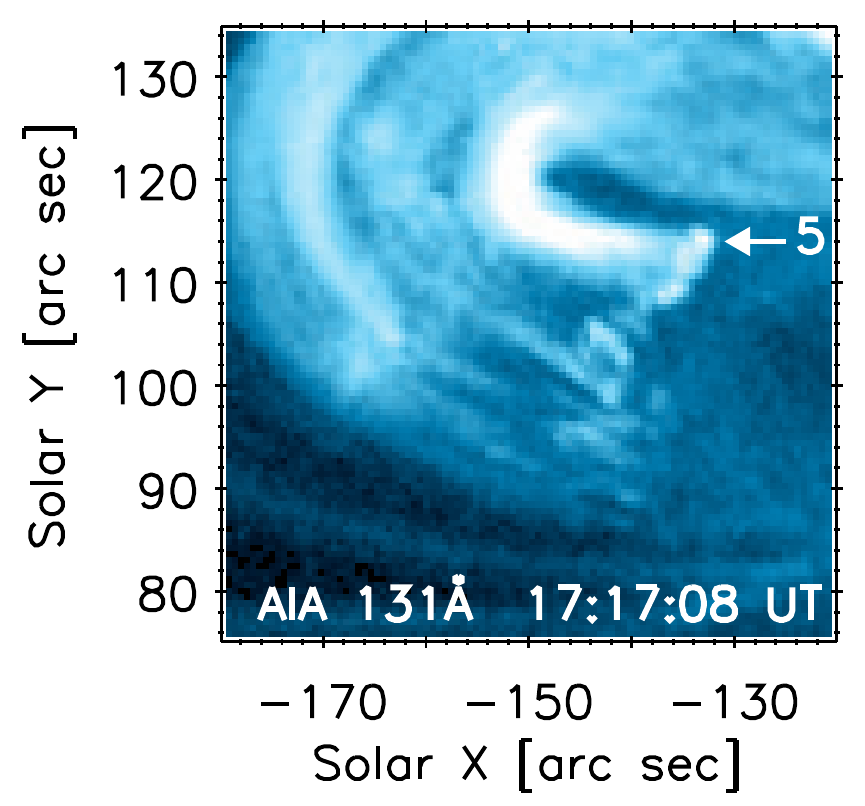}
	\includegraphics[width=4.42cm,clip,bb= 0 40 243 225]{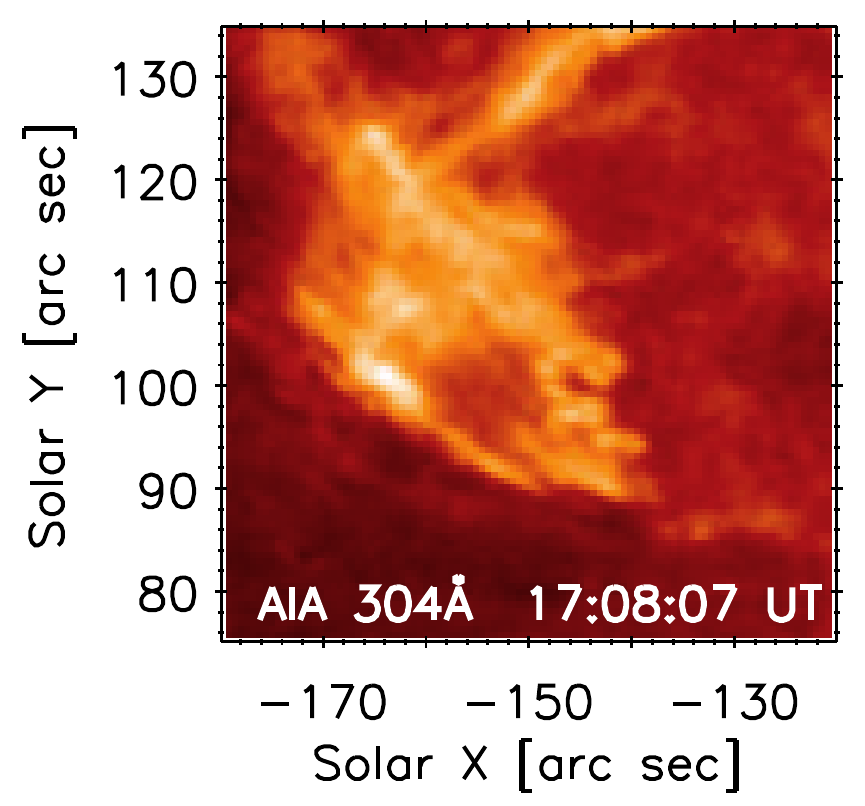}
	\includegraphics[width=3.29cm,clip,bb=62 40 243 225]{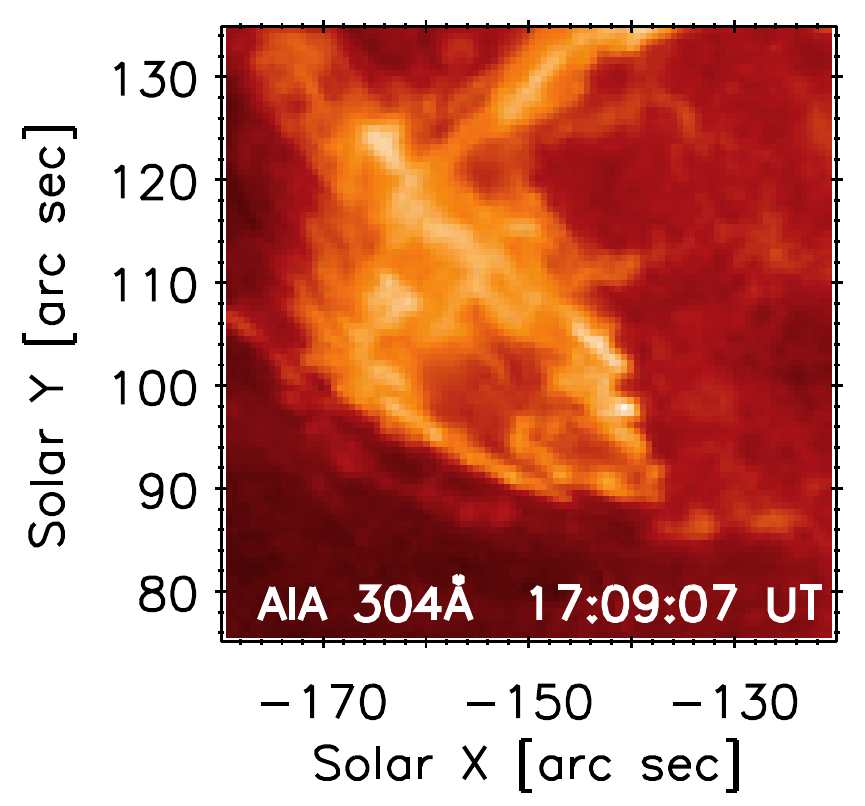}
	\includegraphics[width=3.29cm,clip,bb=62 40 243 225]{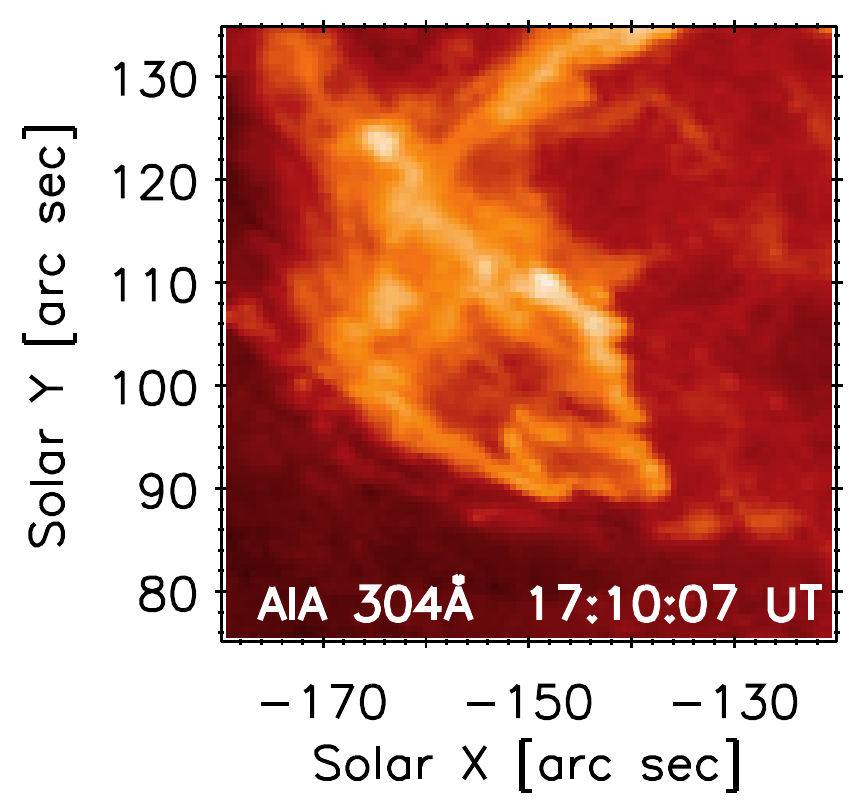}
	\includegraphics[width=3.29cm,clip,bb=62 40 243 225]{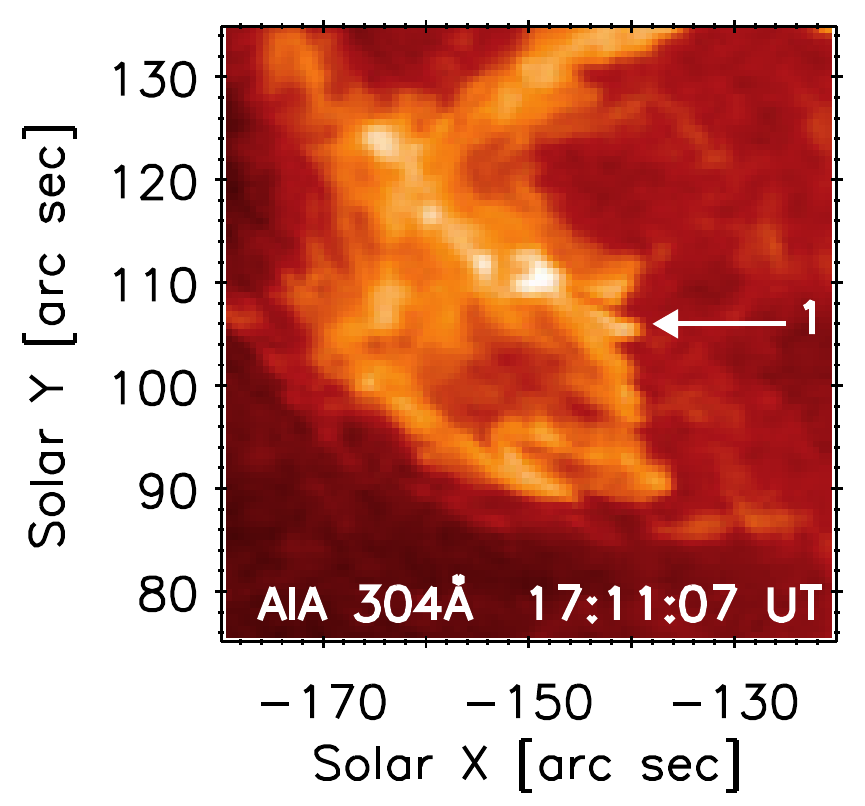}
	\includegraphics[width=3.29cm,clip,bb=62 40 243 225]{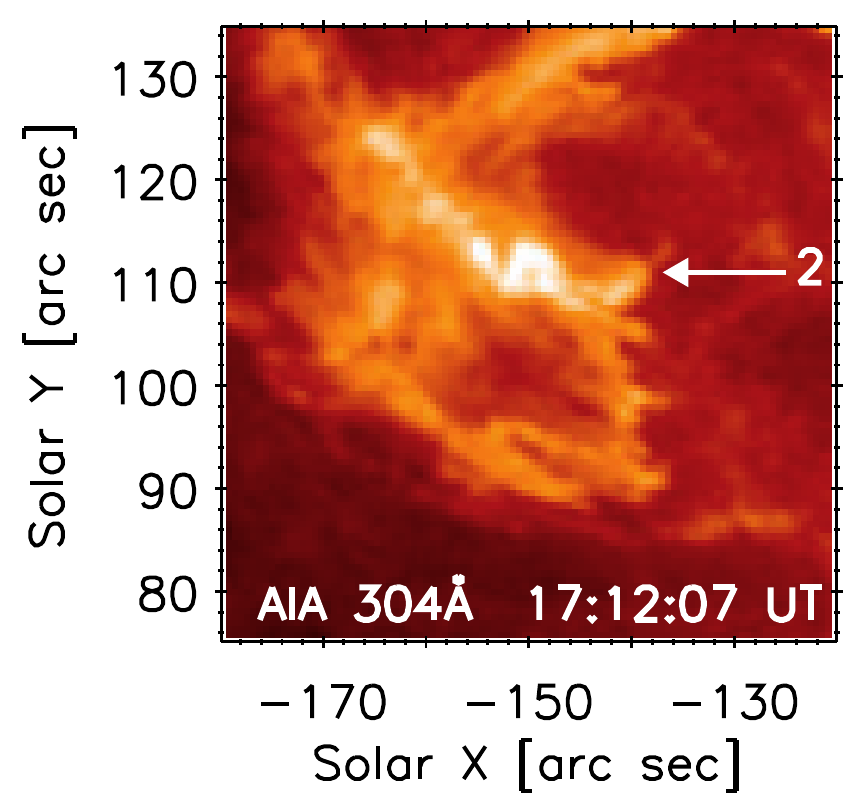}
	\includegraphics[width=4.42cm,clip,bb= 0  0 243 225]{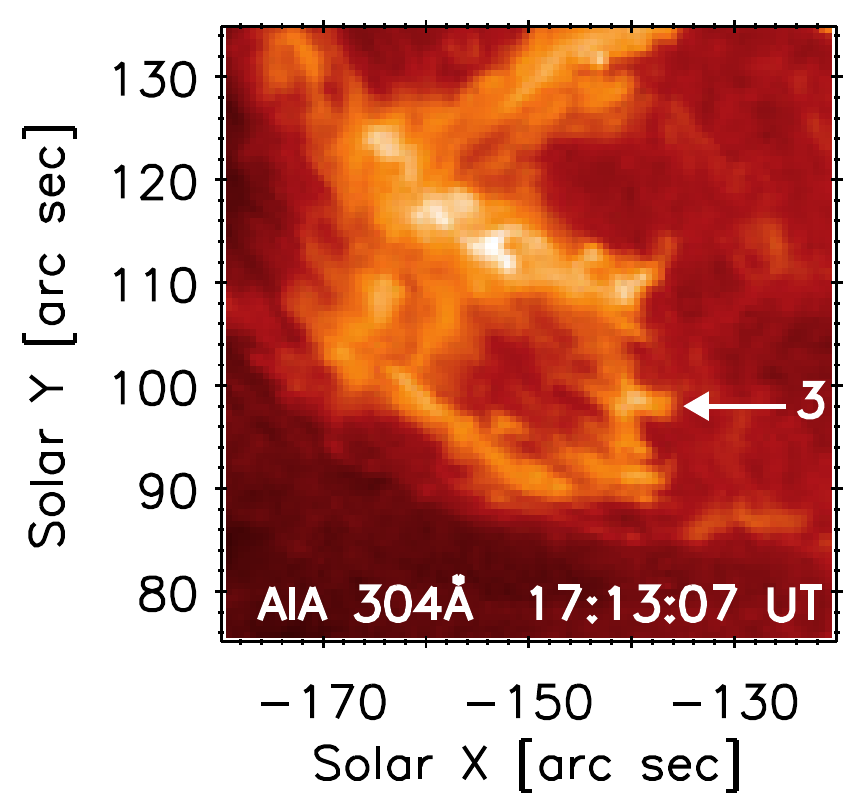}
	\includegraphics[width=3.29cm,clip,bb=62  0 243 225]{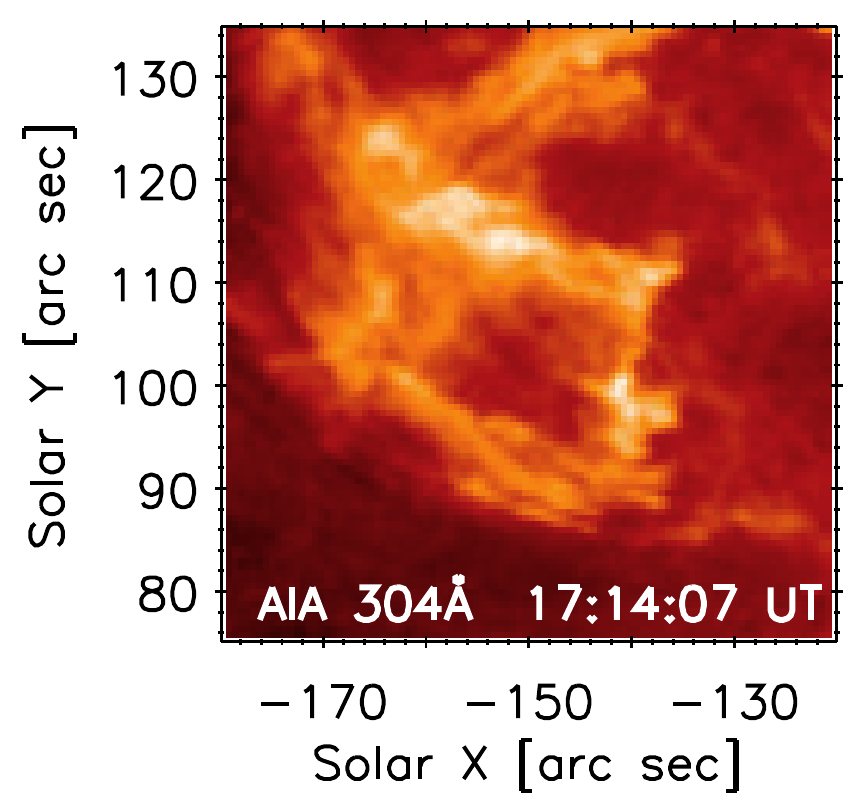}
	\includegraphics[width=3.29cm,clip,bb=62  0 243 225]{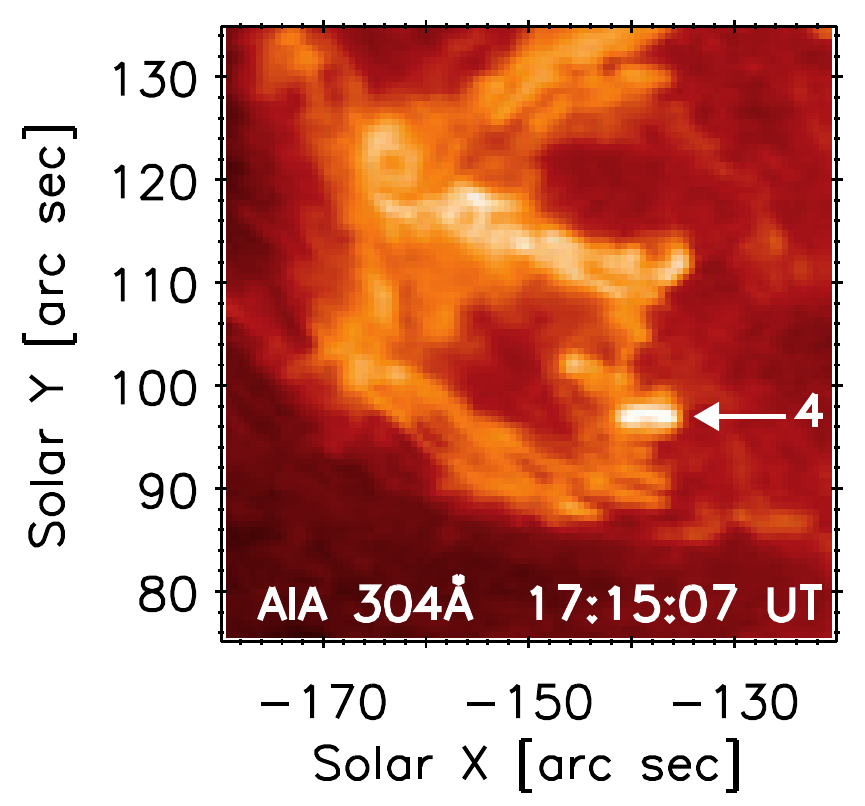}
	\includegraphics[width=3.29cm,clip,bb=62  0 243 225]{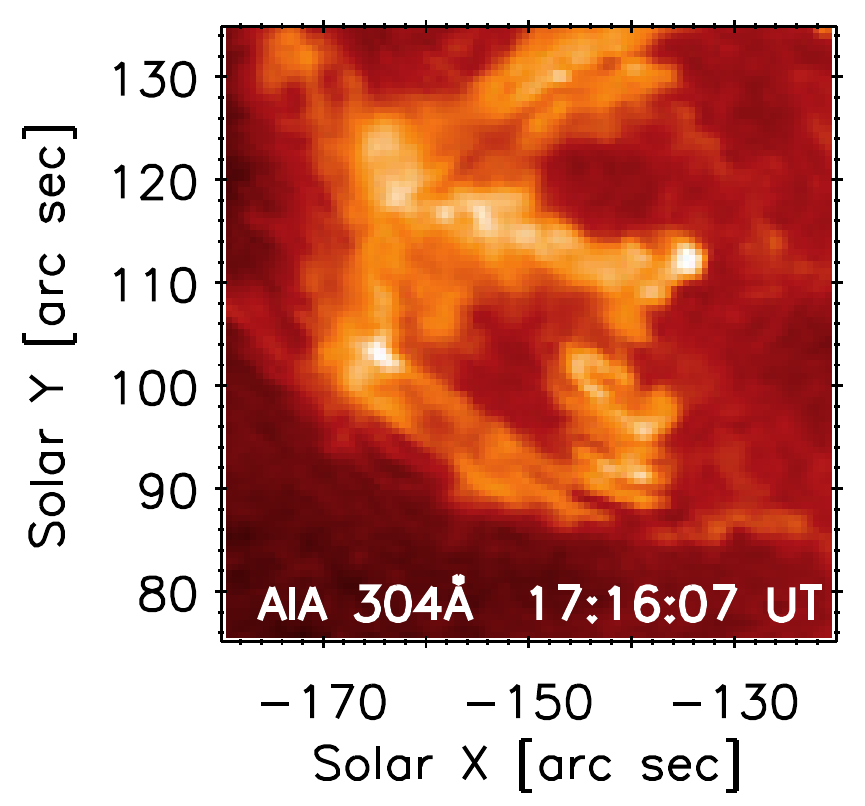}
	\includegraphics[width=3.29cm,clip,bb=62  0 243 225]{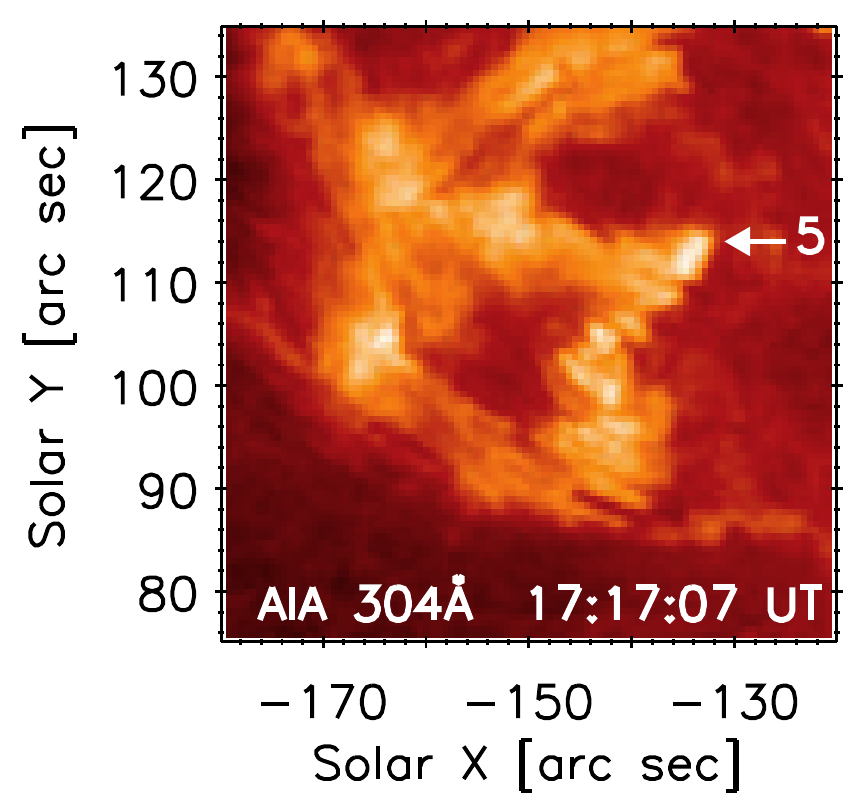}
\caption{Ribbon NR at 17:08 -- 17:17\,UT observed by AIA in the 131~\AA~and 304~\AA~passbands. The dark line represents the cut along which the time-distance plots shown in Fig.\,\ref{Fig:NR_stackplots_slit2a} are produced. The white Arrows 1--5 denote several conspicuous flare loop footpoints. See text for details. \\
A color version of this image is available in the online journal.
\label{Fig:NR_AIA_slit2a}}
\end{figure*}
\begin{figure*}
	\centering
	\includegraphics[width=8.80cm,clip,bb= 0  0 495 385]{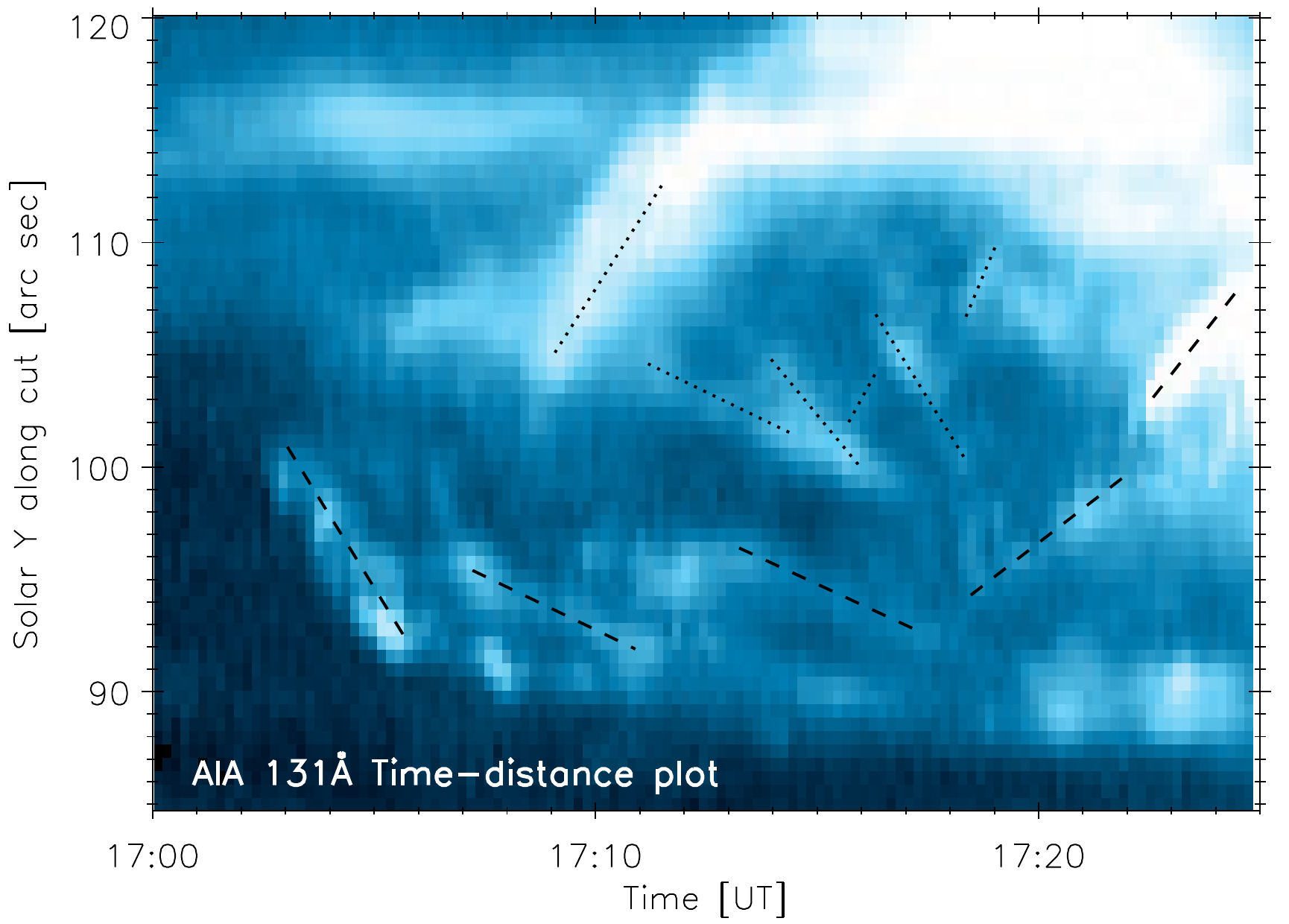}
	\includegraphics[width=8.80cm,clip,bb= 0  0 495 385]{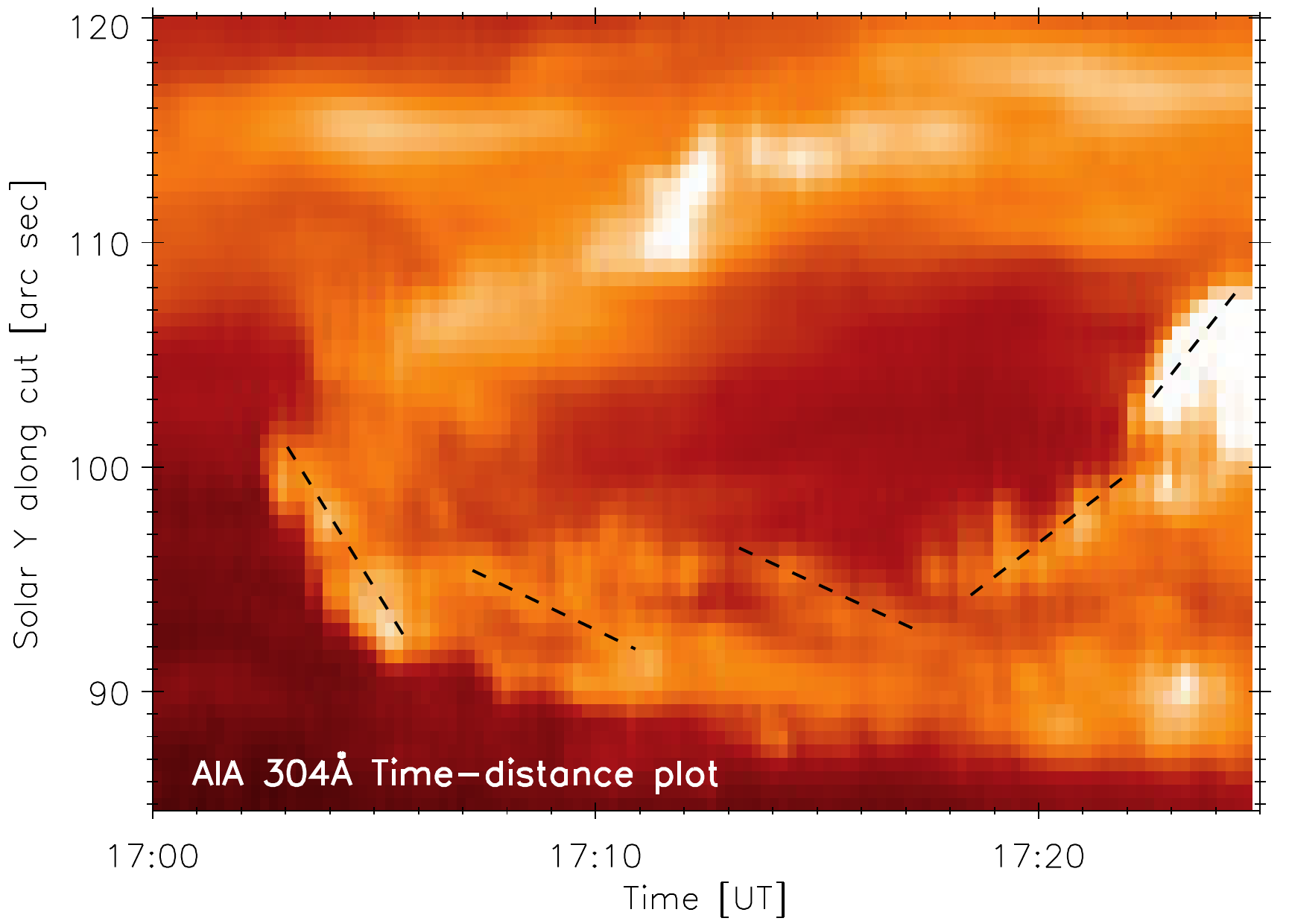}
\caption{AIA 131~\AA~time-distance plots along the cut shown in Fig.\,\ref{Fig:NR_AIA_slit2a}. Individual dotted lines denote some of the brightest slipping loops with velocities of 11 -- 57 km\,s$^{-1}$. The intensity scaling is the same as in Fig.\,\ref{Fig:NR_AIA_slit2a}. The dashed lines indicate some of the slipping bright ribbon knots in the AIA 304\,\AA. See text for details.
\label{Fig:NR_stackplots_slit2a}}
\end{figure*}
\begin{figure*}
	\centering
	\includegraphics[width=6.40cm,clip,bb= 0  0 495 85]{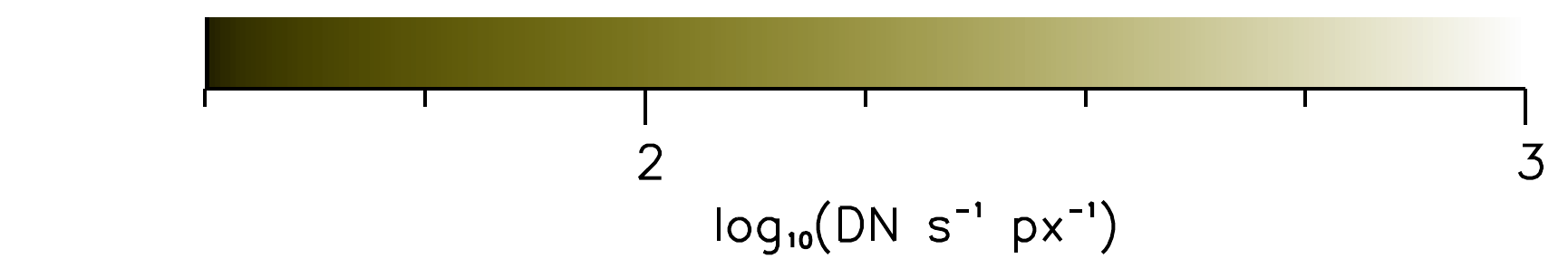}
	\includegraphics[width=5.60cm,clip,bb=62  0 495 85]{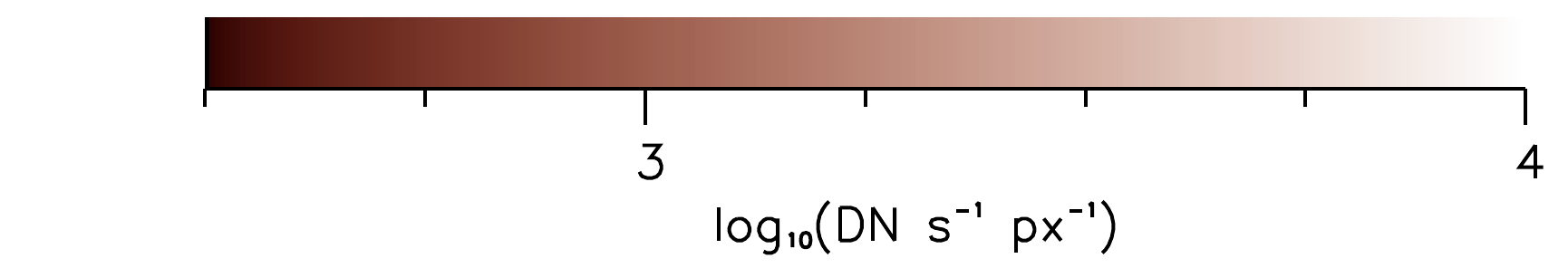}
	\includegraphics[width=5.60cm,clip,bb=62  0 495 85]{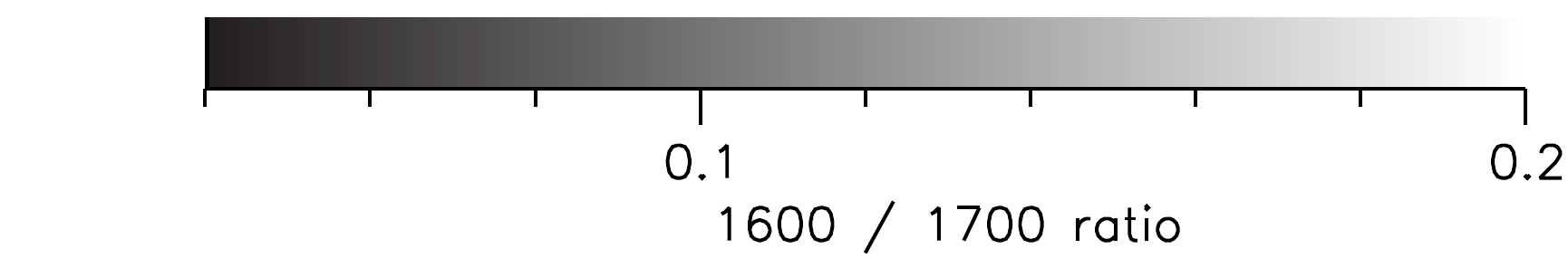}
	\includegraphics[width=4.42cm,clip,bb= 0 40 243 225]{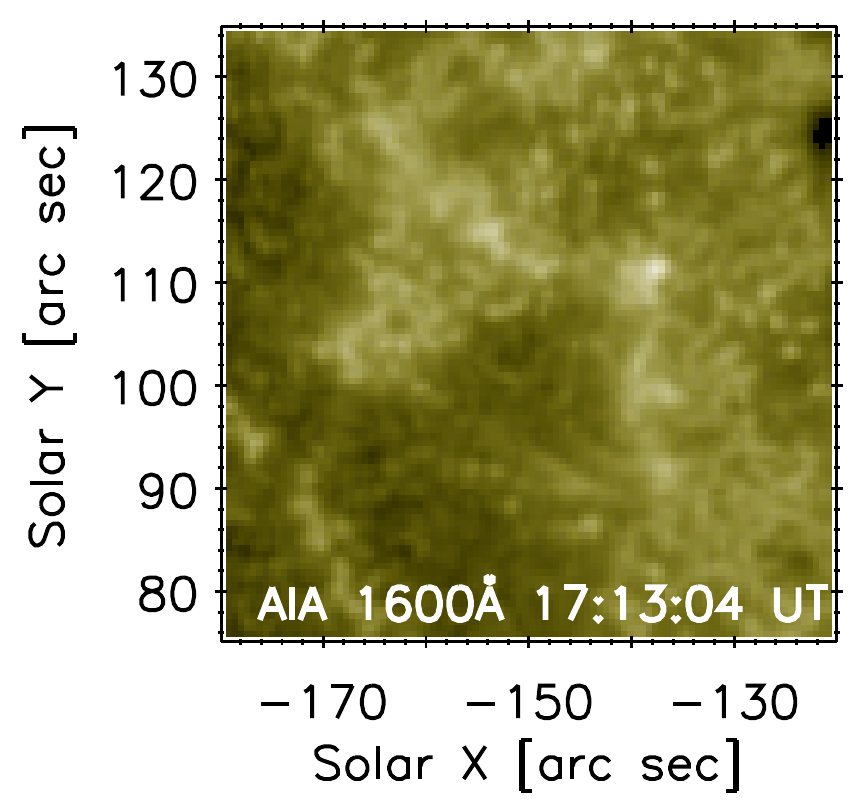}
	\includegraphics[width=3.29cm,clip,bb=62 40 243 225]{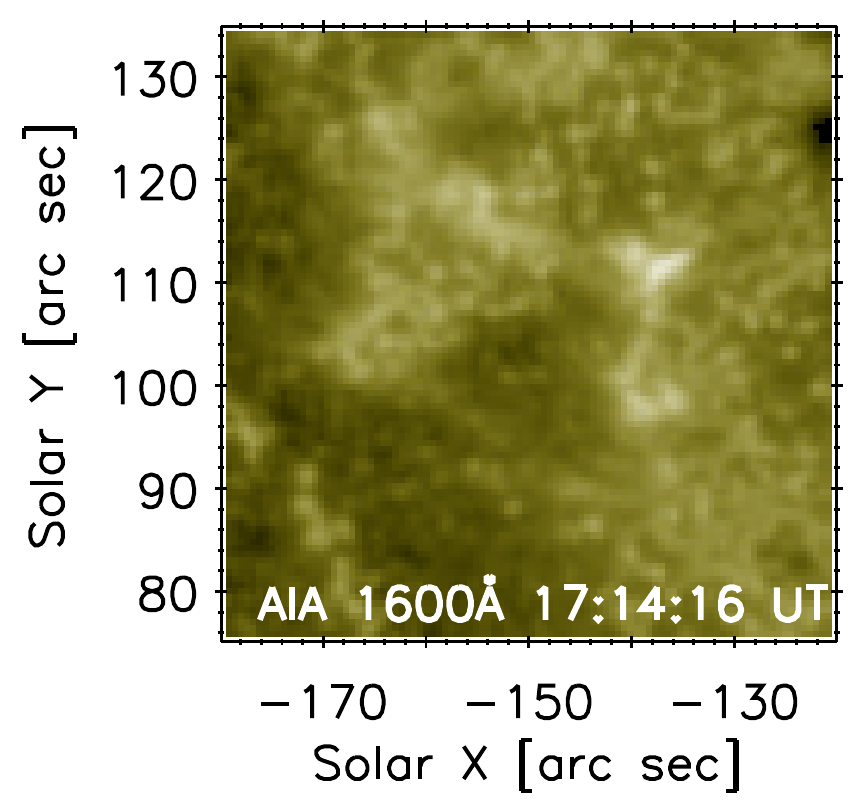}
	\includegraphics[width=3.29cm,clip,bb=62 40 243 225]{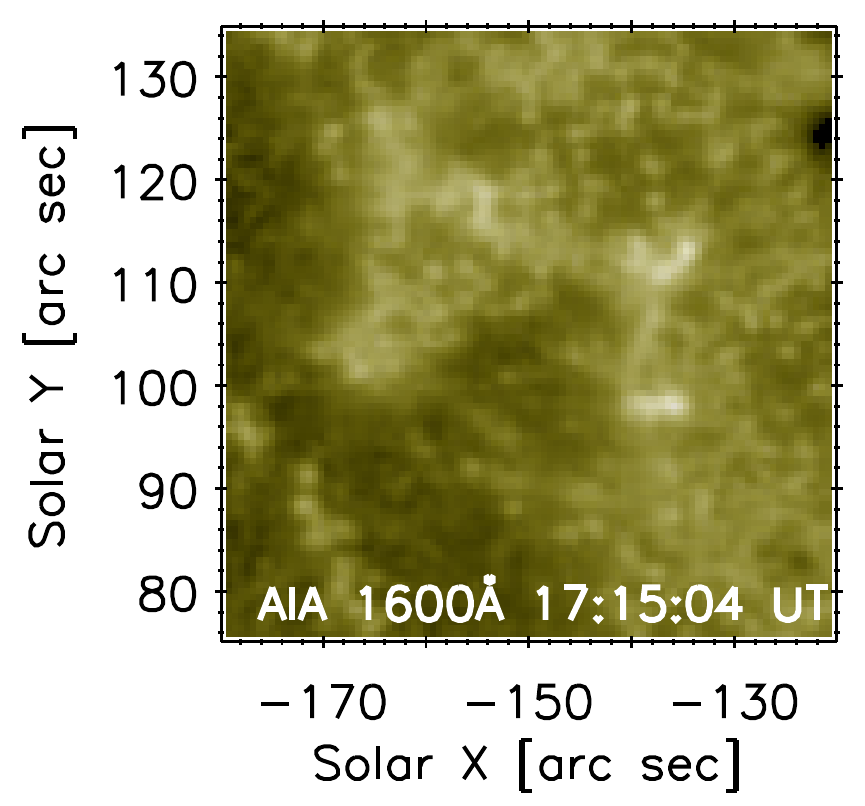}
	\includegraphics[width=3.29cm,clip,bb=62 40 243 225]{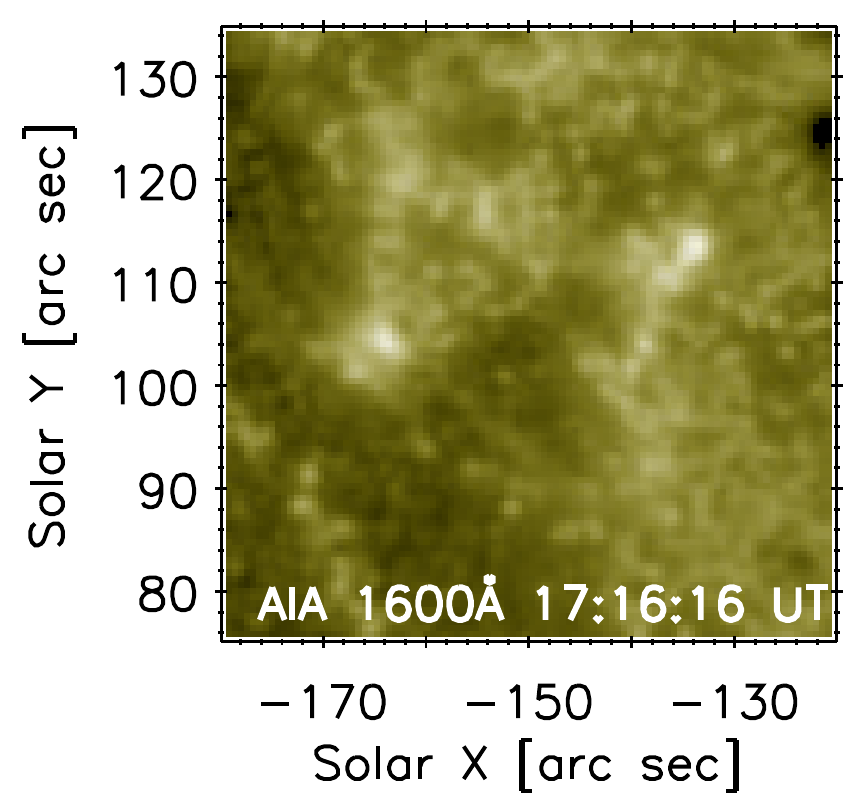}
	\includegraphics[width=3.29cm,clip,bb=62 40 243 225]{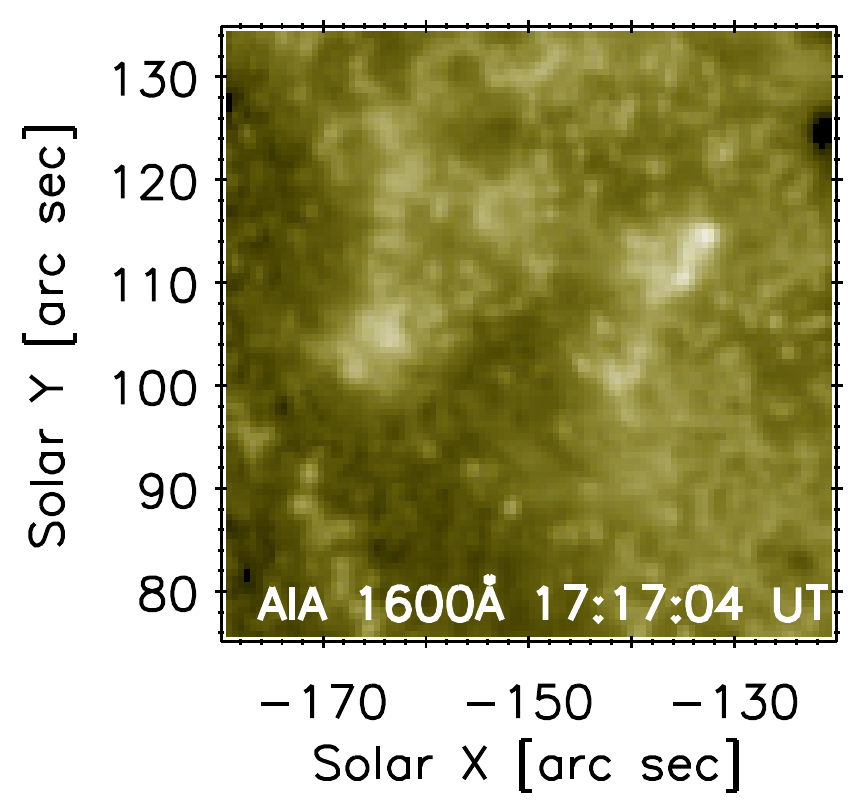}
	\includegraphics[width=4.42cm,clip,bb= 0 40 243 225]{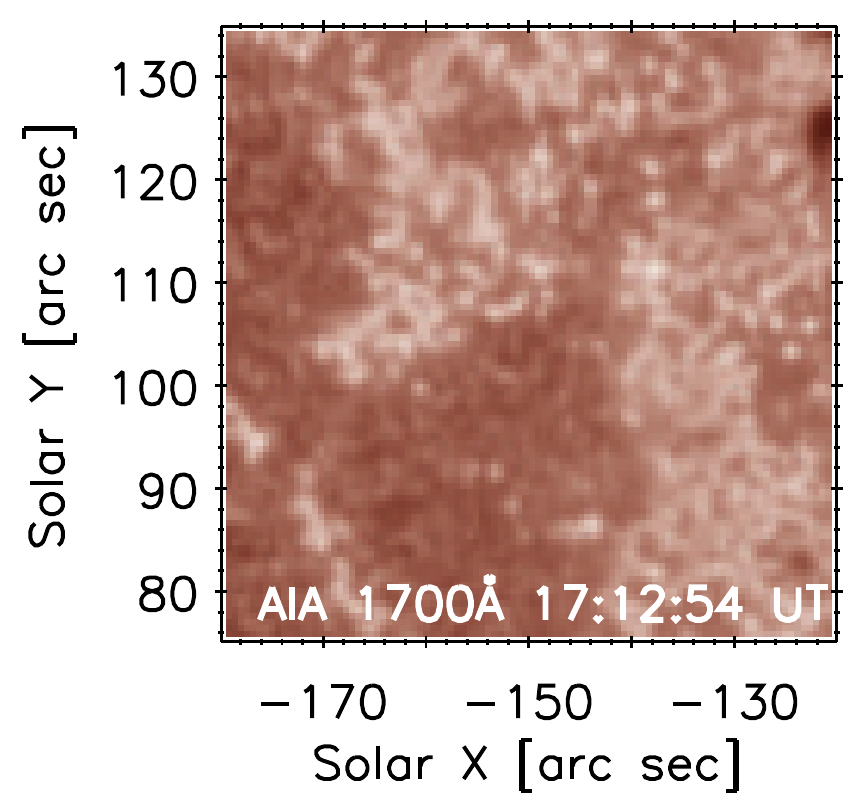}
	\includegraphics[width=3.29cm,clip,bb=62 40 243 225]{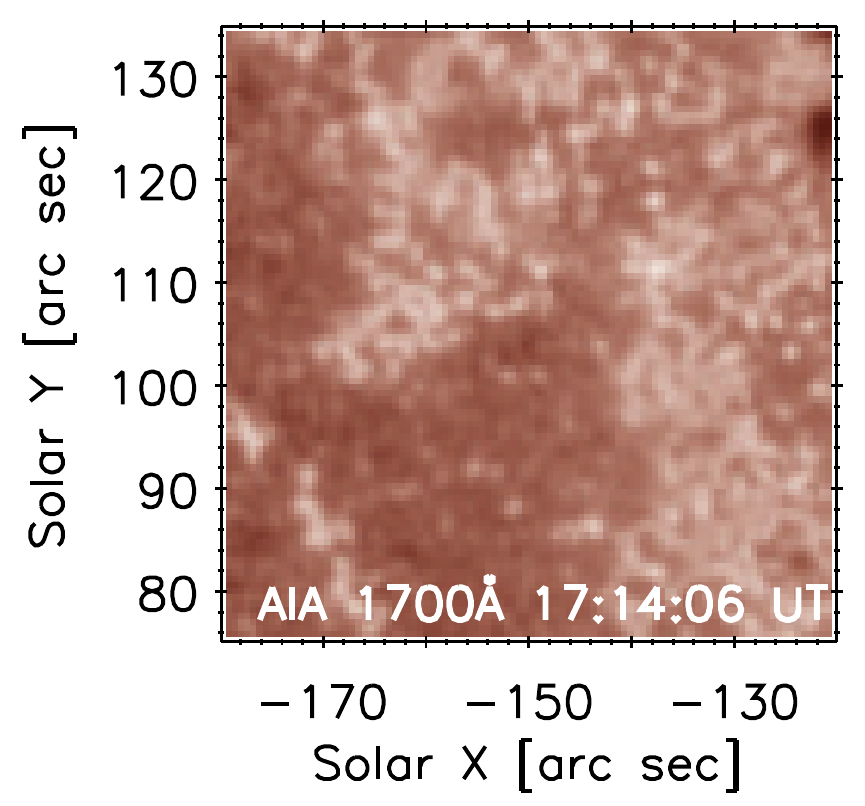}
	\includegraphics[width=3.29cm,clip,bb=62 40 243 225]{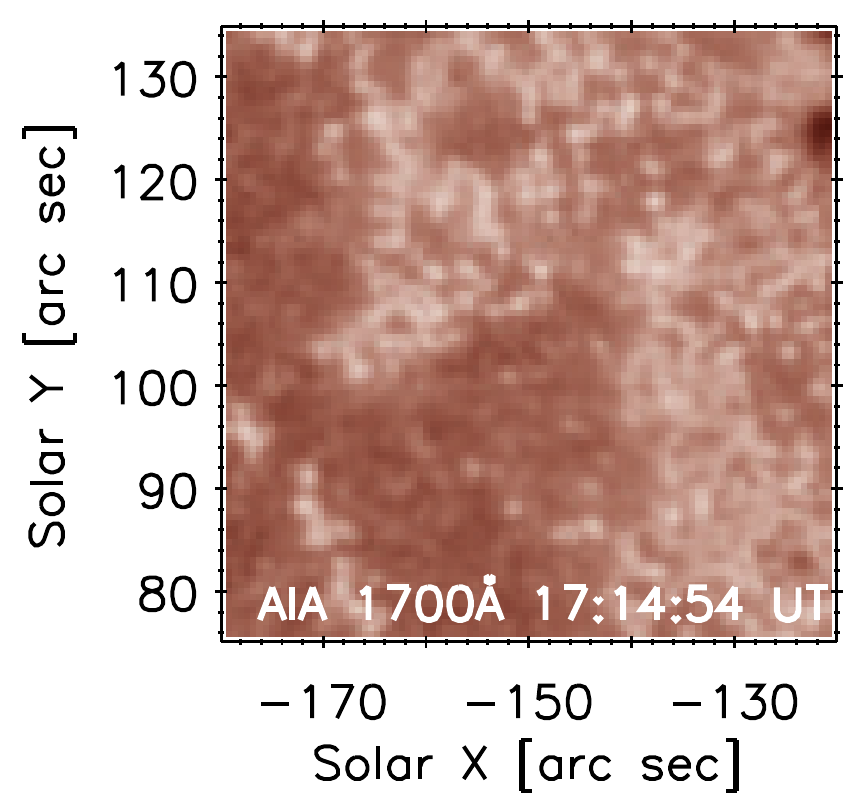}
	\includegraphics[width=3.29cm,clip,bb=62 40 243 225]{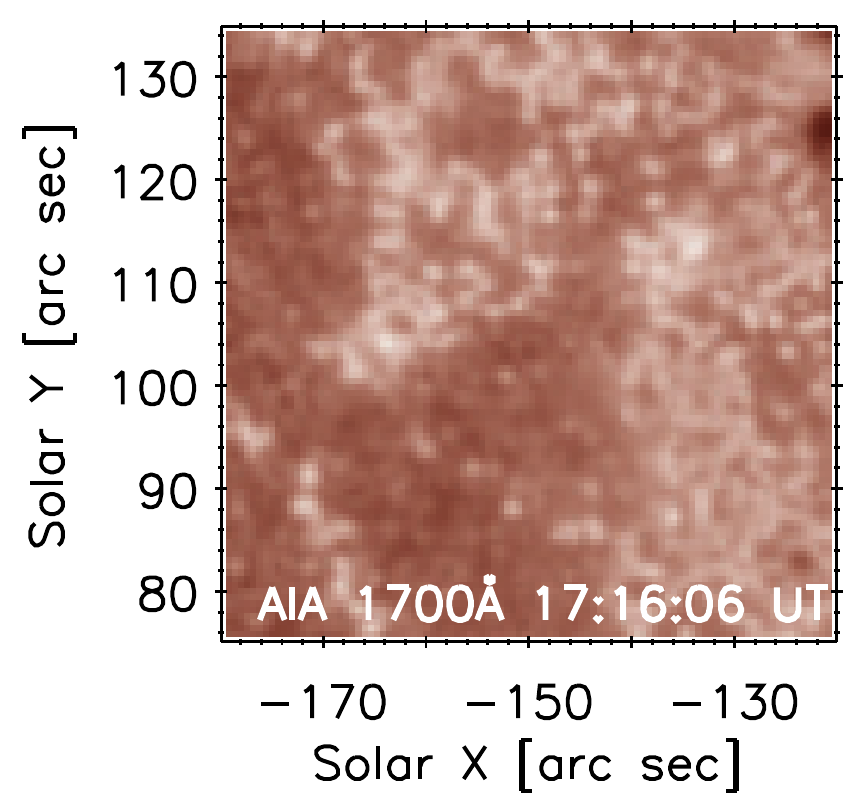}
	\includegraphics[width=3.29cm,clip,bb=62 40 243 225]{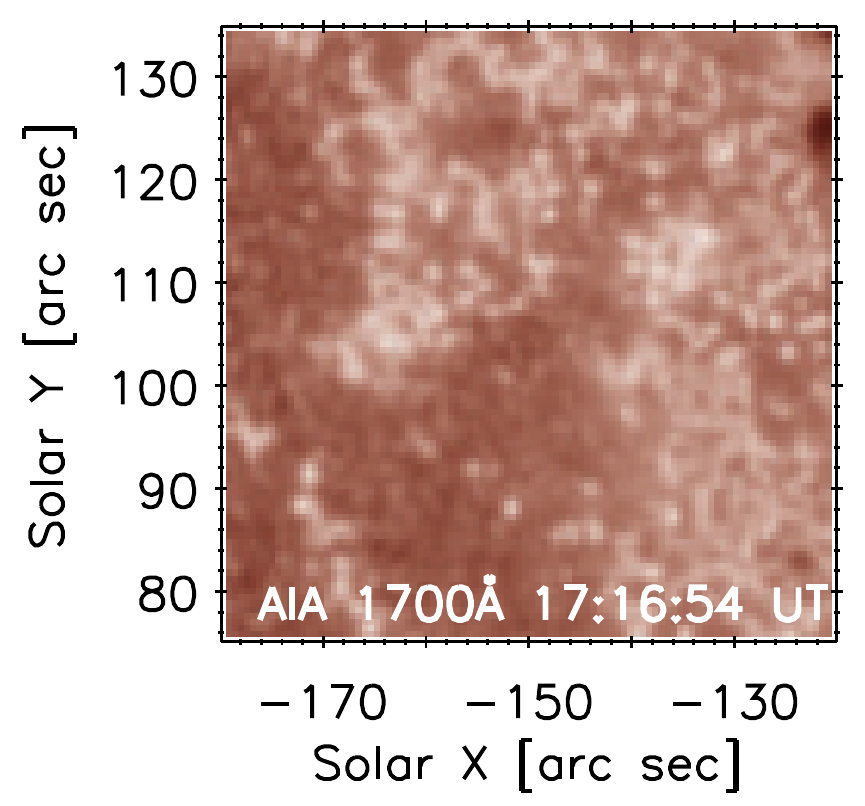}
	\includegraphics[width=4.42cm,clip,bb= 0  0 243 225]{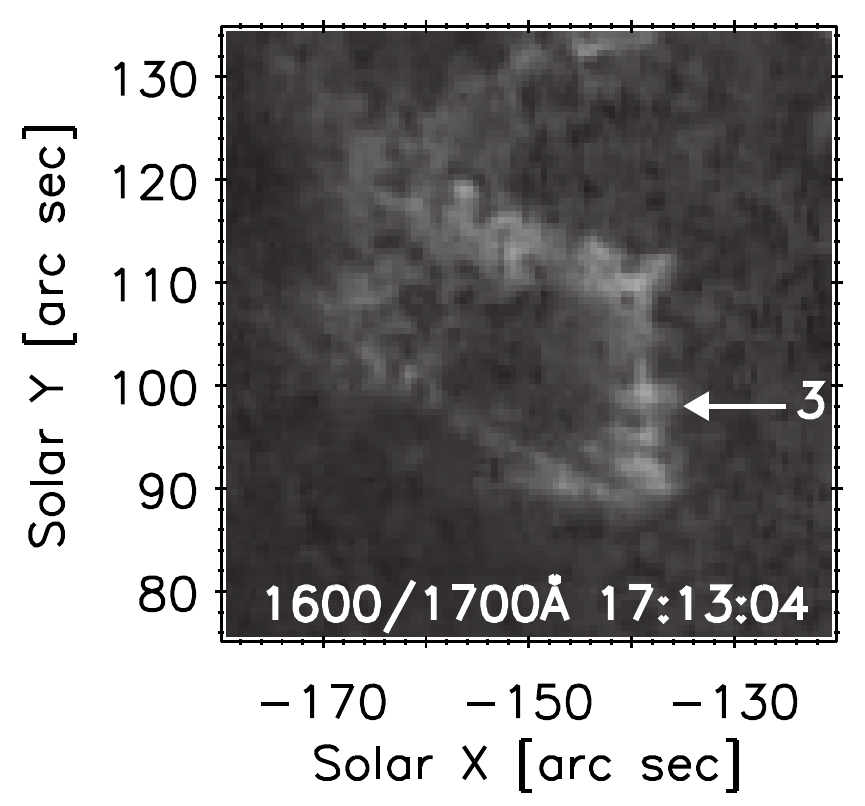}
	\includegraphics[width=3.29cm,clip,bb=62  0 243 225]{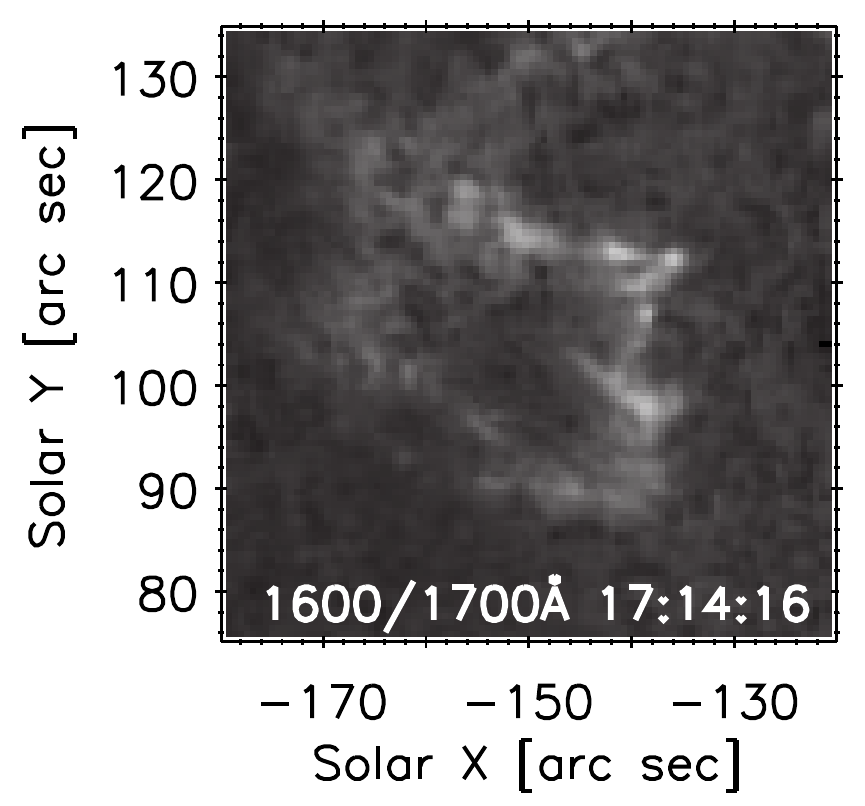}
	\includegraphics[width=3.29cm,clip,bb=62  0 243 225]{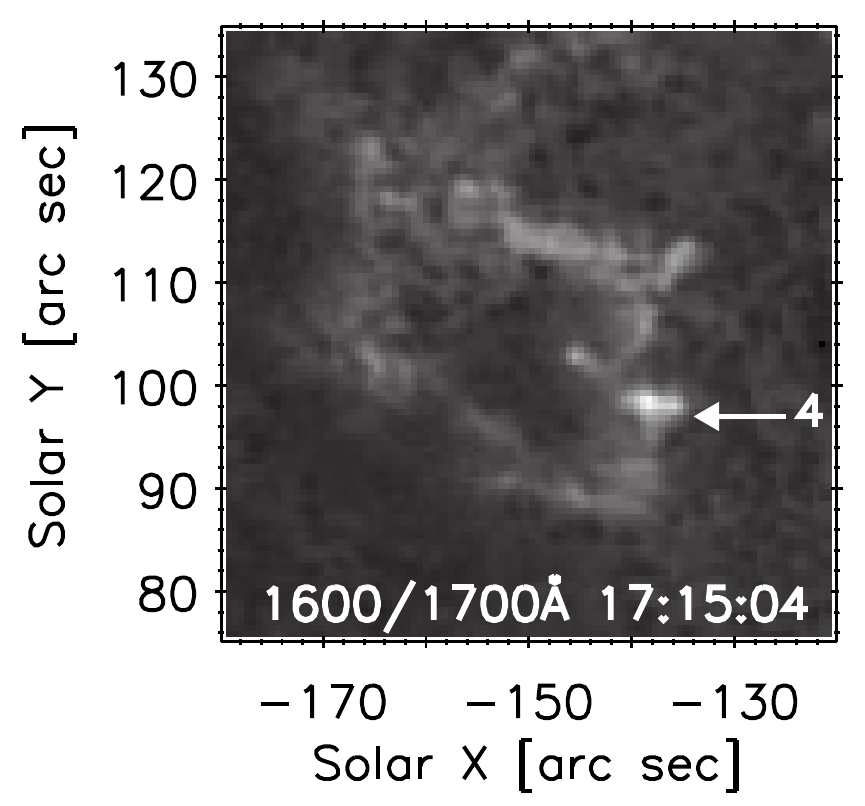}
	\includegraphics[width=3.29cm,clip,bb=62  0 243 225]{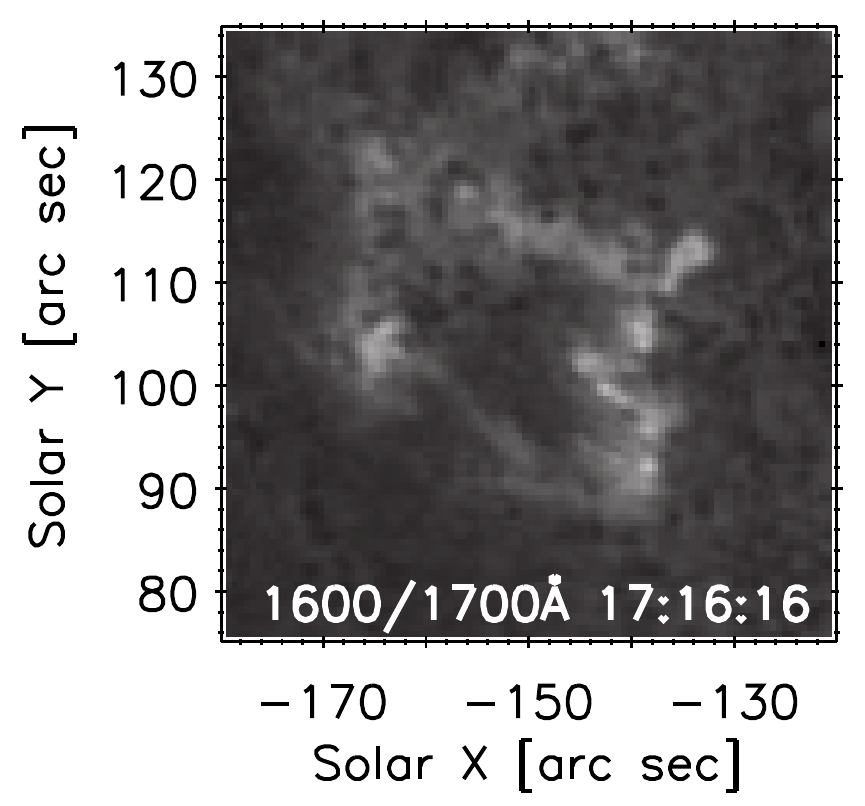}
	\includegraphics[width=3.29cm,clip,bb=62  0 243 225]{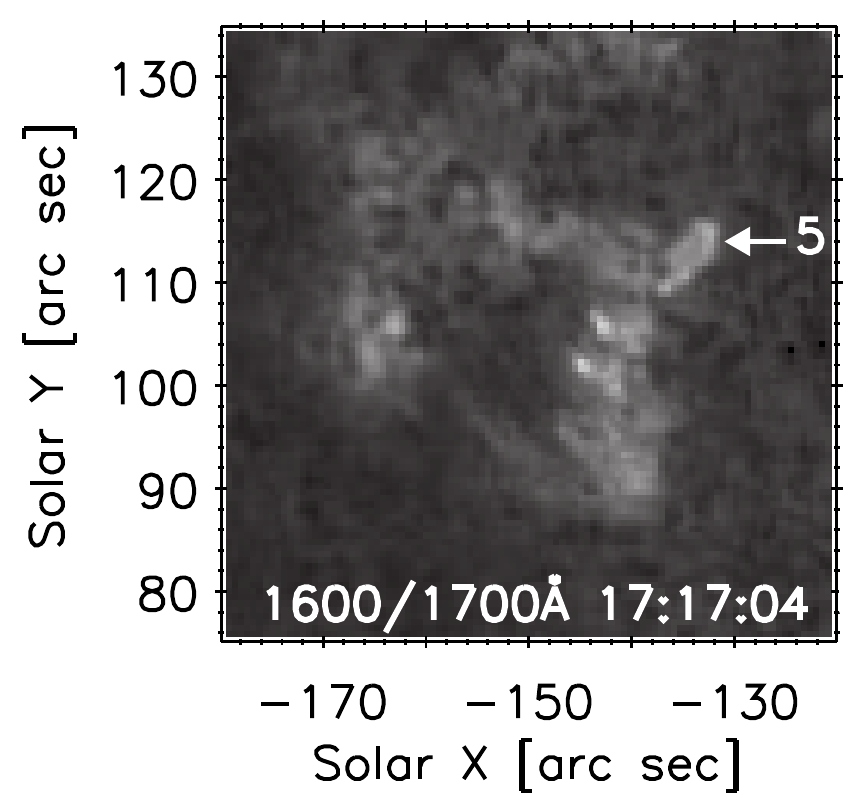}
\caption{AIA 1600~\AA~and 1700~\AA~observations corresponding to the last five times shown in Fig.\,\ref{Fig:NR_AIA_slit2a}. The \textit{bottom} row shows the 1600~\AA\,/\,1700~\AA~ratio, with enhancements of 1600~\AA~emission seen at the location of the flare ribbon. \\
A color version of this image is available in the online journal.
\label{Fig:NR_AIA_slit2a_TR}}
\end{figure*}
\begin{figure*}
	\centering
	\includegraphics[width=9.39cm,clip,bb= 0  0 495 85]{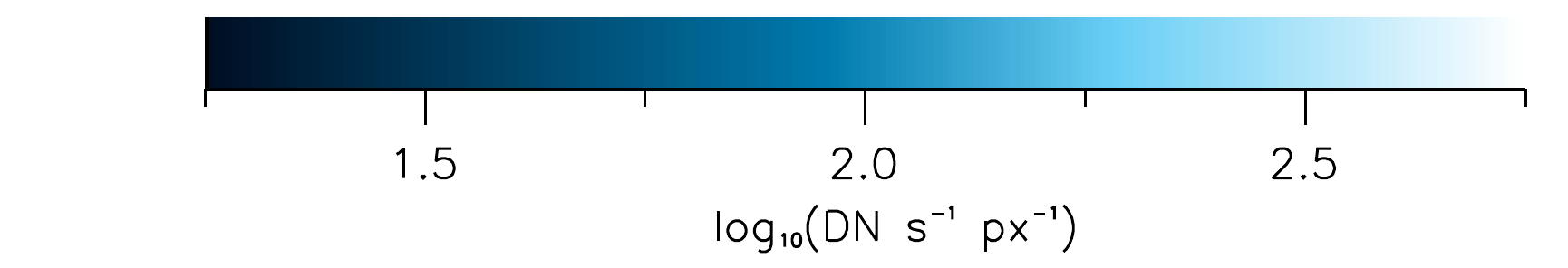}
	\includegraphics[width=8.21cm,clip,bb=62  0 495 85]{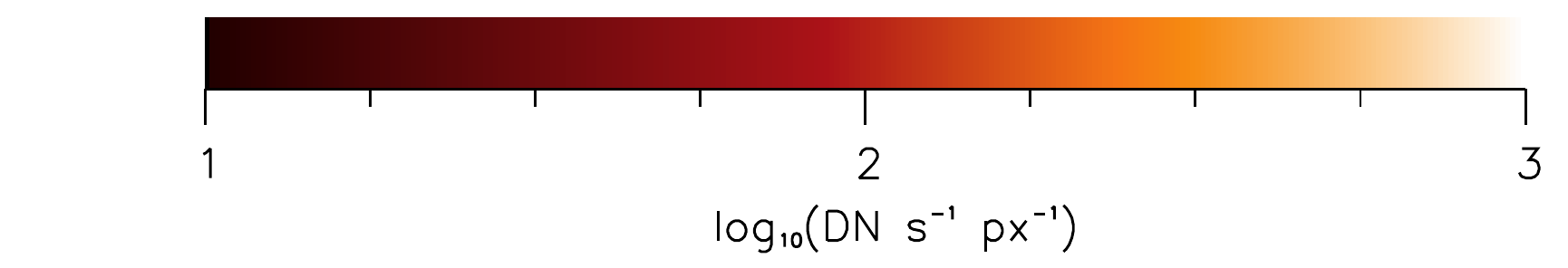}
	\includegraphics[width=4.38cm,clip,bb= 0 40 245 270]{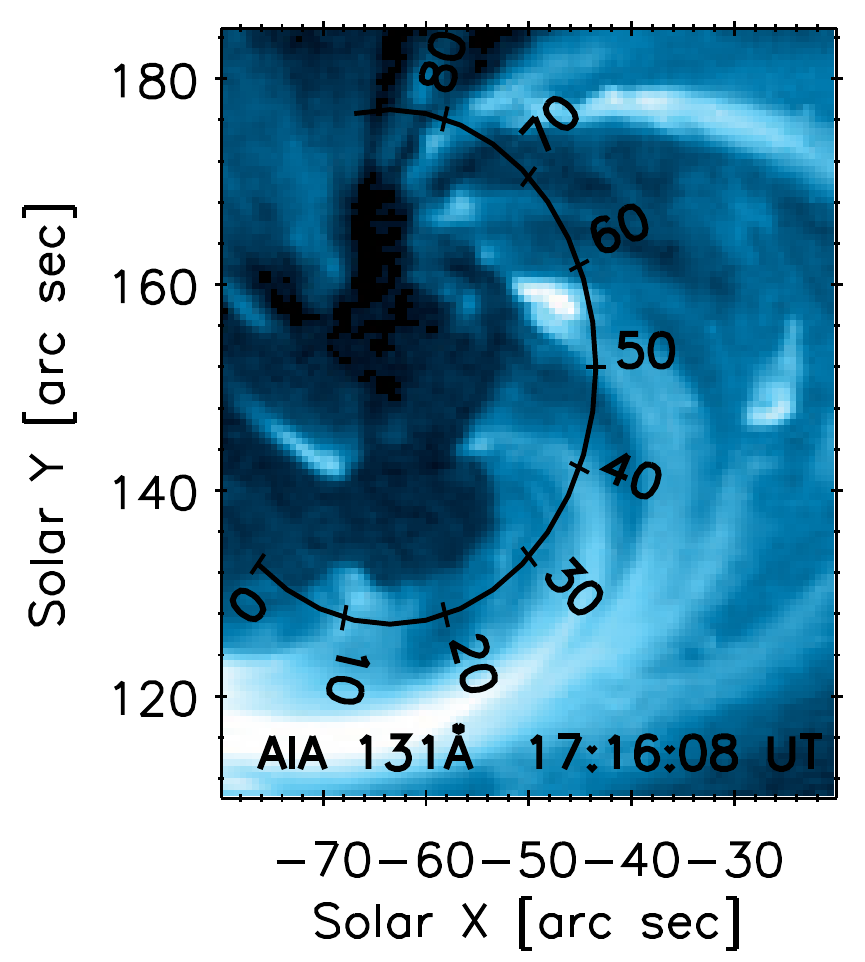}
	\includegraphics[width=3.30cm,clip,bb=60 40 245 270]{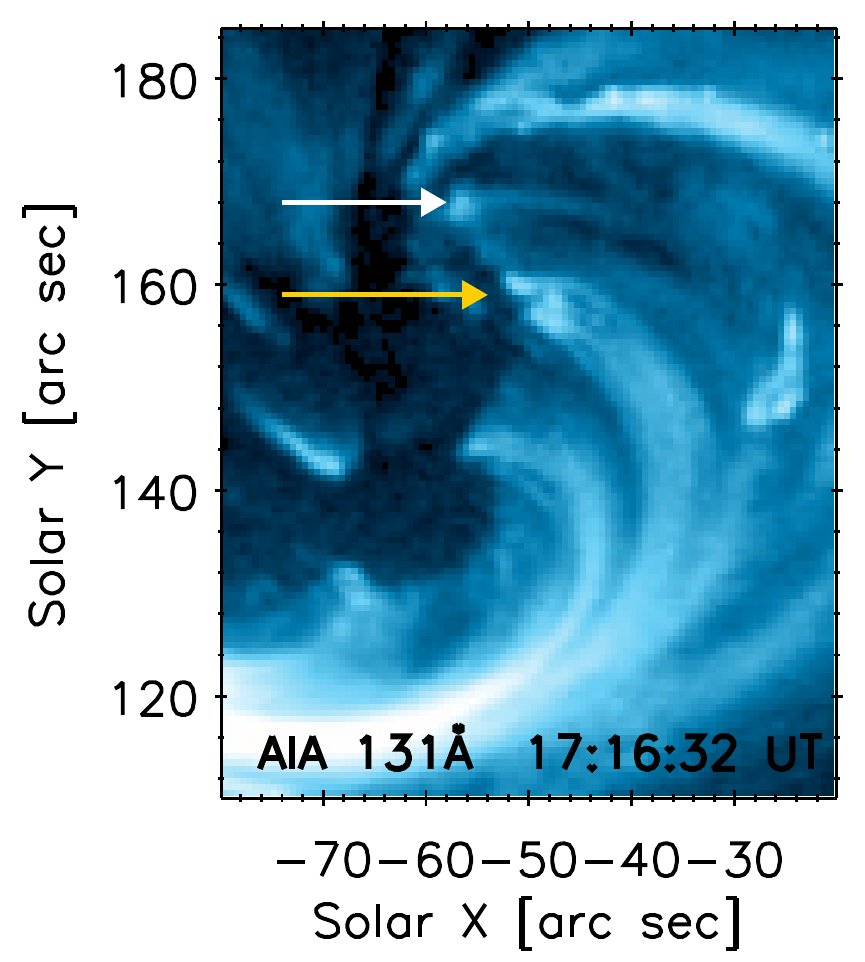}
	\includegraphics[width=3.30cm,clip,bb=60 40 245 270]{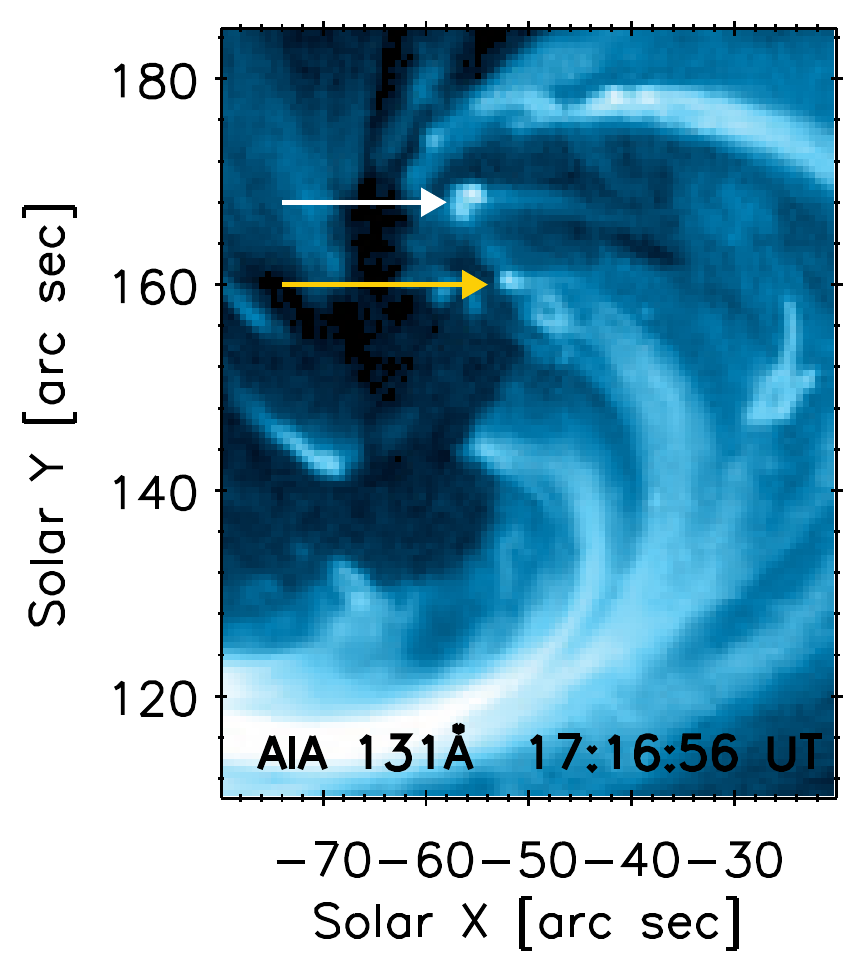}
	\includegraphics[width=3.30cm,clip,bb=60 40 245 270]{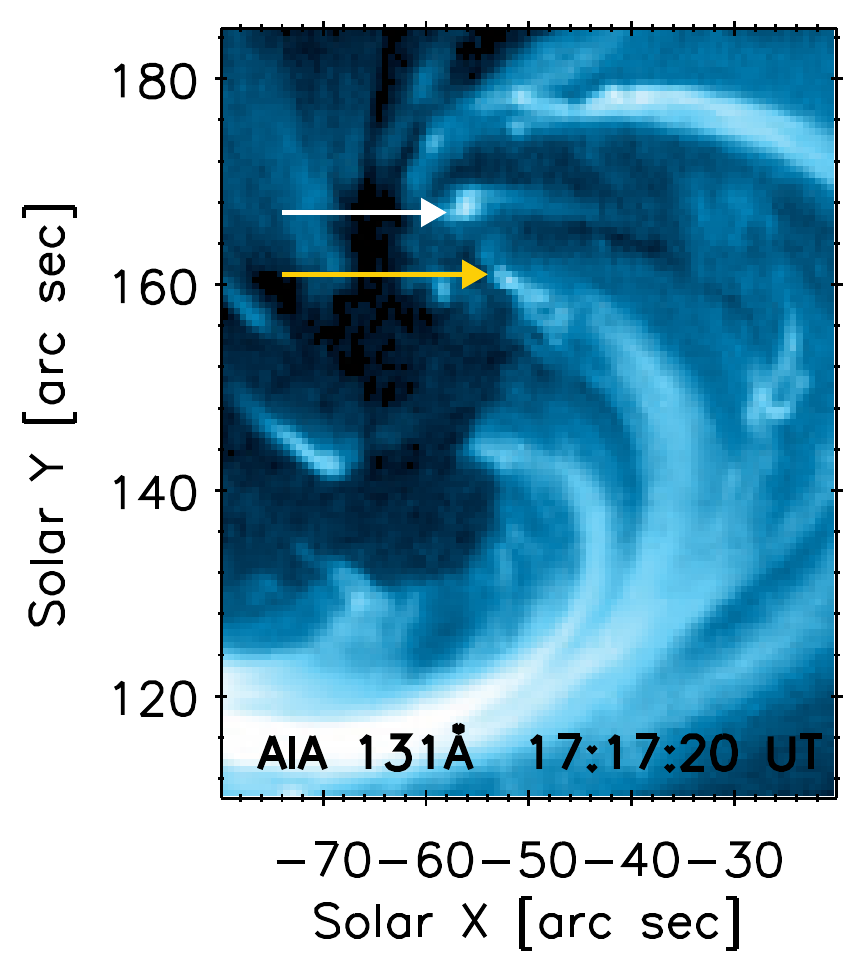}
	\includegraphics[width=3.30cm,clip,bb=60 40 245 270]{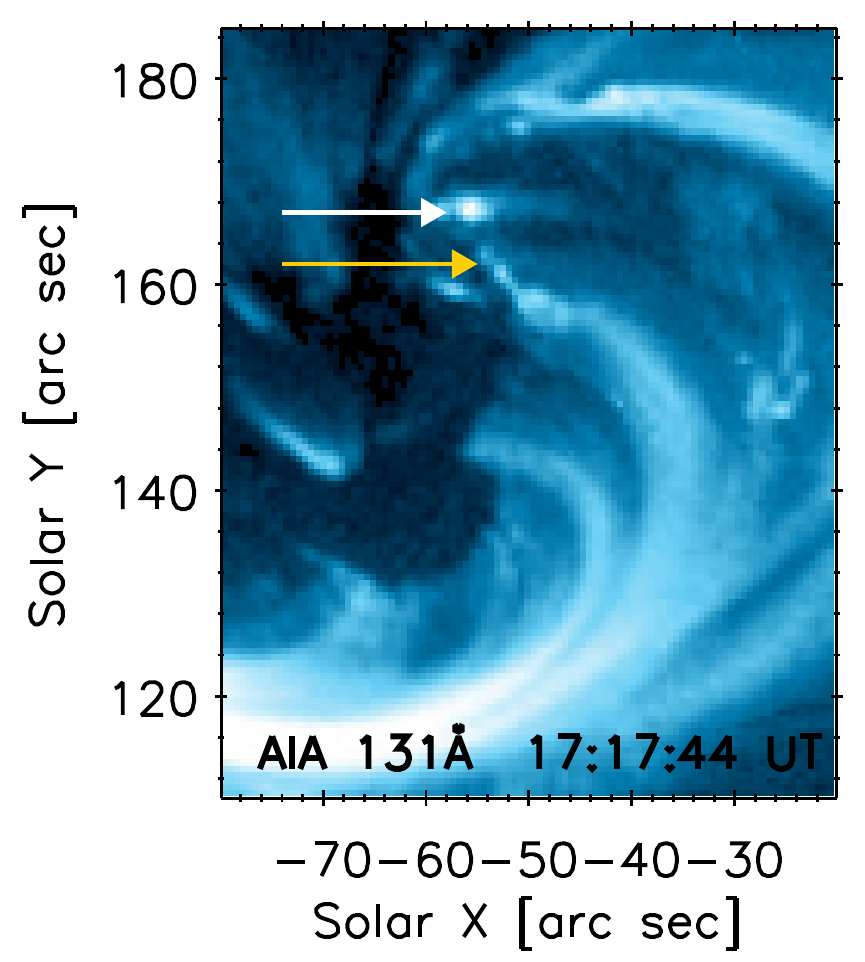}
	\includegraphics[width=4.38cm,clip,bb= 0  0 245 270]{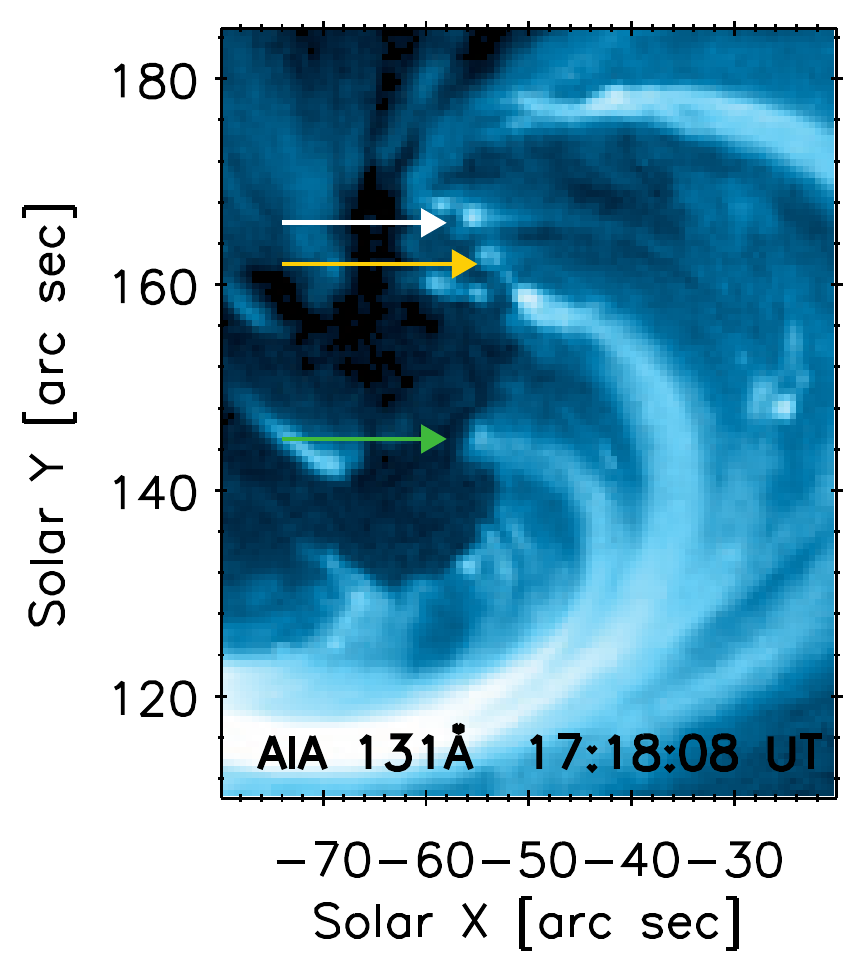}
	\includegraphics[width=3.30cm,clip,bb=60  0 245 270]{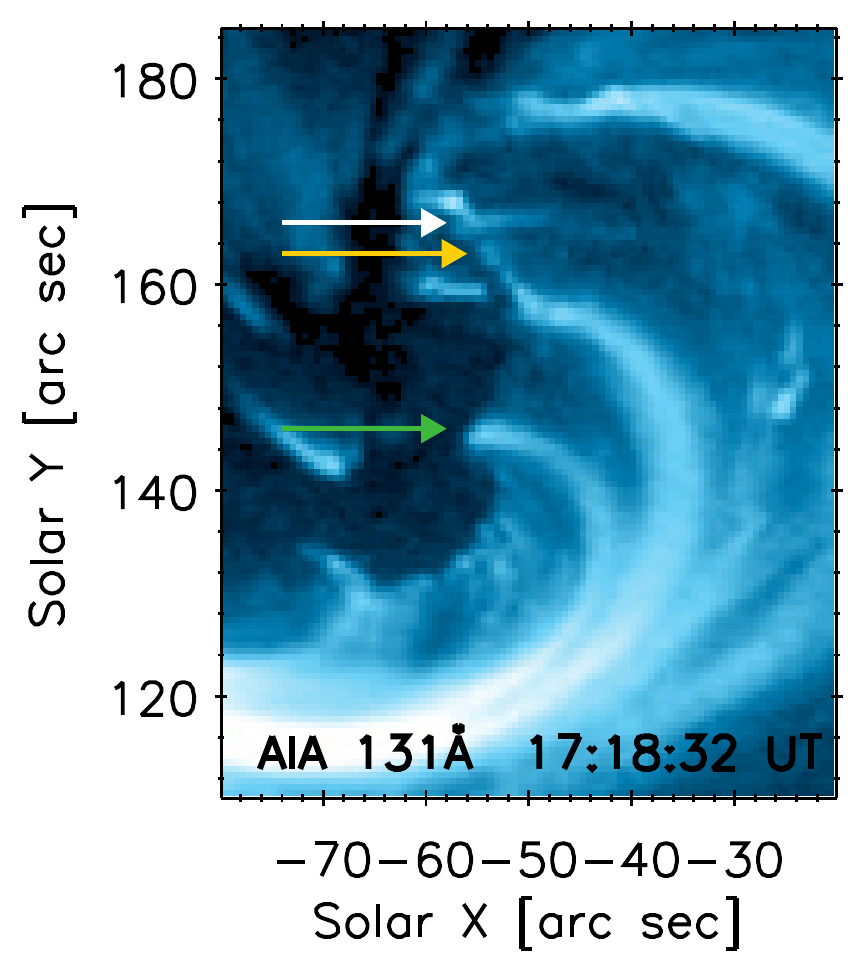}
	\includegraphics[width=3.30cm,clip,bb=60  0 245 270]{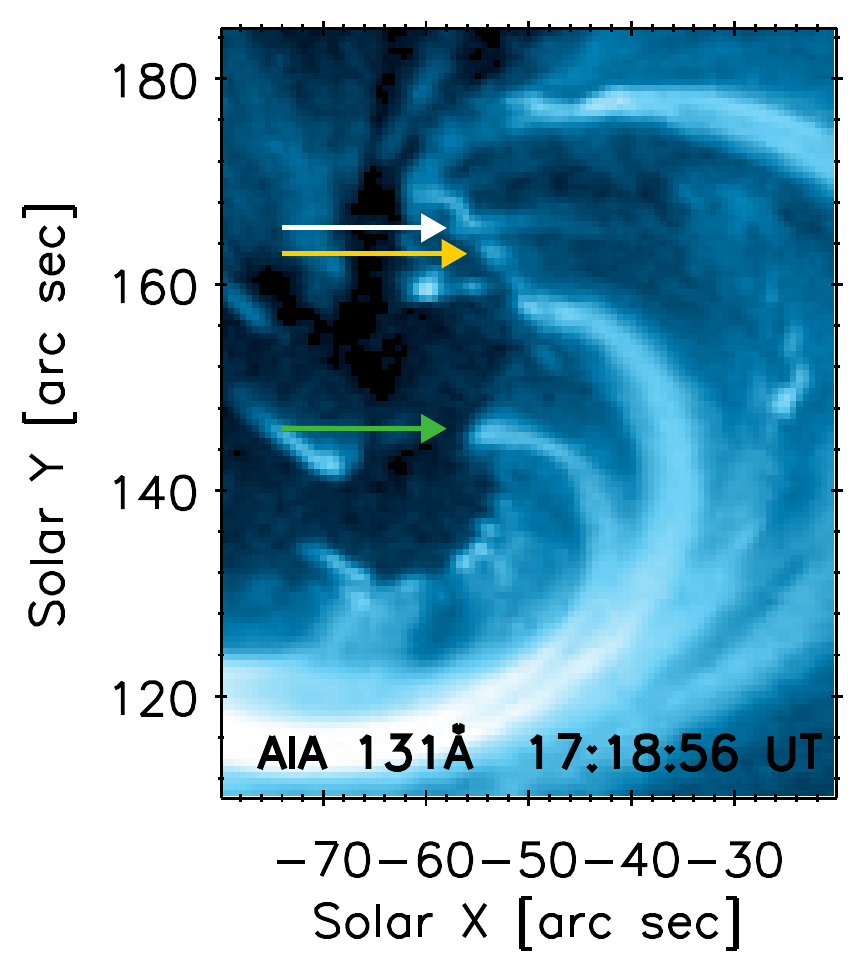}
	\includegraphics[width=3.30cm,clip,bb=60  0 245 270]{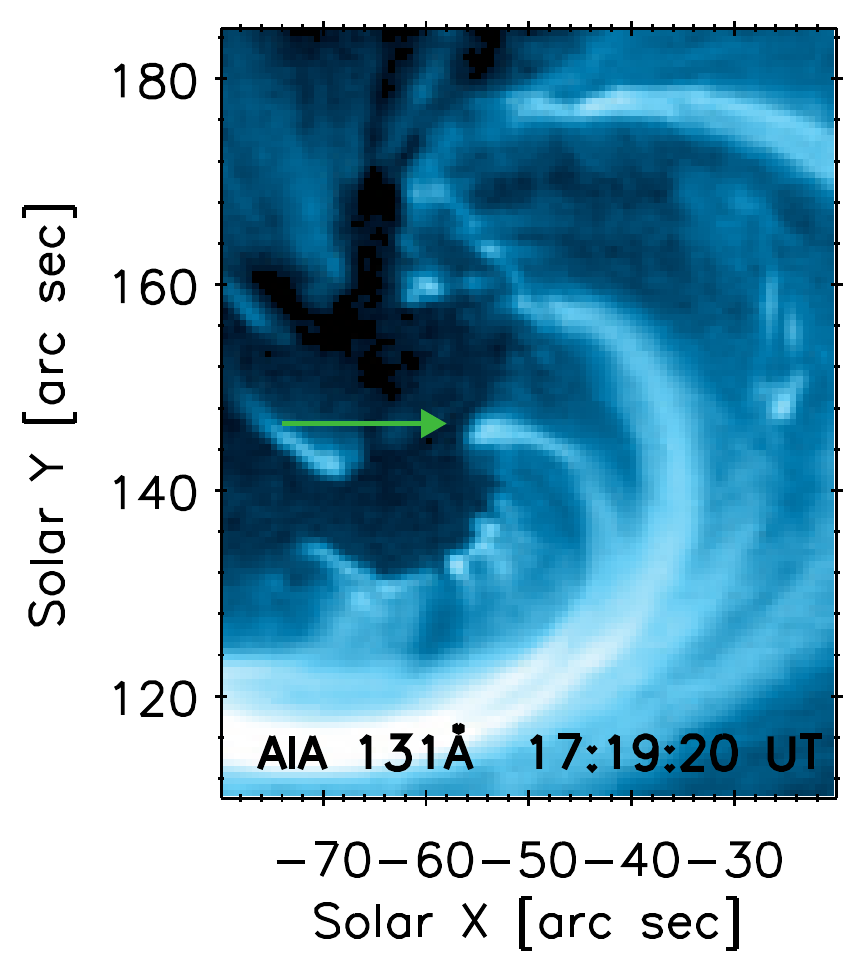}
	\includegraphics[width=3.30cm,clip,bb=60  0 245 270]{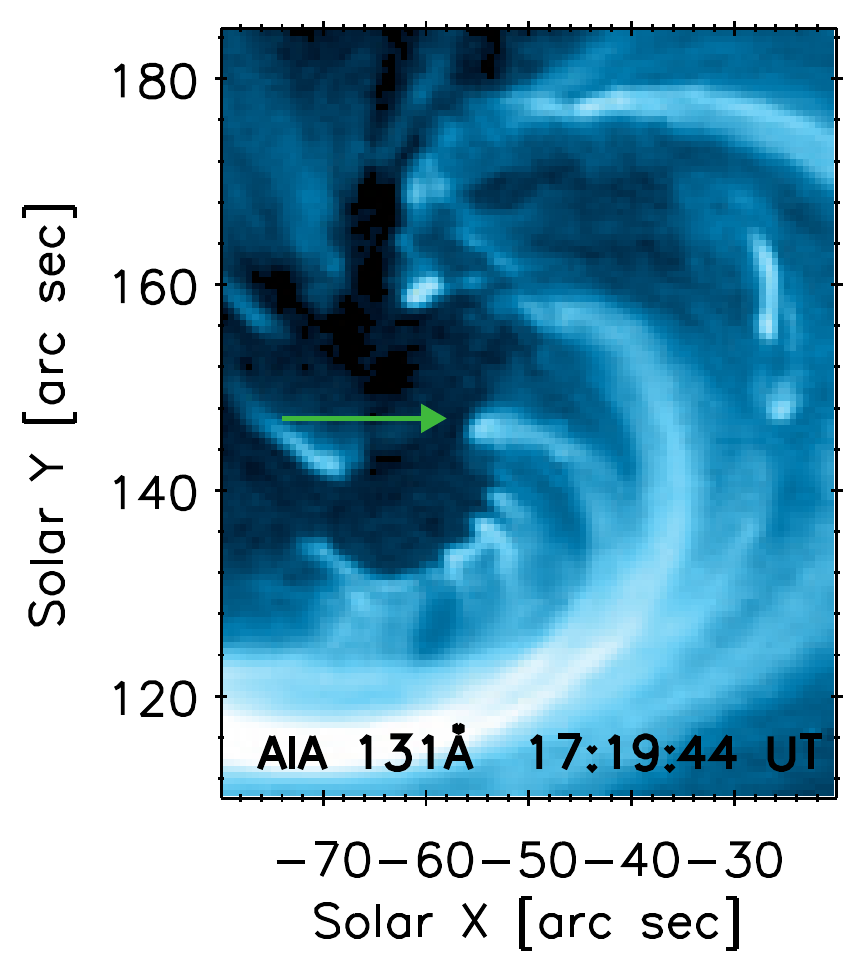}
	\includegraphics[width=4.38cm,clip,bb= 0 40 245 270]{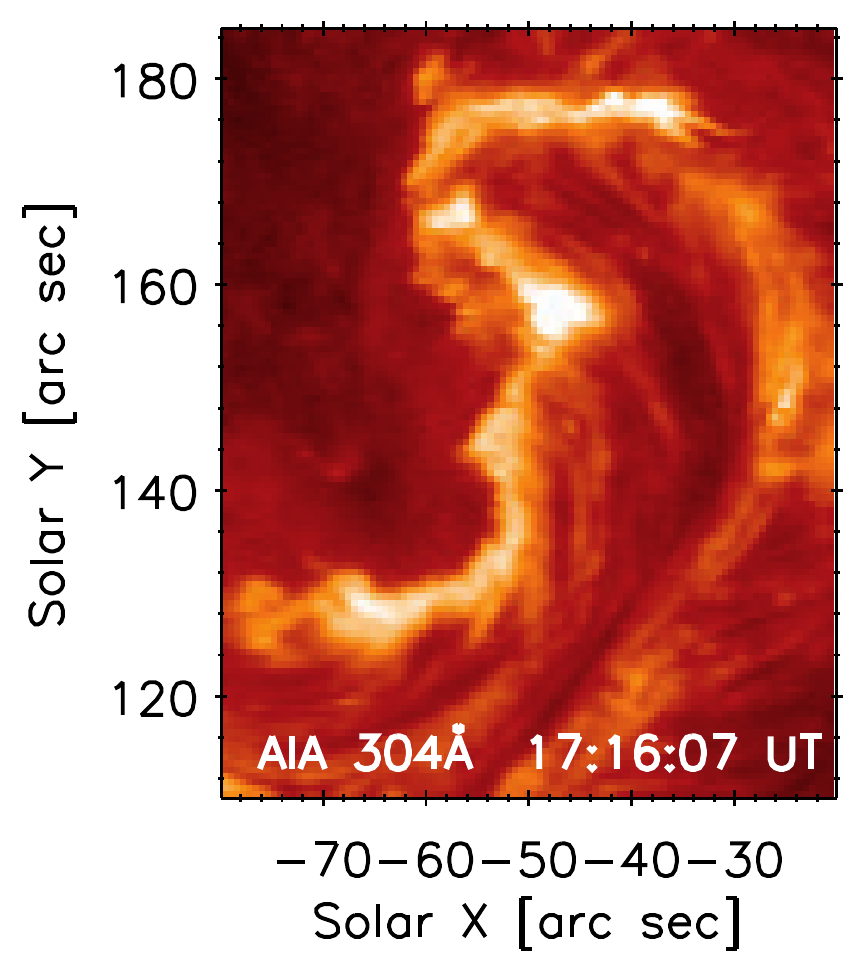}
	\includegraphics[width=3.30cm,clip,bb=60 40 245 270]{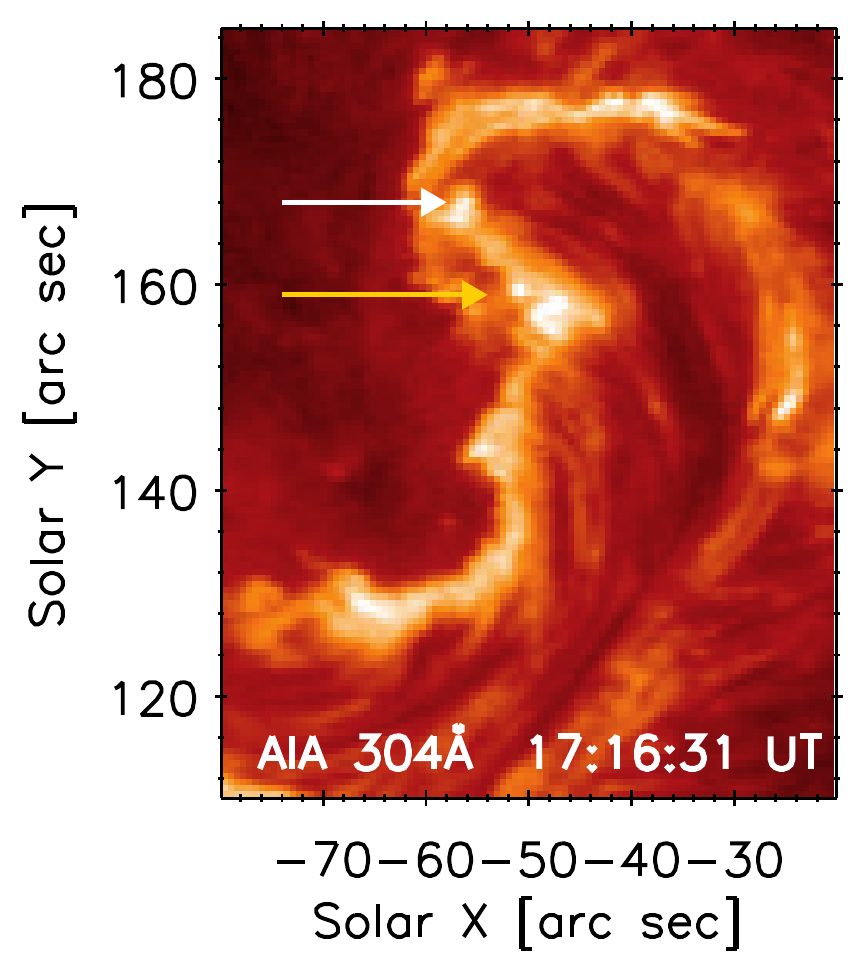}
	\includegraphics[width=3.30cm,clip,bb=60 40 245 270]{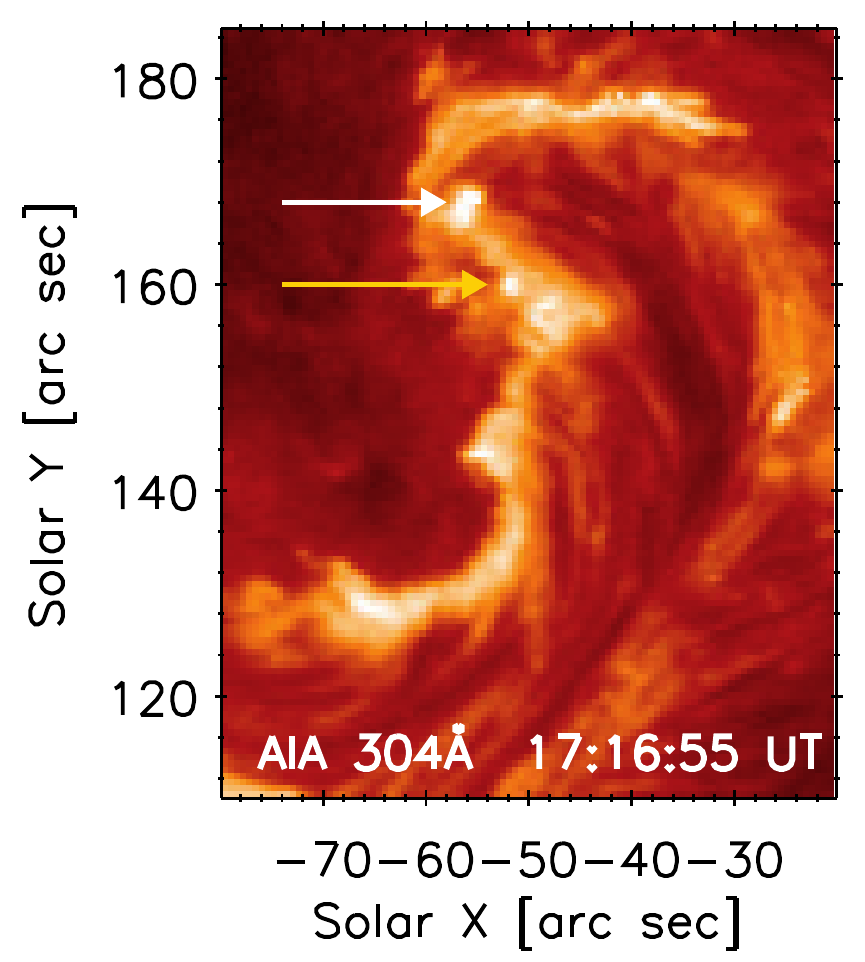}
	\includegraphics[width=3.30cm,clip,bb=60 40 245 270]{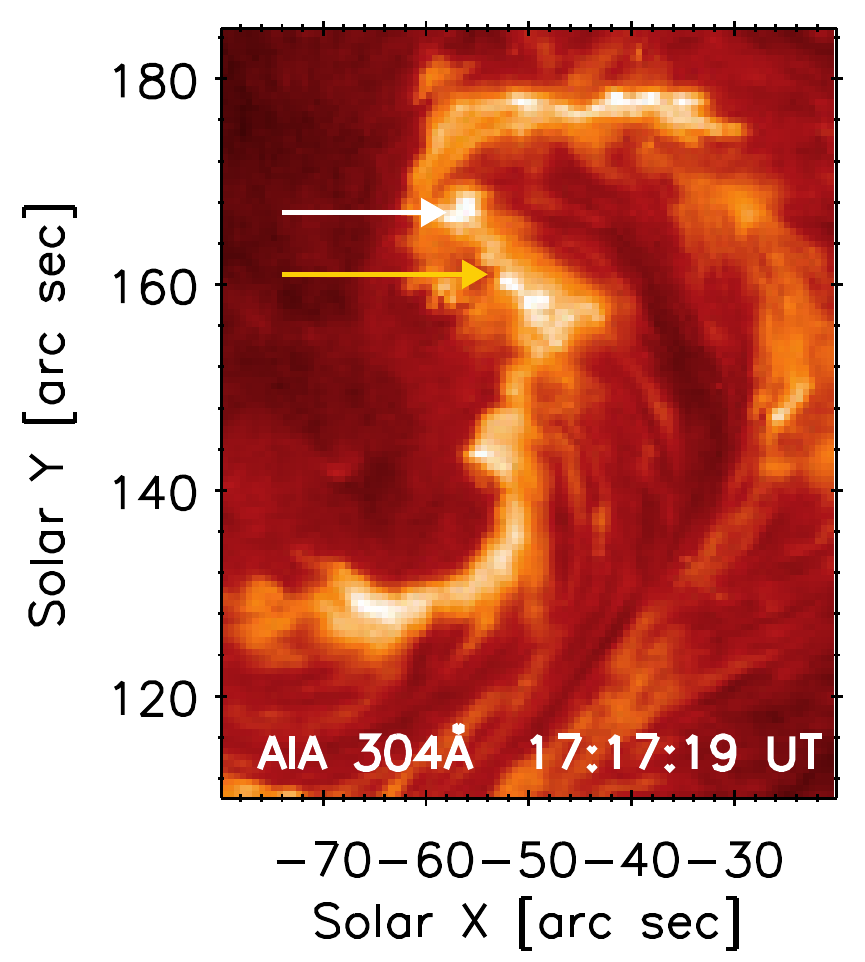}
	\includegraphics[width=3.30cm,clip,bb=60 40 245 270]{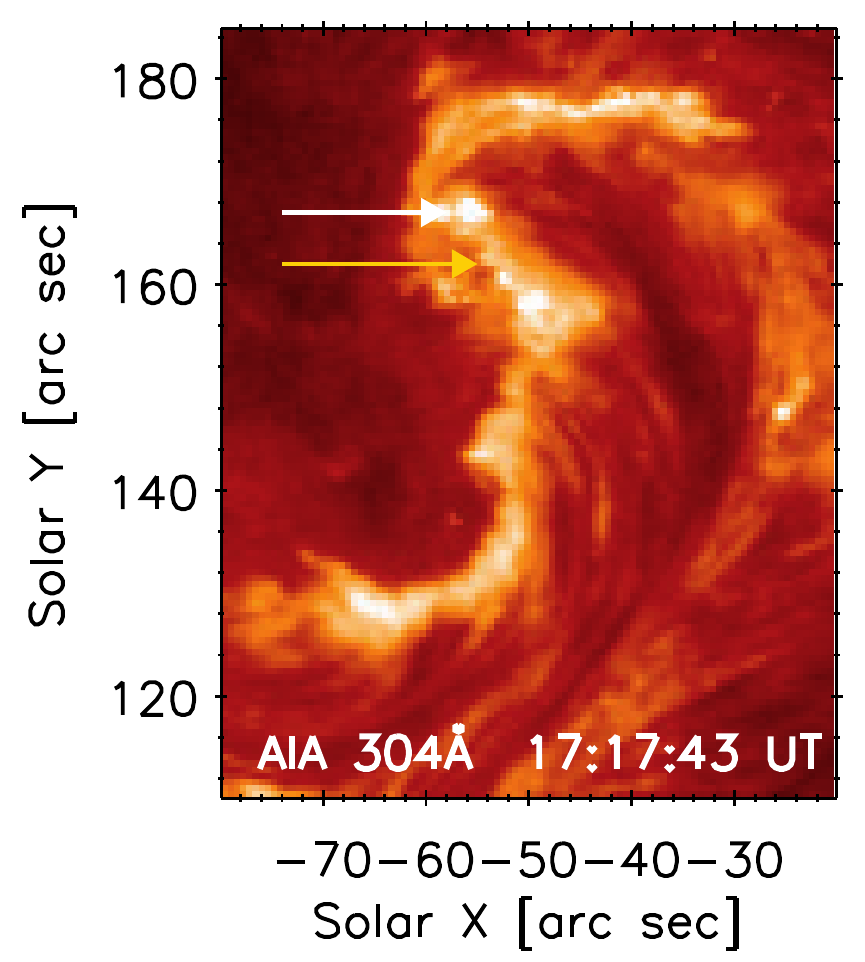}
	\includegraphics[width=4.38cm,clip,bb= 0  0 245 270]{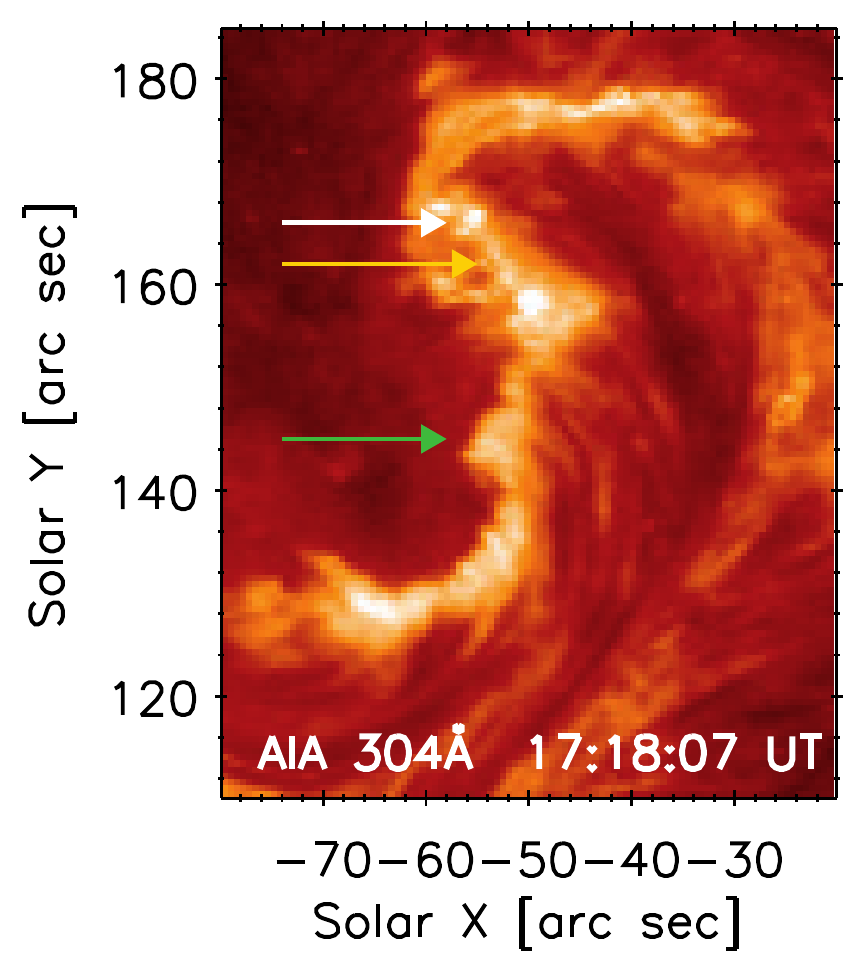}
	\includegraphics[width=3.30cm,clip,bb=60  0 245 270]{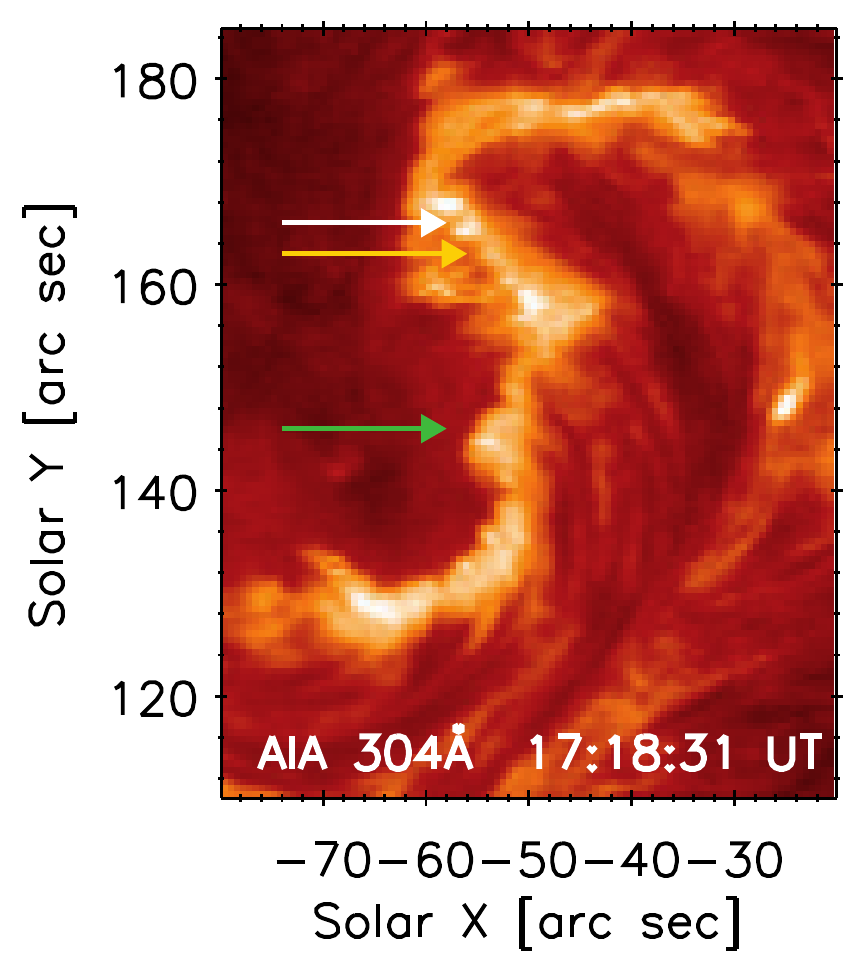}
	\includegraphics[width=3.30cm,clip,bb=60  0 245 270]{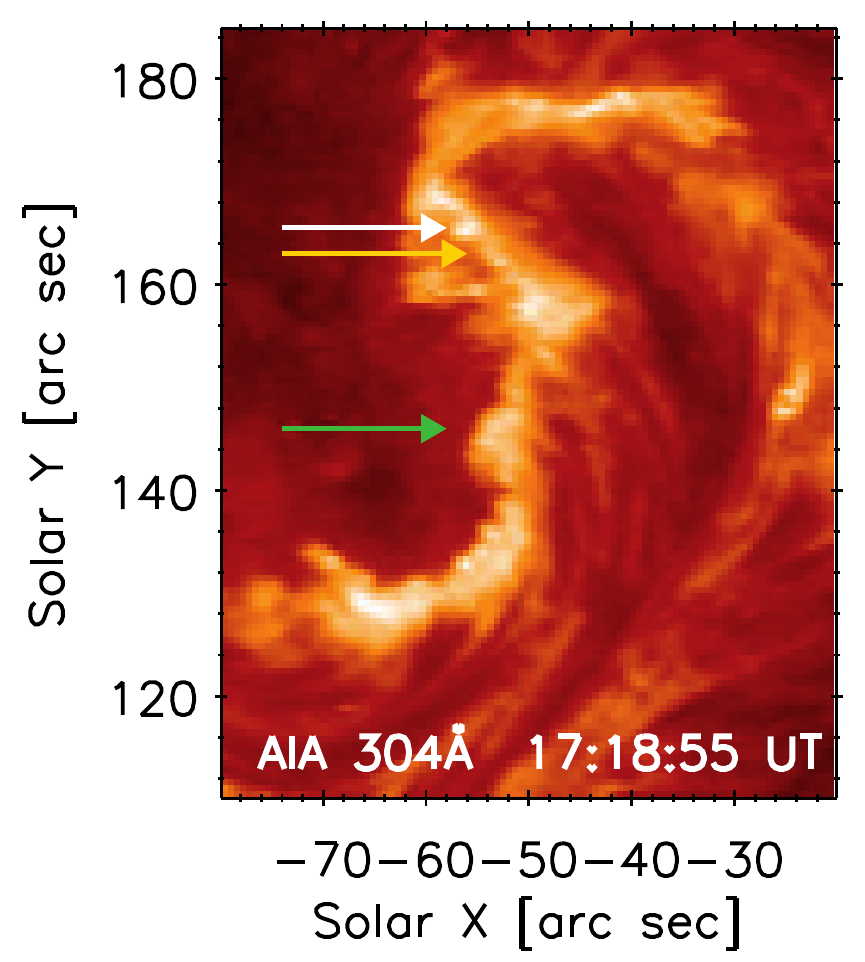}
	\includegraphics[width=3.30cm,clip,bb=60  0 245 270]{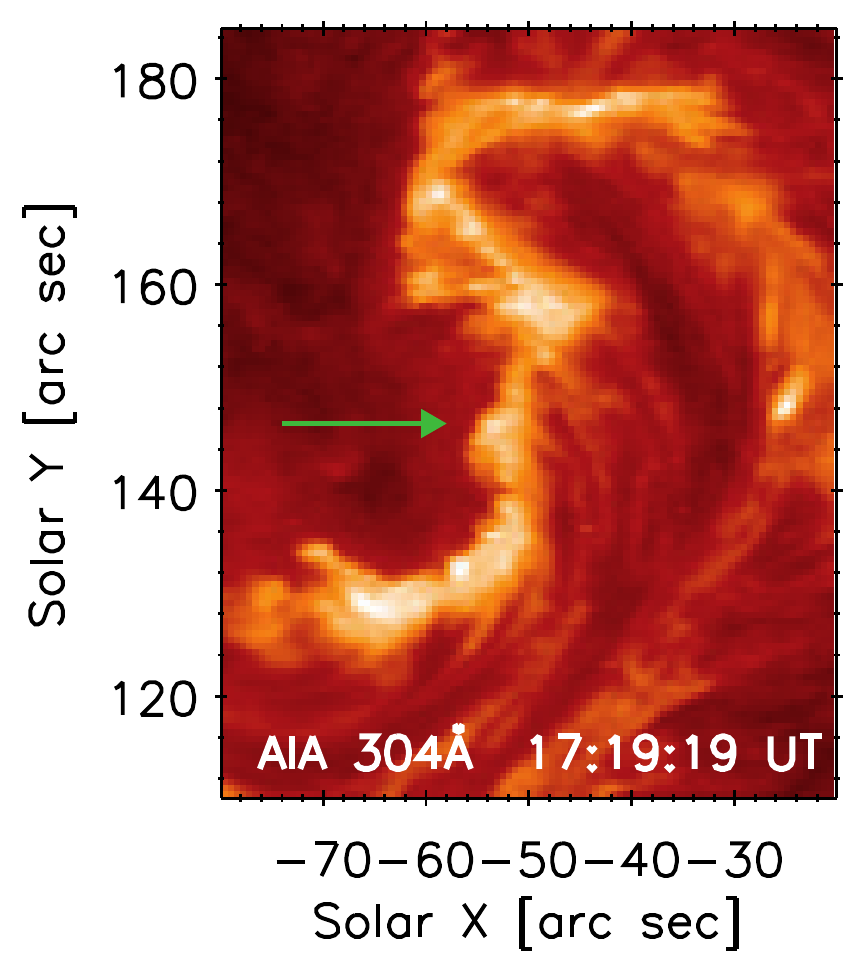}
	\includegraphics[width=3.30cm,clip,bb=60  0 245 270]{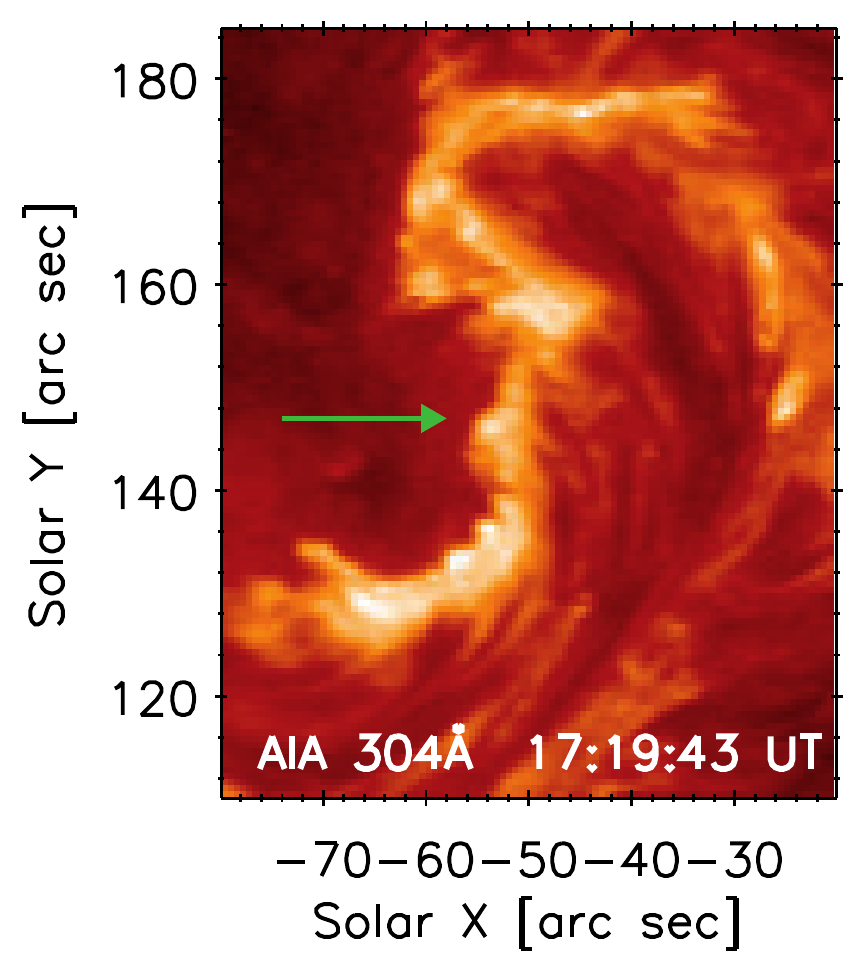}
\caption{Apparently slipping flare loops anchored in the PR ribbon as observed by AIA 131~\AA~and 304~\AA.The ellipse shown in the \textit{top left} image is the curved cut used to construct the time-distance plot shown in Fig.\,\ref{Fig:PR_stackplots}. Positions along the cut are marked. \\
A color version of this image is available in the online journal.
\label{Fig:PR_AIA}}
\end{figure*}
\begin{figure*}[!ht]
	\centering
	\includegraphics[width=8.80cm,clip,bb= 0  0 495 385]{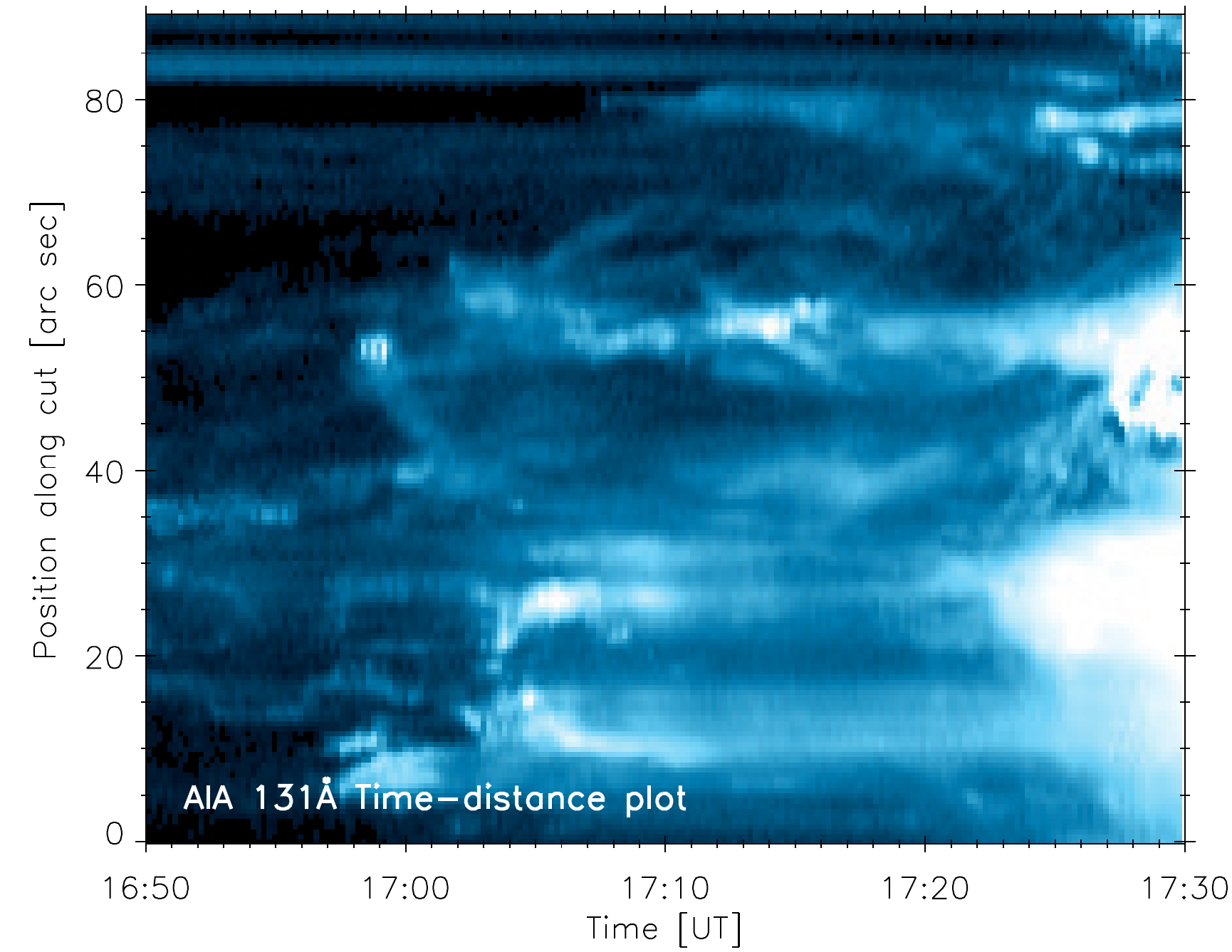}
	\includegraphics[width=8.80cm,clip,bb= 0  0 495 385]{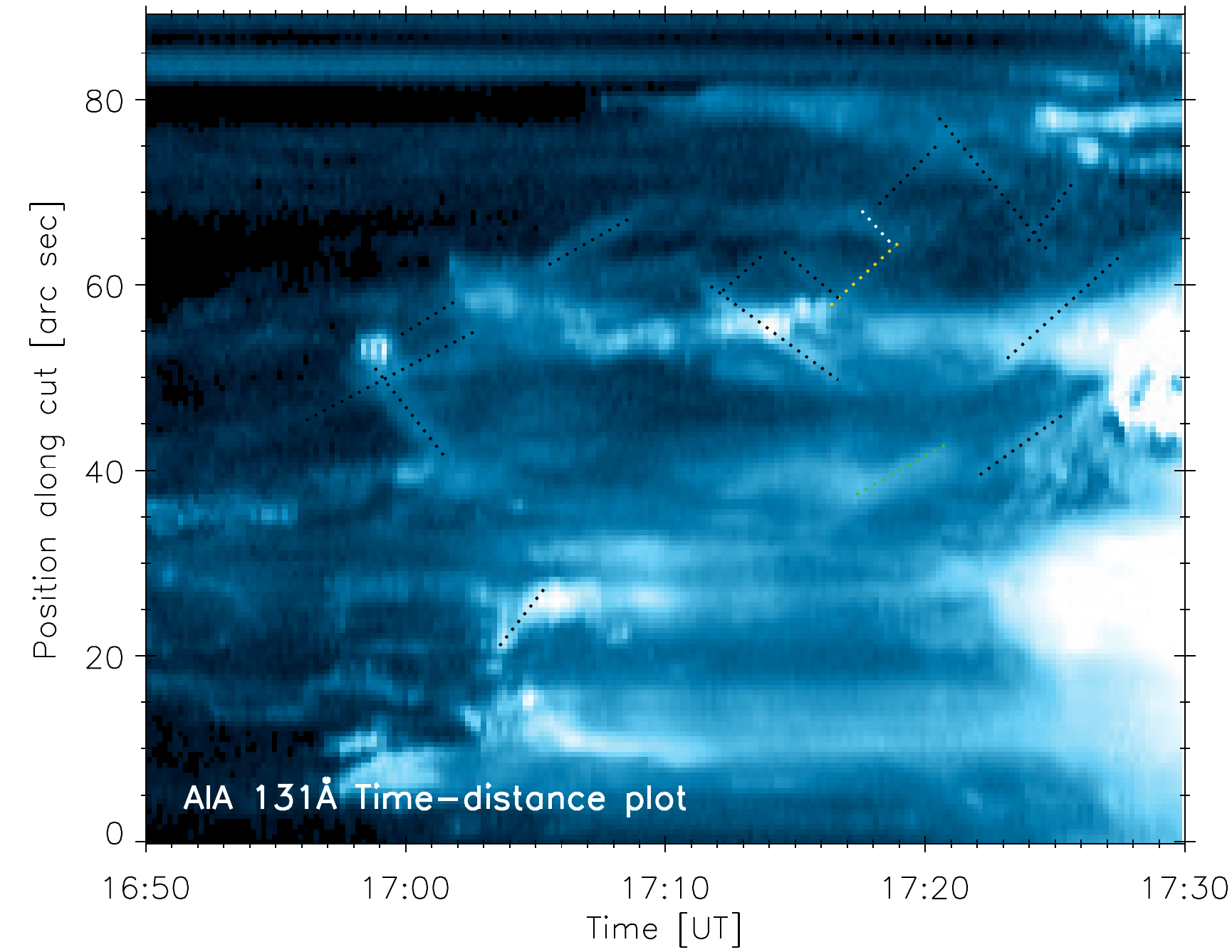}
\caption{AIA 131\,\AA~time-distance plot along the cut shown in Fig.\,\ref{Fig:PR_AIA}. The intensity scaling is the same as in Fig.\,\ref{Fig:PR_AIA}. The right panel is the same as the left hand one, but individual dotted lines denote some of the brightest slipping loops with velocities of 18 -- 44 km s$^{-1}$. See text for details.
\label{Fig:PR_stackplots}}
\end{figure*}

%
\subsection{Slipping reconnection during the flare}
\label{Sect:2.3}

The occurrence of the apparently slipping flare loops in this flare has already been reported by \citet{Li15}, who focused on the NR during the impulsive phase (from about 17:25 onwards) in the vicinity of the \textit{IRIS} slit. Upon reviewing the evolution of the AIA 131~\AA~observations, we found that the slipping reconnection is present during the entire flare. It is noticeable from the very beginning at about 16:50\,UT through the impulsive phase, similarly as in \citet{Dudik14a}. The following gradual phase is characterized by the strong-to-weak shear transition, which can also be explained by the standard solar flare model in 3D \citep[see][]{Aulanier12}. We also note that \citet{Zhao16} calculated the photospheric traces of the QSLs in an extrapolated non-linear force-free field \citep{Gilchrist14} and found that these correspond well with the observed shape of the flare ribbons, thus indirectly supporting the idea of slipping reconnection occurring in QSLs.

Unlike the flare reported in \citet{Dudik14a} however, we note that the early flare phase analyzed here exhibits several instances of apparent counter-motions of slipping flare loops. That is, the sytem of flare loops exhibit apparent slippage in both directions toward both ends of the developing ribbons. These apparent counter-motions are not easy to track however, mainly due to the complicated evolution of the ribbons. For example, the NR exhibits local squirming motions, during which the slipping motion is seen to proceed even in an almost transversal direction with respect to the general direction of the ribbon extension during the next few minutes (see the online Movies 1 and 4). A similar, but less pronounced evolution happens in the PR. Nevertheless, the apparent slipping motions of flare loops can be discerned during particular time intervals. Two such cases are reported on in the remainder of this section.

\subsubsection{Slipping reconnection along the NR}
\label{Sect:2.3.1}

An example of the squirming nature of the evolution of the NR and the associated slipping reconnection can be seen in the online Movie 4 and the corresponding Fig.\,\ref{Fig:NR_AIA_slit2a}, where the time interval of 17:08\,--\,17:17\,UT is shown. Although the NR generally extends in the south-west direction, which is the same direction as reported during the impulsive phase by \citet{Li15}, during the time interval shown in Fig.\,\ref{Fig:NR_AIA_slit2a}, the slipping motion occurs predominantly in the north-south direction. To analyze these slipping motions, we place an artifical cut at Solar $X$\,=\,$-$150$\arcsec$ (see Fig.\,\ref{Fig:NR_AIA_slit2a}, \textit{top left}). This cut is used to produce the time-distance plots in AIA 131~\AA~and 304~\AA~shown in Fig.\,\ref{Fig:NR_stackplots_slit2a}. We chose these two AIA filters, since the 131~\AA~shows the slipping loops emitting \ion{Fe}{21}, while the corresponding footpoints are very bright in the 304~\AA, making it a useful bandpass to study the evolution of the ribbon itself.

At about 17:03\,UT, the footpoints of the apparently slipping loops first reach the location of the cut. The dominant slipping motion is in the southern direction, with a speed of about $-38$\,km\,s$^{-1}$. This velocity corresponds to the leftmost dashed line in Fig.\,\ref{Fig:NR_stackplots_slit2a}. Several other slipping loops enter the cut later on. Some of them are highlighted by the dashed and dotted lines in Fig.\,\ref{Fig:NR_stackplots_slit2a}. After 17:09\,UT, a prominent extension of the ribbon in the opposite (northern) direction occurs at the position of Solar $Y$\,=\,110$\arcsec$. Following this time, a series of loops is seen to be apparently slipping in both directions along the cut. In Fig.\,\ref{Fig:NR_stackplots_slit2a}, the apparently slipping loops are denoted by dotted lines, while the apparently slipping bright knots in the ribbon are denoted by dashed lines. The distinction between the two is easily made by their presence in both the 131~\AA~and 304~\AA~time-distance plots. This is because the 304~\AA~passband shows only the flare loop footpoints, with higher portions of these flare loops being visible in 131~\AA~and not in 304~\AA. The typical apparent slipping velocities found are 11\,--\,57 km\,s$^{-1}$, similar as in \citet{Dudik14a}.

We next examined the relationship between the morphology of flare emission in the hot AIA bandpasses (131\,\AA~and 94\,\AA) and the bandpasses registering the transition-region emission (304~\AA, 1600~\AA, as well as 1700~\AA). We note that the 304~\AA~bandpass is nearly co-temporal with the 131~\AA, with only a 1\,s difference. We find nearly a one-to-one correspondence (Fig.~\ref{Fig:NR_AIA_slit2a}) between the locations of the footpoints of the 131~\AA~loops and the 304~\AA~bright kernels within the evolving ribbon. Several conspicuous examples are pointed out by Arrows 1--5 in Fig.~\ref{Fig:NR_AIA_slit2a}. This relationship was already reported by \citet{Dudik14a} and is confirmed here.

The nearest 1600~\AA~or 1700~\AA~image is usually taken at least several seconds earlier or later compared to the 131~\AA~(Fig.~\ref{Fig:NR_AIA_slit2a_TR}). The close relationship between the 131~\AA~footpoints and the 304~\AA~bright kernels also holds for the kernels observed in AIA 1600~\AA. An example is shown in Fig.\,\ref{Fig:NR_AIA_slit2a_TR}. This figure contains the AIA 1600~\AA~and 1700~\AA~observations, as well as their ratio, for the five last times shown in Fig.\,\ref{Fig:NR_AIA_slit2a}. We see that the brightest kernels in 304~\AA~are also present in 1600~\AA, with faint brightenings being present in the 1700~\AA~bandpass as well. The ribbon is however best seen in the 1600~\AA\,/\,1700~\AA~ratio, the morphology of which appears similar to the 304~\AA (compare Figs. \ref{Fig:NR_AIA_slit2a} and \ref{Fig:NR_AIA_slit2a_TR}), although we note that the difference between acquiring the 1600\AA~and 304\AA~images is several seconds.

We note that both the 1600~\AA~and 1700~\AA~bandpasses have broad spectral response \citep[see Fig.\,9 in][]{Boerner12}. These bandpasses contain a strong contribution from the photospheric continuum being formed near the temperature minimum region. The 1600~\AA~bandpass however also contains the prominent \ion{C}{4} 1548.19\AA~and 1550.77\AA~doublet. By taking the AIA 1600~\AA\,/\,1700~\AA~ratio, we find that the locations of the bright flare kernels show strongly enhanced 1600~\AA~emission, up to a factor of $\approx$3.3 compared to the average of a nearby plage and $\approx$3.8 compared to a quiet Sun region containing both network and internetwork. This result points to a strong \ion{C}{4} emission being present in the flare kernels. Strong increase of transition-region line intensities is commonly observed in a solar flare \citep[e.g.,][]{Cheng81,Poland82,Woodgate83,Schmieder96b,DelZanna13a}, often associated with hard X-ray bursts. Note that a strong increase of \ion{C}{4} emission is expected for non-Maxwellian distributions \citep{Dzifcakova05,DzifcakovaKarlicky08}, however, without direct \textit{RHESSI} observations, or modeling of the optically-thick photospheric continuum, the presence of these non-Maxwellians cannot be unambiguously confirmed from the AIA 1600~\AA~observations, though high-energy tails are routinely observed in flares \citep[e.g.,][]{Veronig10,Fletcher11,Battaglia13,Simoes13b,Simoes15,Milligan14,Oka13,Oka15}.

\subsubsection{Slipping reconnection along the PR}
\label{Sect:2.3.2}

The slipping reconnection including the apparent motion of flare loops in both directions is best visible in the vicinity of the PR, located near the leading positive-polarity sunspot. An example of evolution of the flare loops is given in the online Movie 5 and the corresponding Fig.\,\ref{Fig:PR_AIA}, where the interval of 17:16 -- 17:20\,UT is shown. The first part of this image shows the AIA 131~\AA, while the second part shows the corresponding AIA 304~\AA~images. It can be seen that the one-to-one relationship between the footpoints of the flare loops as seen in AIA 131~\AA~and the bright kernels in 304~\AA~is confirmed for the PR as well.

To identify the apparently slipping loops in both directions along the developing ribbon, we produced a time-distance plot along a curvilinear cut shown in Fig.\,\ref{Fig:PR_AIA}. This cut has the shape of an ellipse, centered on the location $X$\,=\,$-63.5\arcsec$, $Y$\,=\,$152\arcsec$. The elliptical shape of the cut reflects approximately the shape of the ribbon PR and was determined as the best-fit to the manually placed knot points using a trial-and-error method. Using a curvilinear cut is necessary, since a straight cut would only allow a measurement of one velocity component along the ribbon, and could lead to significant underestimates depending on the position along a straight cut.

The time-distance plot along this cut is shown in Fig.\,\ref{Fig:PR_stackplots}. The slipping motion spans nearly the entire pre-flare phase. Several well-defined slipping loops are denoted by the dotted lines. At several times, loops are observed to slip in both directions along the cut. We note especially that in the vicinity of the position 60$\arcsec$ along the cut, a series of loops is observed to slip apparently in successively changing directions, creating a ``criss-cross'' pattern. The apparent velocities of these loops are approximately 30\,km\,s$^{-1}$. In particular, a pair of flare loops slipping in opposite directions is observable at about 17:17--17:19\,UT and is denoted by white and orange arrows in Fig.\,\ref{Fig:PR_AIA}. These loops exhibit converging motion until about 17:19\,UT, after which time they are no longer visible. The corresponding velocities along the curvilinear cut (see Fig.\,\ref{Fig:PR_stackplots}) are $-37.5 \pm10.9$ km\,s$^{-1}$ and $31.2 \pm 5.0$ km\,s$^{-1}$ for the loops denoted by the white and orange arrow, respectively. Another loop, denoted by a green arrow, is seen to be slipping from about 17:18\,UT onwards. Its velocity along the curvilinear cut, corresponding to the green dotted line in Fig.\,\ref{Fig:PR_stackplots}, is $19.0 \pm 3.7$ km\,s$^{-1}$.

Generally, the slipping velocities measured using the time-distance technique during the early flare phase (16:50 -- 17:30\,UT) are in the range of 18--44\,km\,s$^{-1}$. This is is consistent with the apparent slipping velocities determined for the NR in Sect. \ref{Sect:2.3.1}. We however note that the apparent slipping velocities measured here are only lower limits because of the changing angle of the loops with respect to the cut due to the evolution of the ribbon itself.

%
\begin{figure*}[!ht]
	\centering
	\includegraphics[width=13.7cm]{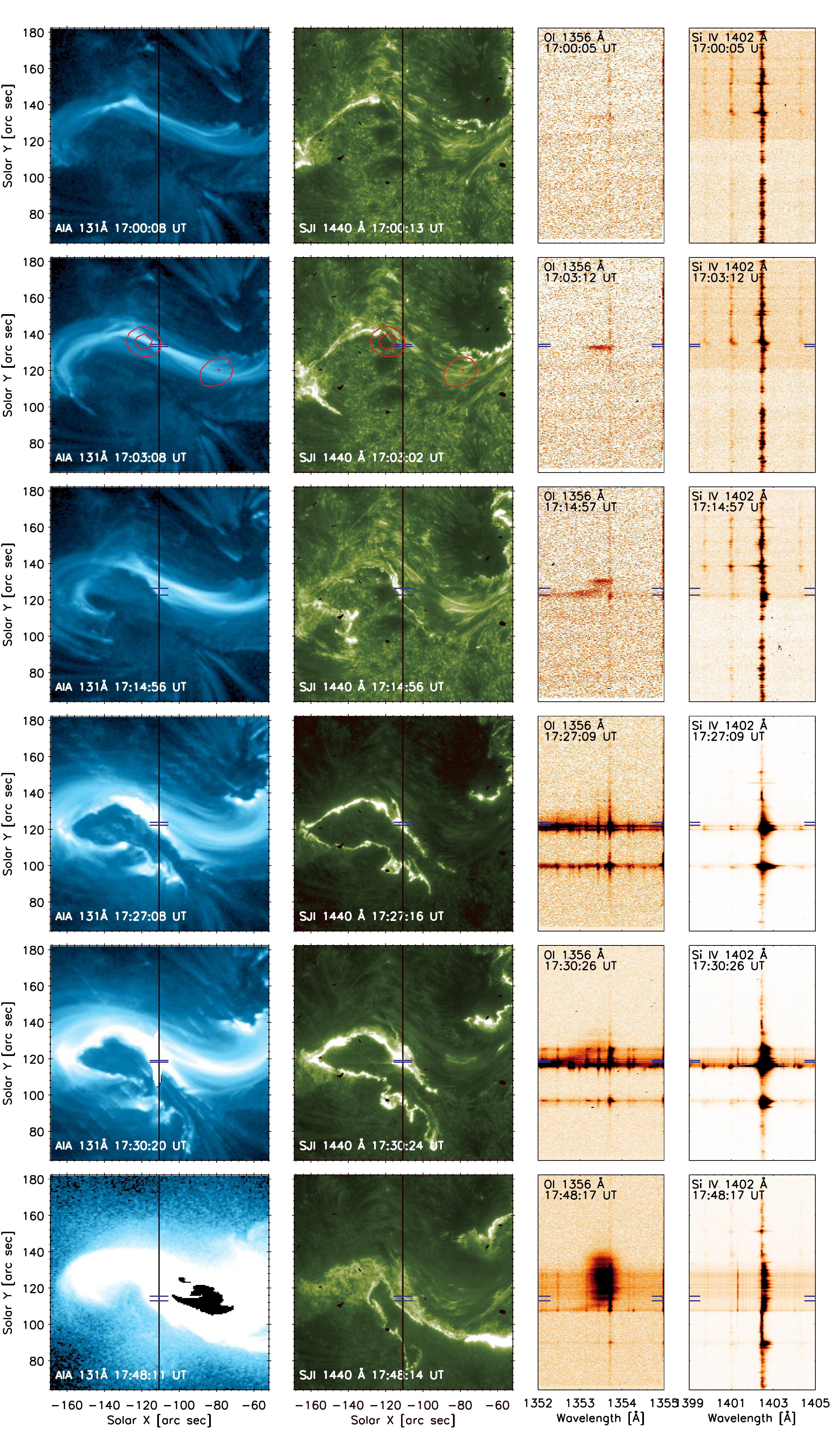}
\caption{Overview of the \textit{IRIS} observations. The 1400\AA~slit-jaw image is shown together with the two \textit{IRIS} FUV spectral windows containing the \ion{Fe}{21}~and \ion{Si}{4}~lines. AIA 131~\AA~is shown for context in the \textit{left} column. The red contours denote \textit{RHESSI} 6--12\,keV sources observed at 17:03\,UT. The blue horizontal lines indicate the locations where we observe the \ion{Fe}{21}~spectra shown in Fig.\,\ref{Fig:Fits_FeXXI}.\\
A color version of this image is available in the online journal.}
\label{Fig:Overview_IRIS}
\end{figure*}

\begin{figure*}[!ht]
	\centering
	\includegraphics[width=17.6cm]{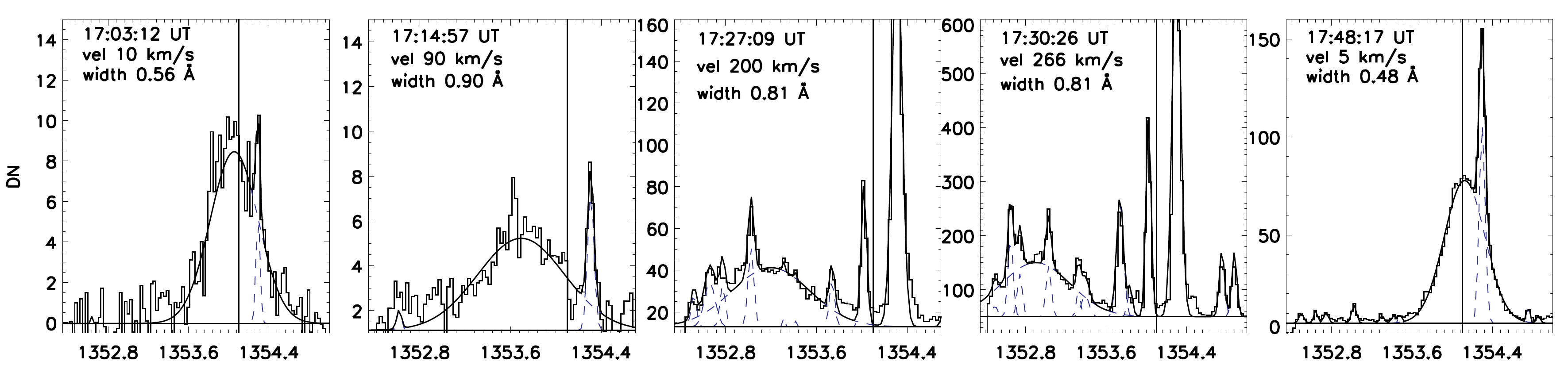}
\caption{Evolution of the \ion{Fe}{21}~1354.07\AA~emission during the flare. The times shown correspond to Fig.\,\ref{Fig:Overview_IRIS}. The fit parameters (centroid velocity and FWHM) are reported in each spectrum. The vertical line represents the expected rest wavelength position. See text for details. \\
A color version of this image is available in the online journal.}
\label{Fig:Fits_FeXXI}
\end{figure*}
%

%
\section{IRIS observations of the flare}
\label{Sect:3}

\subsection{\textit{IRIS} data and context}
\label{Sect:3.1}

Since its launch in 2013, the \textit{Interface Region Imaging Spectrograph} \citep[\textit{IRIS}, ][]{DePontieu14} has provided simultaneous imaging and spectroscopy of the solar atmosphere with unprecedented spatial resolution, (0.33$\arcsec$--0.4$\arcsec$), cadence (up to 2\,s) and spectral accuracy (allowing measurements of Doppler shifts of $\approx$3 km\,$s^{-1}$). The IRIS Slit Jaw Imager (SJI) acquires high resolution images in four different passbands (\ion{C}{2}~1330~\AA~, \ion{Si}{4}~1400~\AA, \ion{Mg}{2}~k 2796~\AA~and \ion{Mg}{2}~wing 2830~\AA), allowing to study the plasma dynamics in great details. Simultaneously, the IRIS spectrograph (SP) observes several emission lines formed over a broad range of temperatures ($log(T)[K] = 3.7 - 7$). Of particular interest is the \ion{Fe}{21}~1354.08~\AA~line formed at $\approx$~11~MK, which represents the only flare emission observed by IRIS. This spectral line was first identified by \citet{Doschek75} in solar flare spectra obtained with the Naval Research Laboratory's S082B spectrometer on board Skylab. 

The spatial and spectral characteristics of the \ion{Fe}{21}~line allow us to investigate the plasma response to the heating during flares and provide new insights into the chromospheric evaporation process. Spatially resolved, blueshifted asymmetric \ion{Fe}{21}~profiles indicative of plasma upflows $\approx$~200 km $s^{-1}$ were first observed by \citet{Mason86} during the impulsive phase of different flares with the UVSP instrument on board the Solar Maximum Mission. In contrast, 1D hydrodynamics simulations of a single flare loop \citep{Emslie1978} predict that entirely blueshifted profiles should be observed at the onset of the flare. However, these early observations lacked good spatial information and the discrepancy can be explained by interpreting the asymmetric profiles as a superposition of different plasma upflows from different sub-resolution locations along the line of sight.

Recently, there has been a lot of interest in studying the \ion{Fe}{21}~emission during flares as observed with the unprecedented resolution of \textit{IRIS}  \citep[e.,g.][]{Young15,Tian15,Graham15,Polito15a,Polito16}. One of the important finding from these recent studies is that the \ion{Fe}{21}~is observed to be entirely blueshifted during the impulsive phase of the flare, suggesting that the sites of evaporation are now likely to be resolved by IRIS.

On 2014 September 10, \textit{IRIS} was running a flare watch observation of the AR 12158 from 11:28 UT to 17:58 UT. Therefore, it captured the flare from onset well into the gradual phase. The observing mode was sit-and-stare with an exposure time of 8\,s and a total cadence of 9.4\,s for the FUV channel. The slit of the \textit{IRIS} spectrograph (SP) crossed two locations along the ribbon NR during all the impulsive and part of the gradual phase of the flare (see Fig.\,\ref{Fig:Overview_AIA}). The Slit-Jaw Imager (SJI) obtained 19\,s cadence images in the 1400~\AA~and 2796~\AA~passbands over a field of view of 119$\arcsec$\,$\times$\,119$\arcsec$ on the Sun. For each spectrograph exposure, one context SJI image was provided alternatively in one of the two filters. In this work, we focus on the \ion{O}{1}~and \ion{Si}{4}~spectral windows included in the spectrograph FUVS and FUVL channels, respectively. We used \textit{IRIS} level 2 data, obtained from level 0 after flat-field, geometry calibration and dark current subtraction. The cosmic rays removal was performed by using the \emph{SolarSoft} routine \emph{despik.pro}.

As a result of the temperature variation of the detectors during the \textit{IRIS} satellite orbit, the FUV channel wavelength scale drifts by about 8\,km\,s$^{-1}$ during one orbit. The orbital variation was corrected by using the strong \ion{O}{1} 1355.60~\AA~line neutral line included in the FUVS CCD. The Doppler shift of this line is often less than 1\,km\,s$^{-1}$ and thus represents a suitable reference line for wavelength calibration purposes. We measured the periodic variation of the \ion{O}{1}~line position over time and subtracted it from both FUVS and FUVL wavelength arrays, assuming that the same wavelength correction can be applied for different FUV CCD channels, (IRIS TN20 \footnotemark[1]\footnotetext[1]{http://iris.lmsal.com/documents.html}). The absolute wavelength calibration can then be obtained by using the strongest photospheric lines in the same spectral range. For the FUVS channel, the absolute correction was given by the difference between the \ion{O}{1} line position (after the orbital correction) and the expected rest wavelength 1355.5977~\AA~\citep{Sandlin86}. For the FUVL CCD, the \ion{S}{1}~1401.51~\AA~neutral line can be used. Even though this line is usually very weak, during this event it was visible at the ribbon location throughout the impulsive phase. 

An overview of the flare evolution as observed by \textit{IRIS} is shown in Fig.\,\ref{Fig:Overview_IRIS}. Section \ref{Sect:3.2} provides a detailed description of the flaring plasma dynamics as observed in the \ion{Fe}{21}~emission. The \ion{O}{4}~and \ion{Si}{4}~lines are reported on in Sect. \ref{Sect:3.3}. 

%
\subsection{Fe XXI 1354.10 \AA~observation}
\label{Sect:3.2}

The evolution of the flare as observed by \textit{IRIS} is shown in Fig.~\ref{Fig:Overview_IRIS}. In this figure, each row captures a particular time, showing (from left to right) the AIA 131~\AA~together with the closest \textit{IRIS} SJI 1400~\AA~images, as well as the corresponding \ion{O}{1}~and \ion{Si}{4}~detector images. The \textit{IRIS} SJI 1400~\AA~band is dominated by \ion{Si}{4}~1402.77\AA~emission at log$(T$/K)\,$\approx$\,4.9\,K. Therefore, the sequence of SJI images in Fig.\,\ref{Fig:Overview_IRIS} shows the morphology of the low temperature ribbon emission over time. The corresponding \ion{Fe}{21}~spectra are reported in Fig.\,\ref{Fig:Fits_FeXXI}, as observed by the \textit{IRIS} spectrograph slits. The line centroid and width given by the Gaussian fit are plotted in each spectrum.

The expected width of the \ion{Fe}{21}~line observed by \textit{IRIS} is $\approx$ 0.43~\AA, given by the quadratic sum of the \textit{IRIS} instrumental FWHM (0.026~\AA; \citet{DePontieu14}) and the line thermal width as estimated in CHIANTI v7.1 \citep{Dere97,Landi13}, assuming an ion formation temperature of 11\,MK. However, the line width is typically observed to be significantly larger during the impulsive phase of flares \citep{Mason86,Polito15a}. We estimate the non-thermal motions as a velocity parameter $W_\mathrm{nth}$ given by $\sqrt{(4\,\mathrm{ln}2)^{-1}\left(\lambda/c\right)^{-2}\cdot(W^2-W_\mathrm{th}^2-W_\mathrm{I}^2)}$, where $W$ is the line FWHM obtained from the fit, $W_\mathrm{th}$ is the line thermal width, $W_\mathrm{I}$ is the instrumental width $\lambda$ is the \ion{Fe}{21}~rest wavelength at 1354.08~\AA~and $c$ is the speed of light.

We note that the IRIS \ion{O}{1}~spectral window includes some cool temperature lines whose emission is usually enhanced during flares and can blend with the \ion{Fe}{21}. Several authors, i.e., \citet{Young15}, \citet{Polito15a}, \citet{Tian15}, and \citet{Graham15} reported a detailed identification of these lines during different flare events. The most important blending is represented by the chromospheric \ion{C}{1}~line at 1534.3\,\AA~\citep{Mason86}. However, the profiles of these low-temperature lines are typically narrow and in most of the cases they can be easily separated from the broad \ion{Fe}{21}~emission.

The first row of Fig.\,\ref{Fig:Overview_IRIS} shows the flare plasma at around 17:00 UT, in the early flare phase. At this time, we observe flare loops connecting the two flare ribbons in the AIA 131~\AA~channel. These are not visible in the 171~\AA~AIA passband (Fig.\,\ref{Fig:Overview_AIA}) and therefore are likely to be hot loops originating from the \ion{Fe}{21}~128.75~\AA~emission contributing to the 131~\AA~band. We note that at this time, \textit{IRIS} does not detect any \ion{Fe}{21}~1354.10~\AA~emission. When there is a flare, the \ion{Fe}{21}~128.75\,\AA~line dominates the 131\,\AA~band \citep[][]{ODwyer10,Petkaki12}, the corresponding  count rates detected in the \textit{IRIS} \ion{Fe}{21}~1354.07\,\AA~line are about 100 times less than the count rates detected in the 131 A band. This is despite the fact that the forbidden \ion{Fe}{21}~1354.07\,\AA~line emits 1.2 more photons than the resonance 128.75\,\AA~line. For example, at 17:03\,UT, the peak counts in the AIA 131\,\AA~band are about 1100~DN\,s$^{-1}$, which are equivalent to about 12\,DN\,s$^{-1}$ in the IRIS \ion{Fe}{21}~line. This estimate was obtained from a full DEM analysis using the six AIA bands via the \citet{Hannah13} method and the current understanding of the in-flight degradation of the IRIS channels. The IRIS study had an exposure time of 8\,s, so 100\,DN in the line correspond to about 5\,DN in the peak value above the continuum, close to the limit of the line being observable, given that the line is normally very broad and blended (cf., Fig.\,\ref{Fig:Fits_FeXXI}).

The \ion{Fe}{21}~line is first observed by \textit{IRIS} only from $\approx$ 17:03\,UT (second row of Fig.\,\ref{Fig:Overview_IRIS}) onwards, during which time the intensity of the flare plasma in the 131~\AA~AIA band becomes more intense. At the location of the \textit{IRIS} slit, the \ion{Fe}{21}~emission originates in upper portions of the hot flare loops visible in the AIA image as the slit cuts these flare loops near their centre. A spectrum of the \textit{IRIS} \ion{Fe}{21}~line at $\approx$~17:03:12\,UT is shown in in the first panel of Fig.\,\ref{Fig:Fits_FeXXI}. The line profile has been obtained by averaging over the slit pixels between the horizontal blue lines indicated in Fig.\,\ref{Fig:Overview_IRIS}. At 17:03\,UT, the intensity of the \ion{Fe}{21}~emission is very weak, around 10 DN (data numbers). The line profile appears almost at rest (10 km\,s$^{-1}$) and is slightly broadened with a width of 0.56\,\AA, corresponding to a non-thermal width of around 48~km\,s$^{-1}$.

The red contours overlaid on the AIA and SJI images represent the intensity contours (70 and 90 $\%$ of the maximum value) of the 6-12 keV sources observed by \textit{RHESSI} \citep{Lin02} during 17:02:48 -- 17:03:00\,UT. These sources coincide with the footpoints of the hot loops rooted in the two flare ribbons, as expected from the thick-target flare heating model \citep{Brown71}. Unfortunately, we are not able to identify any X-ray sources at 17:15\,UT onwards because of the spacecraft night.

At 17:14\,UT, we observe blue-shifted ($\approx$90\,km\,s$^{-1}$) \ion{Fe}{21}~emission in the \textit{IRIS} \ion{O}{1}~detector image as shown in the third panel of Fig.\,\ref{Fig:Overview_IRIS}. At this time, the \textit{IRIS} slit was crossing a bright portion of the ribbon NR visible in both the SJI 1400~\AA~filter and the AIA 131~\AA~bandpass. This portion of the ribbon corresponds to footpoints of a series of hot flare loops connecting PR and NR. That is, the \textit{IRIS} \ion{Fe}{21}~emission is now formed at a ribbon rather than at loop top, meaning that the local magnetic field along which the \ion{Fe}{21}~emission originates is more vertical than at 17:03 UT. The corresponding \ion{Fe}{21} spectrum, shown in the second panel of Fig.\,\ref{Fig:Fits_FeXXI}, is indeed more strongly blue-shifted. We note that the observation of the \ion{Fe}{21}~emission during the early phase of the flare (see Table \ref{Table:1}) has not been reported in the previous studies of the same flare event by \citet{Tian15,Graham15}, which were concerned with the analysis of the strong blueshifts observed from around 17:25~UT onwards.

From about 17:25\,UT, i.e., during the start of the impulsive phase characterized by the onset of the fast eruption (Fig.\,\ref{Fig:Eruption_stackplots}), the \textit{IRIS} slit crosses the NR ribbon at two different locations. Strong \ion{Fe}{21}~emission is detected only at the northern crossing with the NR ribbon, located at around Solar $Y$ $\approx$\,120$\arcsec$. The line profile is plotted in the third panel of Fig.\,\ref{Fig:Fits_FeXXI}. The Gaussian fit shows a strong blue shifted centroid position of about 200 km\,s$^{-1}$ and a very broad line width ($\approx$0.81~\AA~or a non thermal width of $\approx$\,90km\,s$^{-1}$). Few minutes later at about 17:31\,UT, we observe a line profile which is even more blue shifted ($\approx$270\,km\,s$^{-1}$) and extends outside the \ion{O}{1}~spectral window, as reported by \citet{Tian15} and \citet{Graham15}.

In summary, the \ion{Fe}{21}~line is at first weakly blue-shifted and few minutes later becomes more shifted as the ribbon NR enters the position of the \textit{IRIS} slit. We interpet this as the consequence of the slipping reconnection along the NR that is itself developing. We note that although the 1D hydrodynamical simulations of a flare loop predict a rapidly increasing blue shift before the maximum blue shift velocity is reached \citep[][]{Fisher85a,Polito16}, this is strictly true only for one ribbon kernel. In the observations, this effect is compounded by the observed slipping motion of the flare loop footpoints, which are responsible for the changing inclination of the local magnetic field at the position of the \textit{IRIS} slit. Therefore, the evaporating plasma would show different Doppler shifts along the line of sight. Unfortunately, in the present study we cannot reliably distinguish between the geometric effects and the evolution of the evaporation velocities itself. This is due to the lack of the stereoscopic spectral observations.

In addition, we note that the \ion{Fe}{21}~blue-shifted emission is located a few pixels just above the ribbon position where the intensity of the FUV continuum and the cooler emission lines is enhanced, as already observed by \citet{Young15} in the study of another X-class flare. This can be best seen in the third and fourth panels of figure \ref{Fig:Overview_IRIS}, showing that the blue lines (indicating the location of the \ion{Fe}{21}~blueshifts) are clearly above the position where transition region lines included in the \ion{Si}{4}~window are more intense. We note however that such non-cospatiality of only several \textit{IRIS} pixels would not have been resolved by AIA, whose spatial resolution is 1.5$\arcsec$. Therefore, the conclusion reached in Sect. \ref{Sect:2.3} that the AIA ribbon as seen in 304\,\AA~is cospatial with the footpoints of the 131\,\AA~loops remains valid within the AIA resolution.

Later on during the flare, the \ion{Fe}{21}~line profile gradually moves to the rest-position as the flare loops are being filled by the high temperature evaporating plasma. The fourth panel in Fig.~\ref{Fig:Fits_FeXXI} shows the \ion{Fe}{21}~spectrum being almost at rest as observed by the \textit{IRIS} slit at around 17:48\,UT, i.e., after the onset of the gradual phase.

%
\begin{figure*}[!ht]
	\centering
%
	\includegraphics[width=5.90cm,clip]{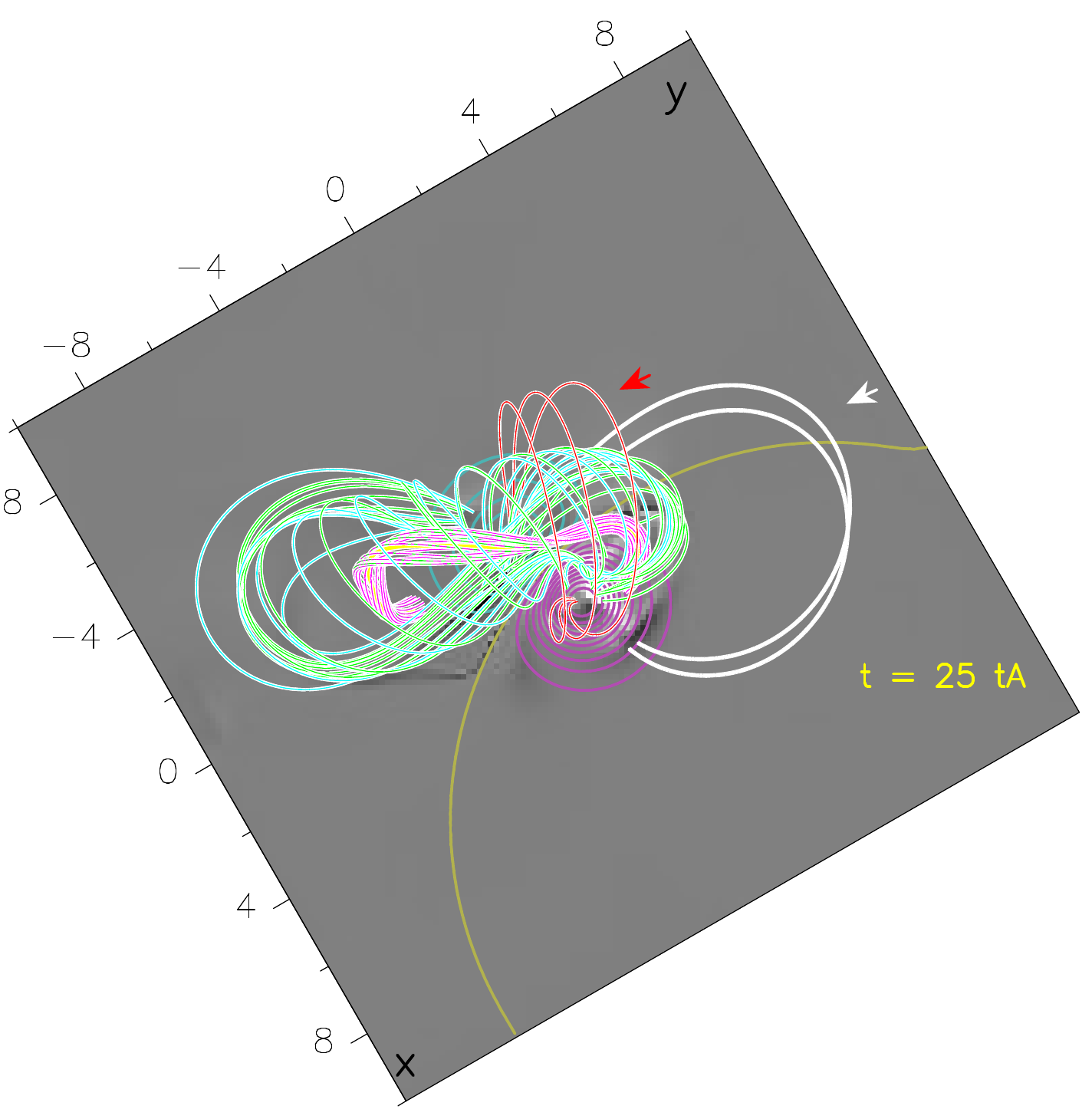}
	\includegraphics[width=5.90cm,clip]{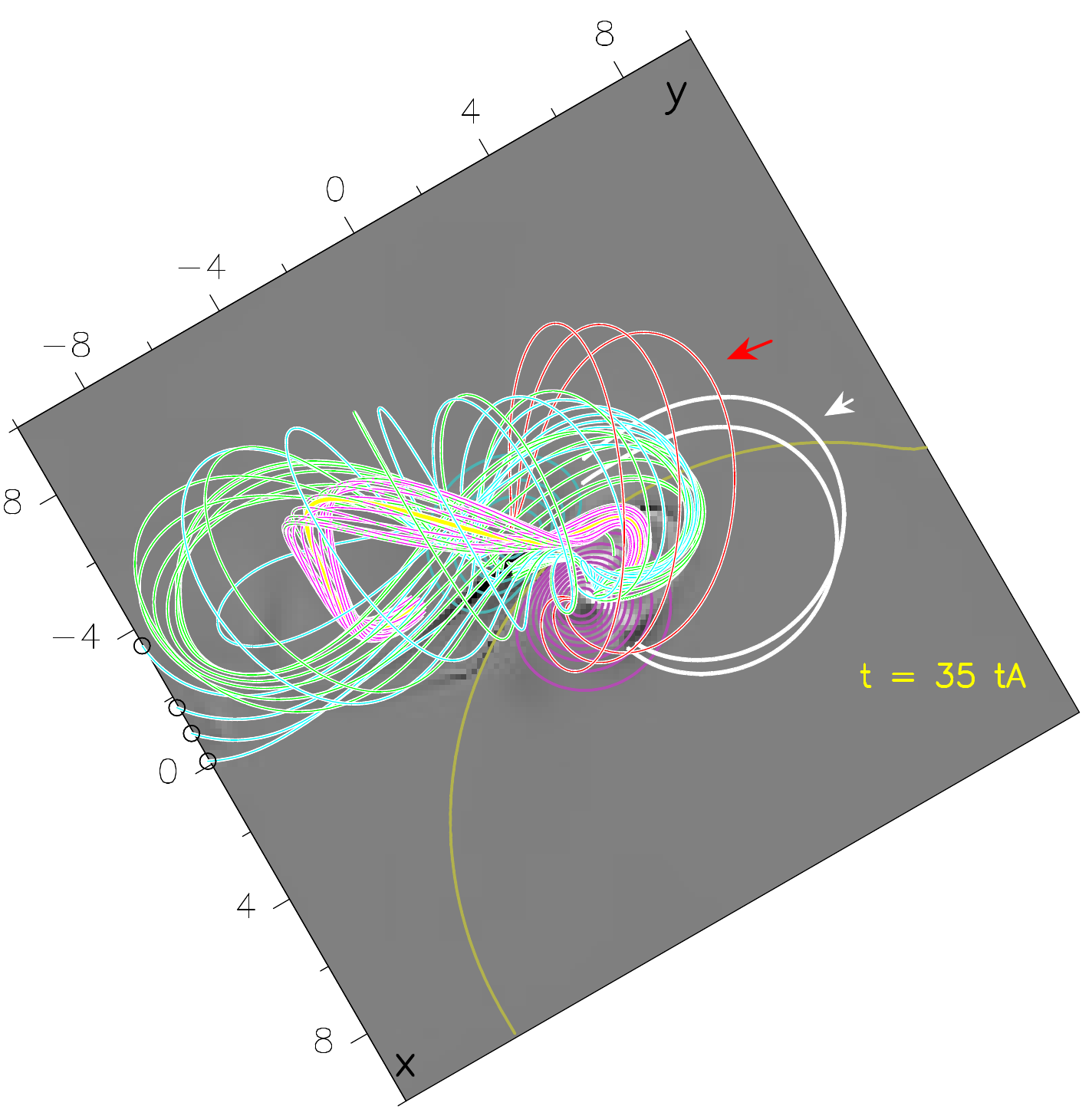}
	\includegraphics[width=5.90cm,clip]{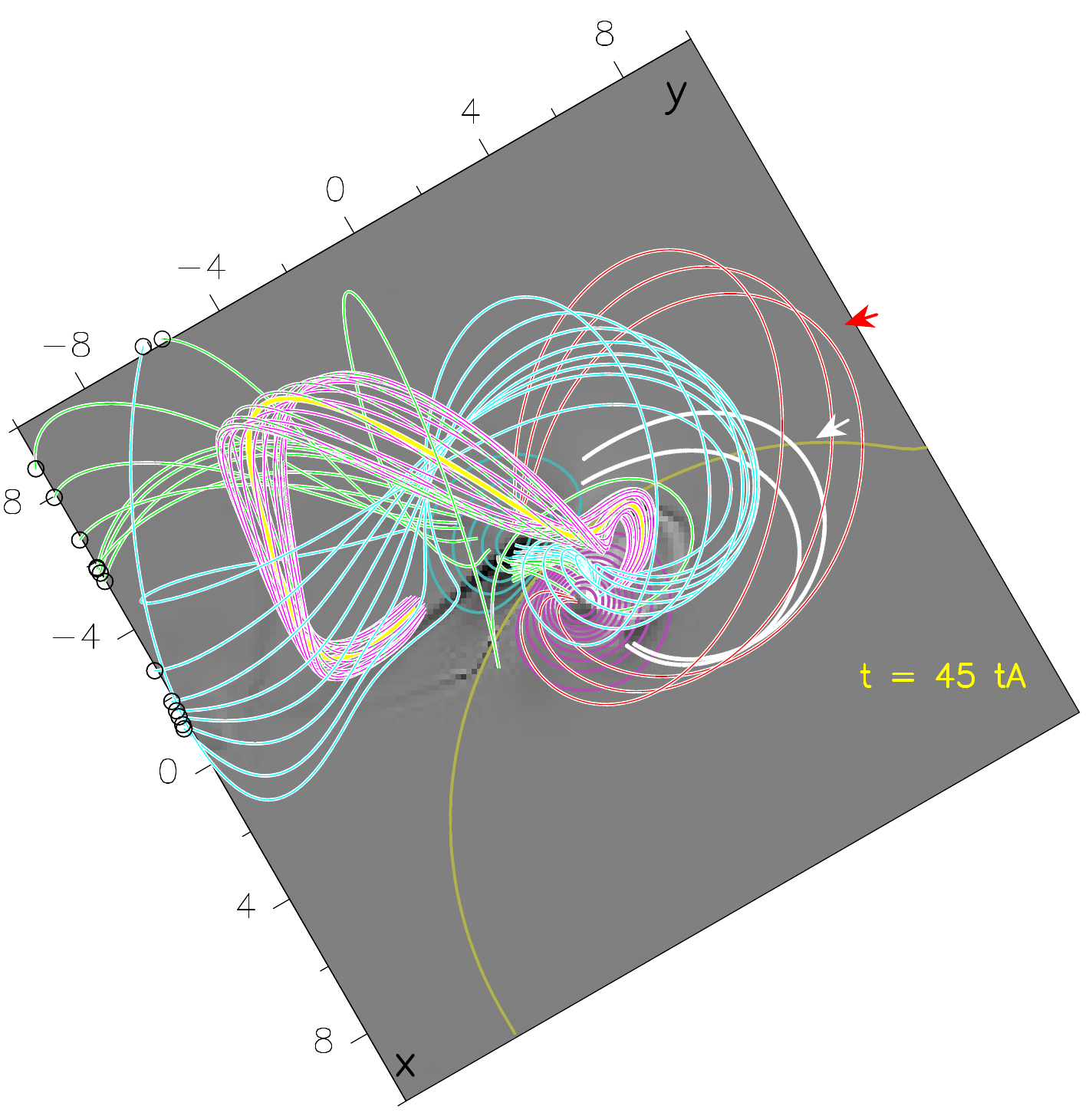}
	\includegraphics[width=5.90cm,clip]{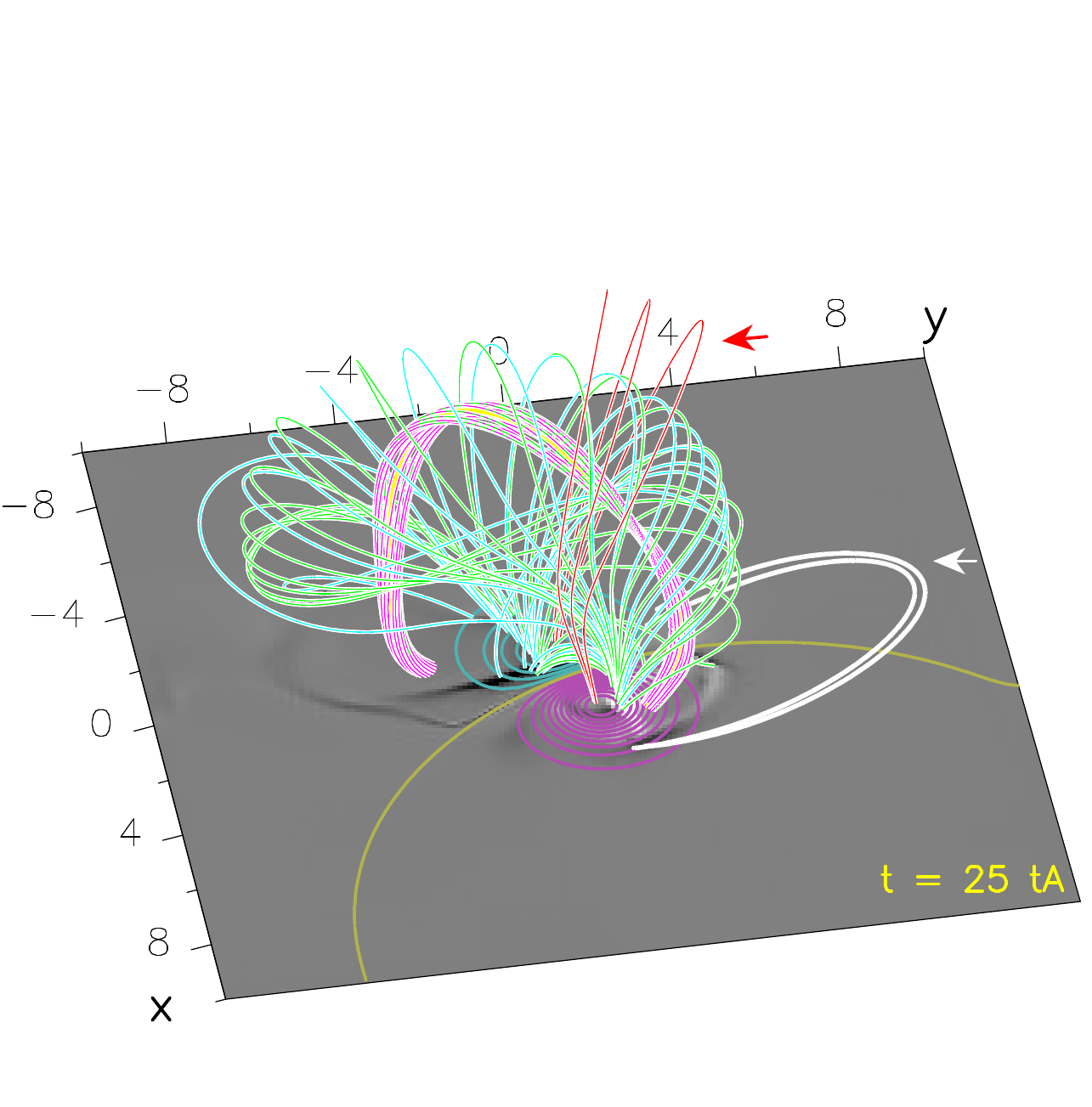}
	\includegraphics[width=5.90cm,clip]{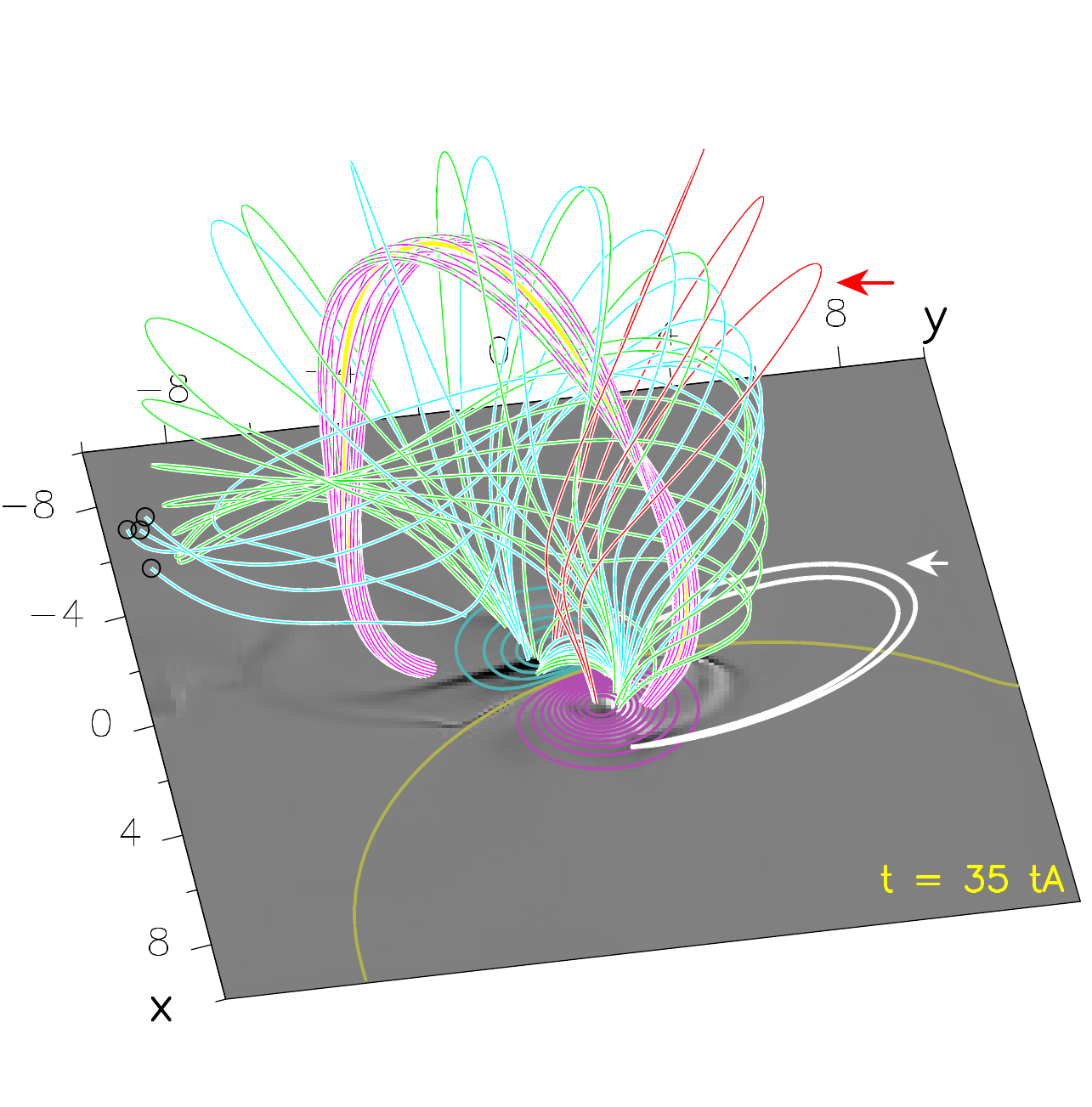}
	\includegraphics[width=5.90cm,clip]{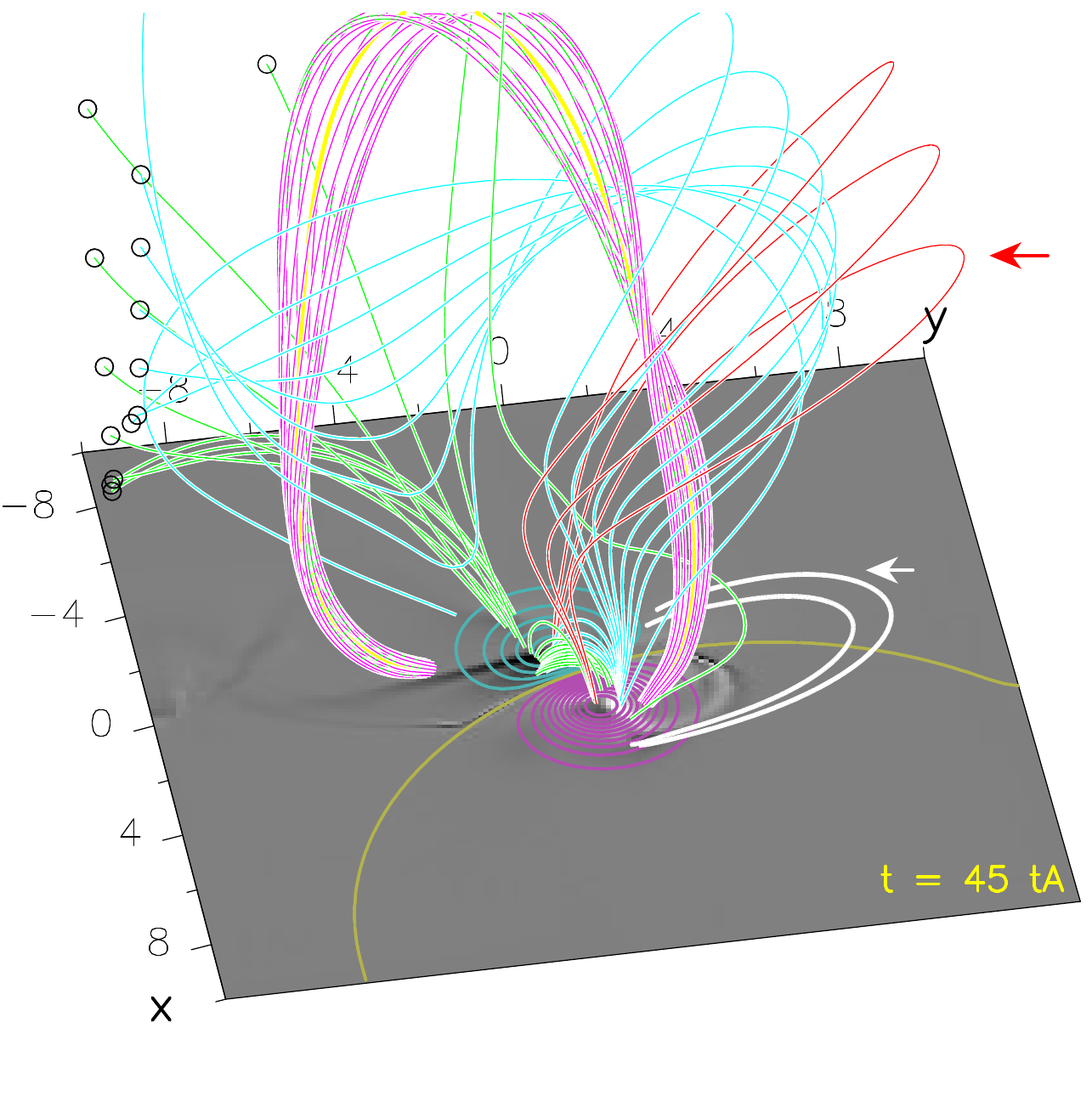}
\caption{Dynamics of loops surrounding the unstable flux rope from the standard solar flare model in 3D of \citet{Aulanier12,Aulanier13}, and \citet{Janvier13}. The flux rope core is depicted by pink, while its envelope is depicted by green and light blue field lines. Series of overlaying loops pushed by the flux rope expansion are shown in red. Highly inclined contracting loops are shown in white. The greyscale shows the $z$-component of the electric current $j_z(z=0)$ in the photospheric plane, while contours stand for the vertical component of the magnetic field $B_z(z=0)$. The time indicated is given in Alfv\'en times $t_\mathrm{A}$ in the model. \\
A color version of this image is available in the online journal.
\label{Fig:Topotr3D}}
\end{figure*}
%

%
\subsection{O IV and Si IV lines}
\label{Sect:3.3}

The \ion{O}{4}~1401.16~\AA, \ion{O}{4}~1399.77~\AA~and \ion{Si}{4}~1402.77~\AA~transition lines observed by \textit{IRIS} provide various diagnostic opportunities, but are affected by several complexities in their interpretation. The response of these transition-region lines in the ribbons during a flare is well-known to be strongly dependent on non-equilibrium ionization effects \citep{Bradshaw04,Doyle13,Olluri13} as well as on the non-Maxwellian electron distributions \citep{Dudik14b}. These lines can also be blended with unidentified photospheric or chromospheric transitions \citep{Polito16} during the impulsive phase. 

The \ion{O}{4}~emission is usually very low in active region spectra but can be enhanced during the impulsive phase of flares. Here, we have estimated the ratio between the \ion{O}{4}~1401.16~\AA~and \ion{Si}{4}~1402.77~\AA~lines at particular times where the \ion{O}{4}~emission was high enough to be reliably measured. The \ion{Si}{4}~line is often saturated during the impulsive phase and thus only an upper limit of the ratio can be obtained. We have integrated the line intensity after background subtraction and calibrated the values in physical units (erg\,s$^{-1}$\,sr$^{-1}$ cm$^{-2}$\,\AA$^{-1}$).
For instance, at 17:30:17 UT we find a ratio of \ion{O}{4}~1401.16~\AA~and \ion{Si}{4}~1402.77~\AA~equal to 0.03 at the slipping footpoint of the NR crossed by the \textit{IRIS} slit at that time (see Movie 6). Similar values are found throughout the impulsive phase.

In addition, the ratio of the \ion{O}{4}~1401.16~\AA~and 1399.77~\AA~lines \textit{IRIS} is sensitive to the electron density of the plasma. Throughout the impulsive phase of the flare we measure a \ion{O}{4}~\AA 1401.16\,/\,1399.77\,\AA~ratio which is below the high density limit of 2.5 reported by CHIANTI v7.1, assuming equilibrium condition. In particular, the ratio is equal to 2.19 at 17:30:07\,UT, and would indicate a density of at least $10^{12}$ cm$^{-3}$ in equilibrium. These line ratios are also consistent with lower densities and non-Maxwellian electron distributions in the flare plasma \citep{Dudik14b}. 


%
\section{Discussion}
\label{Sect:4}

Having demonstrated that the occurrence of slipping reconnection is not inconsistent with a presence of a wide range of dynamical phenomena during the flare, including chromospheric evaporation, hard X-ray emission, loop expansion, contraction, and oscillations, as well as the occurrence of eruptions, we now discuss these observations both in terms of the standard solar flare model in 3D of \citet{Aulanier12} and \citet{Janvier13}, as well as in the light of previous observational results and clues.

%
\subsection{Expanding and contracting loops in the standard solar flare model in 3D}
\label{Sect:4.1}

We first examined the standard solar flare model in 3D for signatures of loop expansion and contraction. Although this model is generic and its photospheric flux distribution does not represent the active region under study here, we found that both processes are present. In Fig.\,\ref{Fig:Topotr3D}, the flux rope is depicted by pink, green, and cyan field lines. The pink field lines represent the flux rope core, while the green and cyan field lines represent the S-shaped envelope that is created as a result of the slipping reconnection \citep{Aulanier12,Janvier13,Dudik14a}. The loops overlying the flux rope are shown in red and white. The white loops are highly inclined, while the red ones are nearly vertical. Both of these loops systems are anchored in the leading positive-polarity spot in the model \citep[see also][]{Aulanier12}. 

During the course of the eruption, the unstable flux rope pushes the red overlying loops, causing them to expand and move sideways. This behaviour persists for several tens of Alfv\'en times. Although the simulation is dimensionless \citep{Aulanier12,Janvier13},  taking an order-of-magnitude estimate of the Alfv\'en speed $v_\mathrm{A}$\,$\approx$\,10$^3$\,km\,s$^{-1}$, and the loop length of 100\,Mm, the duration of the expansion would be about 10$^3$\,s, in broad agreement with the observed duration of the loop expansion during the slow rise of the flux rope. Contrary to the red field lines, the highly inclined white field lines undergo a contraction. This is probably because the stretching of the legs of the flux rope during its eruption leads to a local decrease of magnetic pressure in the legs the flux rope, leading to a contraction of the neighbouring loops.

The expanding and contracting behaviour which we have observed is therefore consistent with the predictions of the standard solar flare model in 3D. This also confirms our earlier conclusion (Sect. \ref{Sect:2.2.3}) that the coronal implosion does not occur during this flare.

Finally, the observed oscillatory behaviour of the field lines (Sect. \ref{Sect:2.2.3}) after the fast eruption is not reflected in the simulation. The reason for this is the relative shortness of the calculation that does not allow the flux rope to leave far 
enough for the overlying loops to have time to reach and bounce from the central part of AR. Additionally, the viscosity in the model may be too large in the locations of the loops, where the computational mesh is stretched.

%
\begin{figure}
	\centering
	\includegraphics[width=0.5\textwidth]{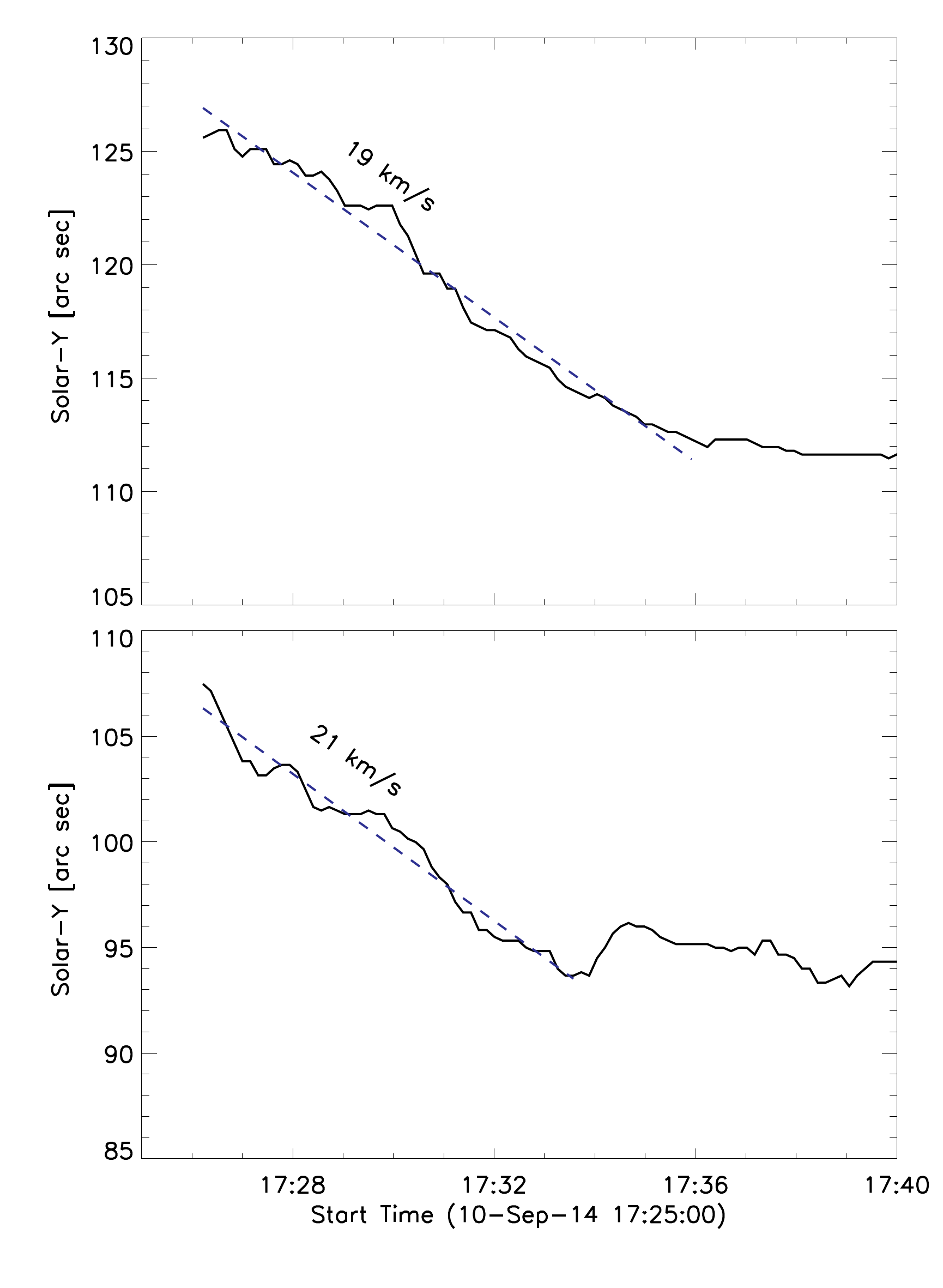}
\caption{Velocity of the footpoints slipping motion at the NR ribbon as estimated by using the \textit{IRIS} FUVS and FUVL spectra. The \textit{top} (\textit{bottom}) image corresponds to the northern (southern) portion of NR crossed by the \textit{IRIS} slit.}
\label{Fig:ribbon_position}
\end{figure}
%
\subsection{Connectivity norm in flare-related QSLs}
\label{Sect:4.2}

The standard solar flare model in 3D also predicts that the apparent slipping velocities of the magnetic field lines $v_\mathrm{slip}$ and the outward velocity of the conjugate ribbon $v_\mathrm{QSL}$, i.e., speed of the ribbon movement perpendicular to the polarity inversion line, are related. The relation is linear, with a proportionality constant given to a first-order approximation by $N$, the local norm of the field line mapping 
\begin{equation}
 v_\mathrm{slip} = N v_\mathrm{QSL}\,,
 \label{Eq:Norm}
\end{equation}
see Eq. (4) in \citet[][]{Janvier13}.

In principle, this equation permits measurement of $N$ if $v_\mathrm{slip}$ and $v_\mathrm{QSL}$ are known, at least within a time scale where the QSLs are not so much evolving. In this work, we have measured the $v_\mathrm{slip}$ during the early and impulsive flare phase and found that this velocity is typically $\approx$20--40 km\,s$^{-1}$ in both ribbons (Sect. \ref{Sect:2.3}).

To estimate the $v_\mathrm{QSL}$, we have first determined the evolution of the Solar $Y$ position of the ribbon NR at the location of the \textit{IRIS} slit (Fig.\,\ref{Fig:ribbon_position}). This is done during the time interval 17:25 -- 17:40\,UT, i.e., during the impulsive phase, when the ribbon is well defined. The locations of both the northern and the southern branch of the NR are determined as the locations of maximum intensity of the FUV continuum as observed in the FUVS \ion{O}{1}~and FUVL \ion{Si}{4}~spectral windows. Using the \textit{IRIS} spectra at a given sit-and-stare slit position has the advantage of a very high cadence of about 9\,s compared to the \textit{IRIS} SJI or AIA images. The ribbon displacements determined using this method are not subject to confusion between the motion of the ribbon itself and the slipping motion, which would need to be separated out if the ribbon position were determined from imaging data. The position of both branches of the NR ribbon shows a linear trend during the impulsive phase of the flare, as shown in Fig.\,\ref{Fig:ribbon_position}. The projected velocity $v_\mathrm{QSL,proj}$ along the \textit{IRIS} slit is about 19 km\,s$^{-1}$ for the northern branch of the NR, and has a similar value for the southern branch. The outward ribbon velocity is then obtained as $v_\mathrm{QSL}$\,=\,$v_\mathrm{QSL,proj}$\,/\,sin$(\alpha)$, where $\alpha$\,$\approx$\,41$^\circ$ is the approximate angle between the northern branch of the NR and the \textit{IRIS} slit at 17:30\,UT (Fig.\,\ref{Fig:Overview_IRIS}). This yields $v_\mathrm{QSL}$\,$\approx$\,29 km\,s$^{-1}$.

Using these values, from Eq. \ref{Eq:Norm} we obtain $N$\,$\approx$\,0.7--1.4, which is very low for a QSL. Even using the highest reported value of $v_\mathrm{slip}$\,=\,200~km\,s$^{-1}$ for this flare \citep{Tian15} would yield only $N$\,$\approx$\,6.9. This poses a significant problem, since the typically expected value for a QSL would be at least several tens or hundreds due to the high squashing of the magnetic flux-tubes in the QSL \citep{Titov02}. The presence of a QSL and the associated current density enhancement \citep{Masson09,Wilmot09} is a necessary condition for the occurrence of slipping reconnection.

Why do we then detect the apparent slipping motion of the flare loops and low values of $N$? In a strict sense, Eq. (\ref{Eq:Norm}) is valid only for a given field line in a given instant in time, with one footpoint anchored in one polarity-related QSL moving at speed $v_\mathrm{QSL}$, while the opposite moving end moves in the conjugate ribbon at speed $v_\mathrm{slip}$, with each quantity deduced on a short timescale, as the profile of the local norm $N$ varies strongly along and across the QSL \citep[see Fig.\,10d in][]{Janvier15}. It is difficult to find a satisfactory relation between the analytical expression in its strict sense and the present data as both speed quantities are obtained only as averages over multiple locations, multiple times and multiple field lines, which are themselves anchored in different sections of the QSL photospheric footprints. This is difficult to avoid as the remote-sensing observations generally rely on plasma emission, which in optically thin conditions is always dominated by dense(r) plasma. While the Eq. (\ref{Eq:Norm}) is only valid strictly for a given field line in a given instant in time, in the observations there also exists a time delay between the energy deposition in the ribbon and the filling of the flare loops by hot and dense evaporated plasma. Indeed, in Sect. \ref{Sect:3.2}, it was found that the \ion{Fe}{21}~blue-shifted emission observed by \textit{IRIS} is not exactly co-spatial with the ribbon as seen in transition-region \ion{Si}{4}~emission or the FUV continuum (Fig.\,\ref{Fig:Overview_IRIS}). Therefore, we may be measuring the slipping speed slightly outside of the QSL, which may have already passed to an adjacent location.

Furthermore, the $v_\mathrm{slip}$ determined from the apparent motion of a flare loop may not represent the slipping velocities of a single field line footpoint, which could even be super-Alfv\'enic \citep{Aulanier06,Aulanier12,Janvier13}. Rather, the observed slow, sub-Alfv\'enic slipping velocities may be an illusion resulting from variable rate of the energy deposition to the chromosphere as a result of many slipping field lines along the ribbon. In any case, large, super-Alfv\'enic slipping velocities are out of the reach of current observations, which only have cadence of the order of 10\,s. Sub-second temporal resolution would be needed to distinguish fast slipping motion. An additional complication is that large $v_\mathrm{slip}$ would result in lower  energy deposition per unit time and unit area of the ribbon, resulting in less evaporated plasma, i.e., fainter loops. We note that faint near-vertical stripes are present in the time-distance plots in Figs. \ref{Fig:NR_stackplots_slit2a} and \ref{Fig:PR_stackplots}. E.g., a short, near-vertical strip is located at 17:03:40\,UT at the position Solar $Y$\,=\,100--103$\arcsec$ in Fig.~\ref{Fig:NR_stackplots_slit2a}. Nevertheless, the 12\,s cadence of AIA does not permit measurement of the velocity of such an intermittent stripe.

\begin{figure}[!ht]
	\centering
	\includegraphics[width=8.8cm]{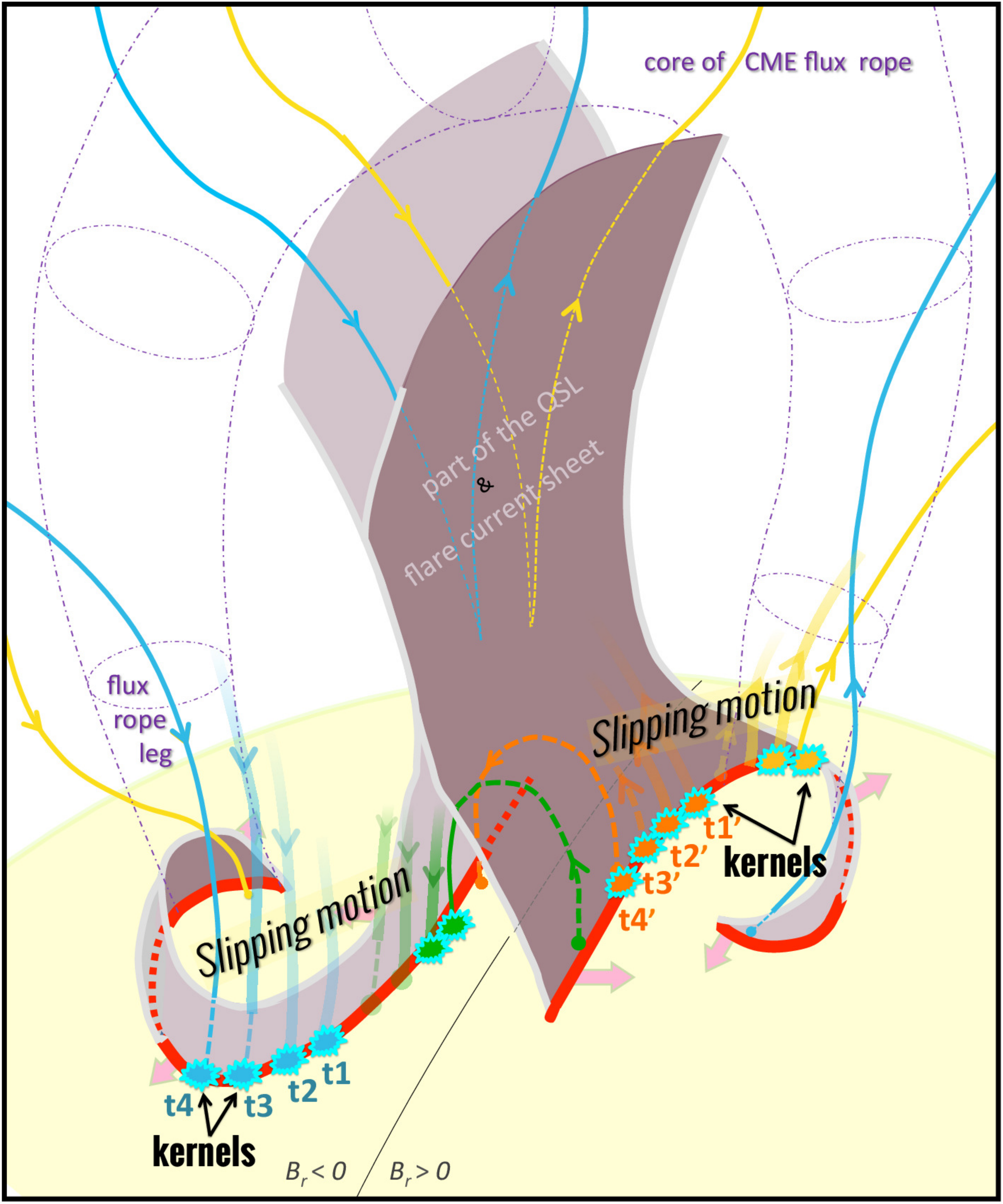}
\caption{Cartoon showing the standard solar flare model in 3D with the tether-cutting nature of the slipping reconnection and the relation to the kernels (green circles) along the flare ribbons (red lines) for different times (t1--t4). Blue and  yellow lines are examples of field lines wrapping the flux rope. Orange and green lines are reconnected field lines (flare loops). See text for more details. \\
A color version of this image is available in the online journal.}
\label{Fig:3D_cartoon}
\end{figure}
%
%
\subsection{Relation of the precursors to the slipping reconnection}
\label{Sect:4.3}

\citet{Chifor07} studied the precursor activity before major flares and found that distinct, localized X-ray brightenings occur 2--50 minutes before the onset of the impulsive phase. These brightenings occured within 10$\arcsec$ of the polarity inversion line (PIL) and had both a thermal and non-thermal component as observed by \textit{RHESSI}. Typically, these X-ray brightenings had also an EUV counterpart observed either by SoHO/EIT \citep{Delaboudiniere95} or TRACE \citep{Handy98}. The main energy release in the impulsive phase occured to within 50$\arcsec$ of the locations of the preflare brightenings. Furthermore, the filament eruption began at the location of the preflare brightenings that also coincided with the presence of emerging or cancelling flux. \citet{Chifor07} interpreted these as the signatures of tether-cutting mechanism \citep[e.g.,][]{Moore92,Moore01,Moore11,Fan12} triggering the main flare and the eruption.

Inspecting the HMI and AIA 304\,\AA~data, we find that \textit{all} these observational characteristics are present during the early flare phase at approximately 16:50--17:20\,UT. A weak X-ray emission is detected in the 6--12\,keV channel by RHESSI, an example of which was shown in Fig.\,\ref{Fig:Overview_IRIS} at 17:03\,UT. The advantage of \textit{IRIS} and \textit{SDO}/AIA is that they offer a good temperature coverage of the solar atmosphere from the chromosphere upwards, including flaring plasmas. The SJI 1400\AA~together with the AIA 1600\AA, 1700\AA, 304\,\AA, and 171\,\AA~images reveal that the EUV brightenings occur within the ribbons involved in the main flare. In fact, these EUV brightenings are the first instances of the ribbon development \citep[similarly as in][]{Dudik14a}, with a one-to-one correspondence to footpoints to the flare loops exhibiting apparent slipping motion (Sect. \ref{Sect:2.3}). This leads us to conclude that at least for the flare studied here, the \textit{precursors are in fact signatures of the flare itself, progressing from its early phase towards the impulsive phase}. This is in line of the results of \citet{Farnik98} who concluded that for the preflare events (precursors) cospatial with the main flare, the soft X-ray emission is present with the same size, shape, and orientation at least 5 minutes before the onset of the impulsive phase.

Furthermore, the standard solar flare model in 3D of \citet{Aulanier12} and \citet{Janvier13} predicts explicitely that the slipping reconnection creating the flare loops and building the flux rope is the tether cutting mechanism itself. An important distinction however is \citep[see Sect. 5 of][]{Aulanier12} that this tether-cutting does not trigger the eruption itself as in the cartoon of \citet{Moore01}; rather, it contributes to the building of the flux rope which then erupts via the torus instability. We note that the presence of tether-cutting reconnection was already suggested independently for this event by \citet{Cheng15}, who analyzed AIA and \textit{IRIS} \ion{Si}{4}, \ion{C}{2}, and \ion{Mg}{2}~observations.

The mechanism of the slipping reconnection and tether-cutting is represented in the cartoon shown in Fig.\,\ref{Fig:3D_cartoon}. This is essentially a 3D representation of the magnetic field configuration in the standard solar flare model in 3D, including the dynamics during the eruption of a flux rope \citep[see also][Fig.\,11 therein]{Janvier15}. The flux rope is represented by dash-dotted purple lines. It is anchored in each polarity in the hooked region of the flare ribbons (red), corresponding to the photospheric signature of the QSLs and the current density volume, represented in purple. The QSL wraps around the flux rope, with the hyperbolic flux tube, where the connectivity distortion is the highest, extending underneath the flux rope. Magnetic field lines entering this high current density volume reconnect successively in the slipping manner. Pairs of such field lines undergoing successive reconnections are represented by colored field lines in Fig.\,\ref{Fig:3D_cartoon}: Blue and yellow field lines have fixed footpoints in the positive and negative ribbons, respectively. Their reconnection counterparts are shown in green and orange, whose footpoints are again fixed in positive and negative ribbons, respectively. The blue field line slip-reconnected with the orange field line. The consequences are represented with the moving end of the blue field line in the negative polarity, away from the PIL and towards the legs of the flux rope. The slip-reconnected counterpart of this blue field line is the orange one, represented with a fixed footpoint anchored in the negative polarity in the left part of the cartoon. Its conjugate footpoint in the positive polarity is slipping and is accompanied by flare kernels moving toward the PIL. The other reconnecting pair are the yellow and green field lines, whose one footpoint is fixed in the negative and positive ribbon, respectively, while the conjugate footpoints move along the conjugate ribbons (Fig.\,\ref{Fig:3D_cartoon}). Together, these four field lines create motions towards both ends of both ribbons. This important prediction of the model is vindicated by the observations reported on in Sect. \ref{Sect:2.3}. These observations are in contrast to the first report of slipping motion of flare loops in another flare \citep{Dudik14a}, where the slipping motion was observed to be predominantly in one direction, towards the hooks of ribbons.

These successive reconnections represented in the cartoon in Fig.\,\ref{Fig:3D_cartoon} come to an end when the two field lines have gone across the current density volume. The orange field line becomes a low-lying flare loop, while the blue field now wraps around the erupting flux rope and becomes its envelope, likely observed as the hot S-shaped erupting loop. The slippage is likely accompanied by particle acceleration in the reconnecting region. These particles propagate along each successively connected field line. Deposition of energy in the chromosphere is seen as the kernel brightenings. These slipping brightenings (Figs. \ref{Fig:NR_AIA_slit2a} and \ref{Fig:PR_AIA}), corresponding to the precursors in the early flare phase, are shown as bright patches appearing at different times at different positions along both ribbons.

In summary, the precursor activity as a result of the tether-cutting reconnection is fully consistent with the standard solar flare model in 3D. The presence of the apparent slipping motion during the precursor phase, clearly identified here as such for the first time, is the tell-tale signature uniting the precursors and the 3D model. 

%
\section{Summary and conclusions}
\label{Sect:5}

We have reported on the observations of the 2014 September 10 X-class flare and the occurrence of slipping reconnection, large-scale dynamics of coronal loops, as well as the chromospheric evaporation and precursors. Our main conclusions are:
\begin{enumerate}
\item The apparent slipping motion occurs throughout the flare from the onset of the early phase at about 16:50 UT. This phase has not been studied in previous publications on this flare. The characteristic velocities of individual apparently slipping loops are 11--57\,km\,s$^{-1}$, typically 20--40\,km\,s$^{-1}$ independently of the ribbon. The slipping reconnection proceeds in both directions along both ribbons, fullfilling the prediction of the standard solar flare model in 3D. This is in contrast to the reports of \citet{Dudik14a} for another X-class flare, where the slipping motion was predominantly in one direction, towards the hooks of both ribbons. We point out however that the model-predicted apparent slipping velocities are faster, well out of the reach of the current instrumentation. It remains to be seen whether such fast velocities exist, or are reduced by some dissipative processes. \item The evolution of the ribbon NR is complex, as it exhibits squirming motions, during which the slipping motion proceeds in almost transversal direction to the direction of the ribbon extension at the later time.
\item The ribbons observed by AIA 304\,\AA, 1600\AA, and 1700\AA~correspond to the footpoints of the 131\,\AA~flare loops exactly. We found that in the bright kernels within the ribbon, the AIA 1600\AA~signal can be enhanced by more than a factor of 3 compared to the AIA 1700\AA, probably due to the strong \ion{C}{4}~component. In the \textit{IRIS} spectra, the strong \ion{Si}{4}~intensities and the relatively weak \ion{O}{4}~lines, together with the \ion{O}{4}~ratios indicate high densities (above 10$^{12}$ cm$^{-3}$). More detailed studies are needed to confirm this. 
\item A failed eruption of the filament F2 is followed by an eruption of hot S-shaped loops observed in the AIA 131\,\AA~channel at the same location later on. This eruption shows non-linear acceleration to projected velocities of more than 270 km\,s$^{-1}$. We interpret this as an eruption of a double-decker flux rope, where the lower deck consists of the F2 and possibly also F1. The F2 which undergoes a failed eruption with possible flux transfer to the upper deck, which is visible as the erupting hot S-shaped loops. In terms of the standard solar flare model in 3D, these S-shaped loops represent the envelope of the torus-unstable erupting flux rope, fed by the ongoing slipping reconnection.
\item In the pre-flare phase before the hot eruption, several of the peripheral warm coronal loops belonging to the same active region exhibit either expanding or contracting motions. The projected velocities of these motions are $-$2.9\,$\pm$0.9 to +21.4\,$\pm$2.1\,km\,s$^{-1}$. In terms of the standard solar flare model in 3D, we interpret these expanding and contracting motions as displacement of the coronal loops by the growing and erupting flux rope.
\item After the hot eruption, a number of coronal loops exhibit contracting motions and subsequent oscillations with periods of several minutes. This behaviour precludes the coronal implosion as the primary energy release mechanism, since the hot flux rope exists and erupts \textit{before} the loops contract and the oscillations set in. Rather, we propose that the apparent implosion is a result of the large-scale dynamics involving the flux rope eruption.
\item The loop oscillations are also detected as a modulated radio flux at the frequency of 350\,MHz. The radio flux is modulated in phase with the loop oscillations. To our knowledge, this is the first such observation.
\item Chromospheric evaporation in the \ion{Fe}{21}~line observed by the \textit{IRIS} instrument shows gradual increase of the blue-shift velocities during the early flare phase. This increase can be at least partially explained by the changing geometry of flare loops at the position of the \textit{IRIS} slit as a result of the slipping reconnection. At first, the \textit{IRIS} slit crosses the top portions of flare loops, while later on, the \ion{Fe}{21}~emission is dominated by bright flare loop footpoints. The highest velocities of $\approx$266\,km\,s$^{-1}$ are detected in the impulsive phase when the IRIS signal is dominated by the footpoint emission. Although the line is visible during most of the early and impulsive phases, a detailed study of the evaporation during the beginning of the early flare phase is somewhat limited by the sensitivity of the \textit{IRIS} instrument, making the \ion{Fe}{21}~line hardly visible or undetectable.
\item In the early flare phase, the precursor activity including \textit{RHESSI} 6--12 keV sources is detected and found to be fully consistent with the standard solar flare model in 3D. These precursors, interpreted previously as signatures of the tether-cutting reconnection, are identified here with the flare itself, progressing from the early phase towards the impulsive phase. In terms of the standard solar flare model in 3D, the tether-cutting mechanism is provided by the slipping reconnection.
\end{enumerate}
In conclusion, this in-depth and comprehensive study of an X-class flare, observed with several different instruments and also in the radio, confirms most of the predictions of the 3D standard flare model.

\begin{acknowledgements}
We thank the referee for remarks that led to improvements of the manuscript. J.D. and E.Dz. acknowledge the Grant P209/12/1652 of the Grant Agency of the Czech Republic. V.P. acknowledge support from the Isaac Newton Studentship and the Cambridge Trust. G.D.Z. and H.E.M. acknowledge STFC funding through the DAMTP astrophysics grant. J.D., V.P., G.D.Z., and H.E.M. also acknowledge support from the Royal Society via the Newton Alumni Programme. M.K. acknowledges Grant Agency of the Czech Republic, Grant No. P209/12/0103. J.D., E.Dz., and M.K. also ackowledge institutional support RVO:67985815 of the Czech Academy of Sciences. M.J. wishes to acknowledge the hospitality provided by DAMTP, University of Cambridge, during her visits.
AIA and HMI data are courtesy of NASA/SDO and the AIA and HMI science teams. \textit{IRIS} is a NASA small explorer mission developed and operated by LMSAL with mission operations executed at NASA Ames Research Center and major contributions to downlink communications funded by the Norwegian Space Center (NSC, Norway) through an ESA PRODEX contract. Hinode is a Japanese mission developed and launched by ISAS/JAXA, with NAOJ as domestic partner and NASA and STFC (UK) as international partners. It is operated by these agencies in cooperation with ESA and NSC (Norway). CHIANTI is a collaborative project involving the NRL (USA), the University of Cambridge (UK), and George Mason University (USA). The authors acknowledge the usage of full disk H$\alpha$ data from the Big Bear Solar Observatory, New Jersey Institute of Technology.
\end{acknowledgements}


\bibliographystyle{apj}         
\bibliography{Slip_Xflare}   

%
%

%

\end{document}